\documentclass[a4paper,fleqn,usenatbib,useAMS]{mnras}

\usepackage[T1]{fontenc} 
\usepackage{ae,aecompl} 

\usepackage{graphicx} 
\usepackage{amsmath} 
\usepackage{amssymb} 
\usepackage{natbib} 
\usepackage{subfig} 
\usepackage{float} 
\usepackage{lscape} 
\usepackage{booktabs} 
\usepackage{color} 
\usepackage{xfrac} 
\usepackage{multirow,bigdelim} 
\usepackage[multidot]{grffile} 
\usepackage{multicol} 
\usepackage{ragged2e} 
\usepackage{siunitx} 
\usepackage{tabulary} 
\usepackage{mathtools} 
\usepackage{cuted} 
\usepackage{graphbox} 

\usepackage{rotating} 
\makeatletter 
\newcommand\footnoteref[1]{\protected@xdef\@thefnmark{\ref{#1}}\@footnotemark}
\makeatother

\hypersetup{draft=False, colorlinks, citecolor=blue, linkcolor=blue, urlcolor=blue} 
\definecolor{green}{rgb}{0.0,0.725,0.024}

\newcommand{\hersc}{{\it Herschel}}
\newcommand{\planck}{{\it Planck}}
\newcommand{\spitz}{{\it Spitzer}}
\newcommand{\montage}{{\sc montage}}

\newcommand{\HI}{H{\sc i}}
\newcommand{\logOH}{$12 + {\rm log}_{10} [ \frac{\rm O}{\rm H}]$}
\newcommand{\kappad}{$\kappa_{d}$}
\newcommand{\kappamu}{$\kappa_{500}$}
\newcommand{\epsilond}{$\epsilon_{d}$}

\newcommand{\comments}[1]{\@bsphack\@esphack}

\title[The First Maps of \kappad]{The First Maps of \kappad\ -- the Dust Mass Absorption Coefficient -- in Nearby Galaxies, with DustPedia}

\author[C.J.R. Clark et\,al.] {\parbox{\textwidth}{C.\,J.\,R.\,Clark$^{\star 1}$, 
P.\,De\,Vis$^{2}$,
M.\,Baes$^{3}$,
S.\,Bianchi$^{4}$,
V.\,Casasola$^{4,5}$,
L.\,P.\,Cassar\`a$^{6}$,
J.\,I.\,Davies$^{2}$,
W.\,Dobbels$^{3}$,
S.\,Lianou,
I.\,De\,Looze$^{3,7}$,
R.\,Evans$^{2}$,
M.\,Galametz$^{8}$,
F.\,Galliano$^{8}$,
A.\,P.\,Jones$^{9}$,
S.\,C.\,Madden$^{8}$,
A.\,V.\,Mosenkov$^{10}$,
S.\,Verstocken$^{3}$,
S.\,Viaene$^{3,11}$,
E.\,M.\,Xilouris$^{12}$,
N.\,Ysard$^{9}$}\\
\\
{\parbox{\textwidth}{$^{1}$\ Space Telescope Science Institute, 3700 San Martin Drive, Baltimore, Maryland, 21218, USA\\
$^{2}$\ School of Physics \& Astronomy, Cardiff University, Queen's Buildings, The Parade, Cardiff, CF24 3AA, UK\\
$^{3}$\ Sterrenkundig Observatorium, Universiteit Gent, Krijgslaan 281 S9, B-9000 Gent, Belgium\\
$^{4}$\ INAF, Osservatorio Astrofisico di Arcetri, Largo E. Fermi 5,I-50125, Florence, Italy\\
$^{5}$\ INAF, Isituto di Radioastronomia, Via Piero Gobetti 101, I40127, Bologna, Italy\\
$^{6}$\ NAF-IASF Milano, Via Alfonso Corti 12, 20133, Milano, Italy\\
$^{7}$\ Department of Physics \& Astronomy, University College London, Gower Street, London, WC1E 6BT, UK\\
$^{8}$\ AIM, CEA, CNRS, Universit\'e Paris-Saclay, Universit\'e Paris Diderot, Sorbonne Paris Cit\'e, F-91191 Gif-sur-Yvette, France\\
$^{9}$\ Institut d'Astrophysique Spatiale, CNRS, Université Paris-Sud, Université Paris-Saclay, Bât. 121, 91405, Orsay Cedex, France\\
$^{10}$\ Central Astronomical Observatory of RAS, Pulkovskoye Chaussee 65/1, 196140, St. Petersburg, Russia\\
$^{11}$\ Centre for Astrophysics Research, University of Hertfordshire, College Lane, Hatfield, AL10 9AB, UK\\
$^{12}$\ National Observatory of Athens, Institute for Astronomy, Astrophysics, Space Applications and Remote Sensing, Ioannou Metaxa and Vasileos Pavlou GR-15236, Athens, Greece\\
$^{\star}$ {\tt \href{mailto:cclark@stsci.edu}{cclark@stsci.edui}}
}}}

\date{}

\pubyear{2019}

\begin{document}
\label{firstpage}
\pagerange{\pageref{firstpage}--\pageref{lastpage}}
\maketitle

\begin{abstract}
The dust mass absorption coefficient, \kappad\, is the conversion function used to infer physical dust masses from observations of dust emission. However, it is notoriously poorly constrained, and it is highly uncertain how it varies, either between or within galaxies. Here we present the results of a proof-of concept study, using the DustPedia data for two nearby face-on spiral galaxies M\,74 (NGC\,628) and M\,83 (NGC\,5236), to create the first ever maps of \kappad\ in galaxies. We determine \kappad\ using an empirical method that exploits the fact that the dust-to-metals ratio of the interstellar medium is constrained by direct measurements of the depletion of gas-phase metals. We apply this method pixel-by-pixel within M\,74 and M\,83, to create maps of \kappad. We also demonstrate a novel method of producing metallicity maps for galaxies with irregularly-sampled measurements, using the machine learning technique of Gaussian process regression. We find strong evidence for significant variation in \kappad. We find values of \kappad\ at 500\,\micron\ spanning the range 0.11--0.25\,${\rm m^{2}\,kg^{-1}}$ in M\,74, and  0.15--0.80\,${\rm m^{2}\,kg^{-1}}$ in M\,83. Surprisingly, we find that \kappad\ shows a distinct inverse correlation with the local density of the interstellar medium. This inverse correlation is the opposite of what is predicted by standard dust models. However, we find this relationship to be robust against a large range of changes to our method -- only the adoption of unphysical or highly unusual assumptions would be able to suppress it.
\end{abstract}

\begin{keywords}
galaxies: ISM -- galaxies: general -- ISM: dust -- ISM: abundances -- submillimetre: ISM -- methods: observational
\end{keywords}

\setcounter{footnote}{0}

\section{Introduction} \label{Section:Introduction}

Interstellar dust provides an indispensable window for studying galaxies and their evolution. Dust, which primarily emits in the Mid-InfraRed (MIR) to Far-InfraRed (FIR) to submillimetre (submm) wavelength regime, can be observed in very large numbers of galaxies very rapidly, with the beneficial effects of negative $k$-correction enhancing our ability to detect dusty galaxies out to high redshift \citep{Eales2010A,Oliver2012A}. This has made dust a standard proxy for studying galaxies' star-formation \citep{Kennicutt1998H,Buat2005A,Kennicutt2009B}, gas mass \citep{Eales2012A,Scoville2014A,Lianou2016A}, and chemical evolution (\citealp{Rowlands2014B}; \mbox{\citealp{Zhukovska2014A};} \mbox{\citealp{DeVis2017A,DeVis2017B,DeVis2019B}}) -- which are otherwise difficult and time-consuming to observe directly.

However, many of the valuable insights that dust can provide rest upon one simple expectation -- that we are able to use observations of dust emission to actually infer physical dust masses. Unfortunately, astronomers remain {\it terrible} at this. This is due to the fact that \kappad\ (variously called the dust mass absorption coefficient, or the dust mass opacity coefficient), the wavelength-dependent conversion factor used to calculate dust masses from FIR--submm dust Spectral Energy Distributions (SEDs), is extremely poorly constrained.

\begin{figure}
\centering
\includegraphics[width=0.475\textwidth]{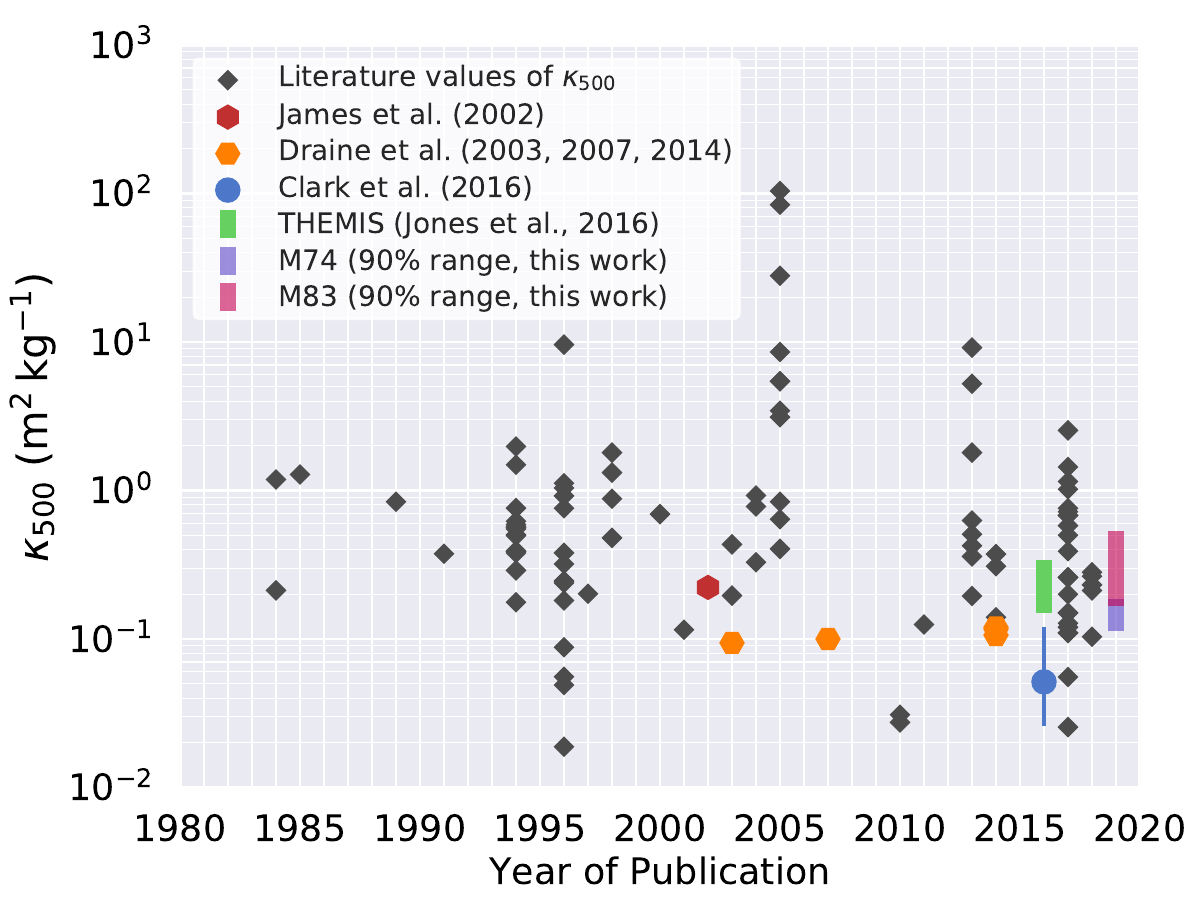}
\caption[]{Literature values of $\kappa_{500}$, plotted against the year in which they were published. This is an updated version of Figure~1 from \citet{CJRClark2016A}, revised to include values published subsequent to that work, plus additional historical values. A full list of references for the plotted values is provided as a footnote to this figure$^{\rm a}$. All values were converted to the 500\,\micron\ reference wavelength$^{\rm b }$ according to Equation~\ref{Equation:Kappa_Wavelength}, assuming$^{\rm c }$ $\beta = 2$. Several prominent values have been highlighted. Rectangular markers indicate the range encompassed by a particular set of values. The 5\textsuperscript{th}--95\textsuperscript{th} percentile ranges we find for M\,83 and M\,74 in this work are also plotted, for later reference (with the overlap between their ranges correspondingly shaded).}
\label{Fig:Year_vs_Kappa}
\footnotesize
\justify
$^{\rm a }$ The plotted values of \kappad\ include the values given in the compilation tables of \citet{Alton2004A} and \citet{Demyk2013A}, along with the values reported by: \citet{Ossenkopf1994B,Agladze1996A,Weingartner2001A,James2002,Draine2003A,Dasyra2005B,Draine2007A,Eales2010C, Ormel2011A,Compiegne2011A,Draine2014A,Gordon2014B,Planck2013XI,Kohler2015A,Jones2016A,Roman-Duval2017B,Bianchi2017A,Demyk2017A,Demyk2017B,Chiang2018A}.\\
$^{\rm b }$ The choice of reference wavelength has negligible (\textless\,0.1\,dex) effect on the standard deviation of the literature \kappad\ values in the plot, as long as $100<\lambda_{0}<1000$\,\micron.\\
$^{\rm c }$ Changing $\beta$ to any value in the standard range of 1--2.5 has negligible (\textless\,0.05\,dex) effect on the standard deviation of the literature \kappad\ values in the plot.
\end{figure}

\kappad\ is essentially a convenience factor, amalgamating the various properties of dust grains that dictate their emissivity -- such as the distributions of size, morphology, density, and chemical composition. These individual properties are extremely hard to constrain observationally, and highly degenerate with each other in their effect upon dust emission \citep{Whittet1992A}; combining them in \kappad\ allows them to be considered in terms of their net effect. Dust emission in the FIR--submm regime is traditionally modelled as a Modified BlackBody (MBB; or, `greybody'), where the observed flux density $S_{\lambda}$ at wavelength $\lambda$ is described by:
\begin{equation}
S_{\lambda} = \frac{1}{D^{2}} \sum^{n}_{i} M_{i} \kappa_{\lambda_{i}} B(\lambda,T_{i})
\label{Equation:Greybody}
\end{equation}

\noindent where $D$ is the distance to the source of the dust emission, $n$ is the number of dust components being modelled, $M_{i}$ is the mass of dust component $i$, $\kappa_{\lambda_{i}}$ is the value of \kappad\ at wavelength $\lambda$ for dust component $i$, and $B_{\lambda}(T_{i})$ is the Planck function evaluated at wavelength $\lambda$ for temperature $T_{i}$ of dust component $i$. While the dust population of a source will in reality span a continuum of temperatures, availability of FIR--submm data typically forces observers to fit their data with only 1 or 2 components (although point-process methods are starting to provide a way to model dust in a more continuous manner; see \citealp{Marsh2015A,Marsh2017B}).

The value of \kappad\ can be estimated in various ways, usually by some combination of: consideration of the elemental constituents of dust (derived from depletions); physical modelling of possible grain structures; chemical modelling of likely dust compositions; radiative transfer modelling; analysis of Ultraviolet (UV) to Near-Infra-Red (NIR) extinction and scattering; laboratory analysis of artificial dust grain analogues; and examination of retrieved grains of interplanetary and interstellar dust. For a fuller summary, and compilation of references, see Section~1 of \citet{CJRClark2016A}.
Troublingly, the various methods that have been employed for estimating \kappad\ yield a very wide range of possible values. In order to directly compare different values of \kappad, they need to be converted to the same reference wavelength. This can be done using the formula:
\begin{equation}
\kappa_{\lambda} = \kappa_{0}\left(\frac{\lambda_{0}}{\lambda}\right)^{\beta}
\label{Equation:Kappa_Wavelength}
\end{equation}

\noindent where $\kappa_{\lambda}$ is the value of \kappad\ at a particular wavelength $\lambda$, $\kappa_{0}$ is the value of \kappad\ at a reference wavelength $\lambda_{0}$, and $\beta$ is the dust emissivity spectral index. Laboratory analysis of dust analogues and chemical modelling  suggest that this relation is reliable in the wavelength range $150 \lesssim \lambda \lesssim 1000$\,\micron; at wavelengths shorter than this the variation of \kappad\ with wavelength becomes much more complex, whilst at longer wavelengths the behaviour of \kappad\ is less clear, with some evidence of an upturn \citep{Demyk2017A,Demyk2017B,Ysard2018A}.

Figure~\ref{Fig:Year_vs_Kappa} compiles a wide range of \kappad\ values that have been reported in the literature (all have been converted to a reference wavelength of 500\,\micron\ as per Equation~\ref{Equation:Kappa_Wavelength}; we only plot values for which the original quoted reference wavelength was in the reliable 150--1000\,\micron\ range). Over 100 values are plotted, with a standard deviation of 0.8\,dex, and spanning a total range of over 3.6 orders of magnitude. Worse still, there is no sign that values of \kappad\ reported in the literature are converging over time.

So, despite the excellent sensitivity and wavelength coverage provided by modern FIR--mm observatories, any dust masses inferred from observed dust emission remain enormously uncertain, stymieing our understanding of the InterStellar Medium (ISM) in galaxies. Moreover, this high degree of uncertainty means that, out of necessity, \kappad\ is often treated as being constant -- even though it is well understood that this can't be true in reality. Even the more complex, multi-phase dust model frameworks, such as those of \citet{Jones2013C,Jones2017A}, usually only incorporate 2 or 3 types of dust, each with a corresponding \kappad.

As such, understanding how kappa varies -- both between different galaxies, and within individual galaxies -- is clearly vital for the field.

In this paper, we use an empirical method for determining the value of \kappad\ -- which we employ on a resolved, pixel-by-pixel basis in two nearby galaxies -- to produce the first maps of how \kappad\ varies within galaxies, as a proof-of-concept study. The theory behind the dust-to-metals method we employ to find \kappad\ is described in Section~\ref{Section:Theory}. The galaxies and data we use in this work are described in Section~\ref{Section:Data}. The application of the technique to produce maps of \kappad\ is Section~\ref{Section:Application}. Our results are presented in Section~\ref{Section:Results}, and are discussed in Section~\ref{Section:Discussion}. For brevity and readability, `flux density' will be termed `flux' throughout the rest of the paper.

\section{Theory} \label{Section:Theory}

Of the many methods proposed for estimating the value of \kappad, one of the most simple is that first proposed by \citet{James2002}. The \citet{James2002} method is entirely empirical, and relies upon just one central assumption -- that the dust-to-metals ratio in the ISM, \epsilond, has a known value. If the ISM mass of a galaxy is known, along with the metallicity of that ISM, it is straightforward to calculate the total mass of interstellar metals in that galaxy; then, by assuming a fixed dust-to-metals ratio, it is possible to infer a galaxy's dust mass {\it a priori}, without any reference to the dust emission. This {\it a priori} dust mass can then be compared to that galaxy's observed dust emission, and hence \kappad\ can be calibrated. Here we use the \epsilond\ notation for the dust-to-metals ratio, instead of $\mathcal{D\,T\mkern-3mu M}$. This maintains consistency with \citet{James2002} and \citet{CJRClark2016A}, and avoids any ambiguity arising from the fact that $\mathcal{D\,T\mkern-3mu M}$ is often used to denote a dust-to-metals ratio {\it normalised by the Milky Way value}, whereas our quoted dust-to-metals ratios are always absolute values.

The vast majority of all reported values of \epsilond\ lie in the range 0.2--0.6 (considering only values of \epsilond\ that are not based upon some assumed value of \kappad: \citealp{Issa1990B,Luck1992A,Whittet1992A,Pei1992A,Meyer1998B,Dwek1998B,Pei1999A,Weingartner2001A,James2002,Kimura2003C,Draine2007C,Jenkins2009B,Peeples2014A,McKinnon2016A,Wiseman2017B,Telford2019B}). As such, it seems fair to conclude that \epsilond\ is significantly better constrained than \kappad\ -- making the former a useful tool for pinning down the value of the latter. And whilst some authors suggest larger values of \epsilond\ (for instance \citealp{DeCia2013C}, who find values in the region of 0.8), we can at least be confident that, by definition, no galaxy has a dust-to-metals ratio greater than 1 -- no such helpful constraint exists for \kappad. Furthermore, thanks to observations of elemental depletions in the neutral ISM, \epsilond\ can be determined far more directly than \kappad.

\citet{CJRClark2016A} built upon the \citet{James2002} method, to correct for a number of systematics that affected that original implementation, and to enable it to take advantage of higher-quality modern FIR--submm data. In this work, we apply the \citet{CJRClark2016A} iteration of the dust-to-metals method on a resolved basis, in nearby galaxies. Therefore, for completeness, we here provide a cursory description of the technique as implemented in this work; for a full derivation and description, refer to Section~2 of \citet{CJRClark2016A}. The final form of the method can be rendered as the following formula for computing $\kappa_{\lambda}$ for the ISM of a source:
\begin{equation}
\kappa_{\lambda} = \frac{ D^{2} }{ \xi\, ( M_{{\rm HI}} + M_{\rm H_{2}} )\, \varepsilon_{d}\, f_{Z} } \sum_{i}^{n} \left( \frac{ S_{\lambda_{i}} }{ B_{\lambda}(T_{i}) } \right)_{i}
\label{Equation:Kappa}
\end{equation}

\noindent where $\xi$ is a correction factor to account for the fraction of ISM mass due to elements other than hydrogen, $M_{\rm HI}$ is the atomic hydrogen mass, $M_{\rm H_{2}}$ is the molecular hydrogen mass, $\epsilon_{d}$ is the dust-to-metals ratio, and $f_{Z}$ is the ISM metal mass fraction. The $\sum_{i}^{n} ( \frac{ S_{\lambda_{i}} }{ B_{\lambda}(T_{i}) } )_{i}$ term corresponds to the model used to fit the observed dust emission of the target source -- in this instance, $n$ MBBs, as per Equation~\ref{Equation:Greybody};  $n$ is the number of dust components being modelled, $S_{\lambda_{i}}$ is the flux emitted at wavelength $\lambda$ by dust component $i$, and $B_{\lambda}(T_{i})$ is the Planck function evaluated at wavelength $\lambda$ for temperature $T_{i}$ of dust component $i$; our SED-fitting procedure is described in Section~\ref{Subsection:SED_Fitting}.

The formulation in Equation~\ref{Equation:Kappa} gives a combined \kappad\ value, that incorporates the contribution from all dust species present, for each temperature component (for $n > 1$). The problem becomes unconstrained if each dust component is treated as having a different \kappad. The potential impact of line-of-sight mixing of dust components at different temperatures is discussed in Section~\ref{Subsection:SED_Fitting}.

The correction factor $\xi$ is required in Equation~\ref{Equation:Kappa}, as the dust-to-metals method is concerned with the {\it total} mass of the ISM, not just the mass of hydrogen. It is standard in the literature to account for mass other than hydrogen by applying a fixed factor of 1.36 -- corresponding to the Milky Way helium abundance. However this fails to consider how helium abundance varies with galaxy evolution, or the contribution of metals to the mass of the ISM. Thus $\xi$ is defined as:
\begin{equation}
\xi =  \frac{ 1 }{ 1 - \left(f_{\rm He_{\it p}} + f_{Z} \left[\frac{\Delta f_{\rm He}}{\Delta f_{Z}}\right] \right) - f_{Z} }
\label{Equation:xi}
\end{equation}

\noindent where $f_{\it He_{p}}$ is the primordial helium mass fraction, and $[\frac{\Delta f_{\it He}}{\Delta f_{Z}}]$ describes the evolution of the helium mass fraction with metallicity. We use $f_{\it He_{p}} = 0.2485 \pm 0.0002$ from \cite{Aver2013A}, and $[\frac{\Delta f_{\it He}}{\Delta f_{Z}}] = 1.41 \pm 0.62$ from \cite{Balser2006D}. Given Equation~\ref{Equation:xi}, $\xi$ can therefore vary from 1.33 (for low-metallicity galaxies where $Z$$\to$0) to 1.45 (for high-metallicity giant ellipticals where $Z = 1.5Z_{\odot}$).

It is important to note that \logOH\ measurements trace gas-phase metallicity in the ionised phase (predominantly H{\sc ii} regions), whereas we are concerned with the metallicity of the ISM at large. This means that we must account for the fraction of interstellar oxygen mass in H{\sc ii} regions depleted onto dust grains, $\delta_{O}$, and hence missed by gas-phase metallicity estimators. We use a value of $\delta_{O} = 1.32 \pm 0.09$ from \citet{Mesa-Delgado2009B}, which is in good agreement with numerous other reported values \citep{Peimbert2012C,Kudritzki2012B,Bresolin2016B}. Whilst the oxygen depletion factor in the ISM at large is known to vary by at least 0.3\,dex \citep{Jenkins2009B}, oxygen depletion in H{\sc ii} regions is found to be remarkably constant, at $\sim$\,1.3 (ie, $\sim$\,0.1\,dex) across nearby galaxies (evaluated by comparing abundances in H{\sc ii} regions to abundances in the atmospheres of nearby B stars; \citealp{Bresolin2016B} and references therein). Additionally, given that the elemental composition of oxygen-rich dust is found to exhibit minimal variation at intermediate-to-high metallicities \citep{Mattsson2019C}, the assumption of a constant $\delta_{O}$ is valid modulo a constant $\epsilon_{d}$ -- which is the central premise of our method. 

Atomic hydrogen mass, $M_{\rm HI}$ (in ${\rm M_{\odot}}$), is determined using observations of the 21\,cm hyperfine structure line, according to the standard prescription:
\begin{equation}
M_{\rm HI} = 2.356 \times 10^{-7}\,S_{\rm HI} D^{2}
\label{Equation:HI_Mass}
\end{equation}

\noindent where $S_{\rm HI}$ is the velocity-integrated flux density of the 21\,cm line (in ${\rm Jy\,km\,s^{-1}}$), and the source distance $D$ is here in units of pc.

The mass of molecular hydrogen associated with a source cannot be determined directly from emission; because the ${\rm H_{2}}$ molecule is non-polar, it does not radiate when in the ground state (which is the case for the bulk of molecular hydrogen in galaxies). Instead, molecular hydrogen masses are typically inferred by treating CO as a tracer molecule, via observations of the ${\rm ^{12}C^{16}O}$(1-0) rotational line (referred to as CO(1-0) hereafter). The mass of molecular hydrogen, $M_{\rm H_{2}}$ (in ${\rm M_{\odot}}$), can thus be calculated using the relation:
\begin{equation}
M_{\rm H_{2}} = I_{\rm CO} \alpha_{\rm CO} (2\,D \tan\left(\frac{\theta}{2}\right))^{2}
\label{Equation:H2_Mass}
\end{equation}

\noindent where $I_{\rm CO}$ is the velocity-integrated main-beam brightness temperature of the CO(1-0) line (in ${\rm K\,km\,s^{-1}}$), $\alpha_{\rm CO}$ is the CO-to-${\rm H_{2}}$ conversion factor (in ${\rm K^{-1}\,km^{-1}\,s\,M_{\odot}\,pc^{-2}}$), $\theta$ is the angular diameter of the target source, and the source distance $D$ is here in units of pc. The value of $\alpha_{\rm CO}$ is a matter of much debate, but the standard Milky Way value is $\alpha_{\rm CO_{\it MW}} = 3.2\,{\rm K^{-1}\,km^{-1}\,s\,M_{\odot}\,pc^{-2}}$, which is treated as uncertain by a factor of 2 (see \citealp{Obreschkow2009A}, \citealp{Saintonge2011A}, \citealp{Bolatto2013B}, and references therein). Note that Equation~\ref{Equation:H2_Mass} is simply the standard ${\rm H_{2}}$ mass surface-density prescription, $\Sigma_{\rm H_{2}} = I_{\rm CO} \alpha_{\rm CO}$ (where $\Sigma_{\rm H_{2}}$ is in units of ${\rm M_{\odot}\,pc^{2}}$), rendered in terms of $M_{\rm H_{2}}$ for consistency with Equations~\ref{Equation:Kappa} and \ref{Equation:HI_Mass}. The CO-to-${\rm H_{2}}$ conversion factor can alternatively be expressed as $X_{\rm CO}$, which is in terms of column number density density of molecules, being related to $\alpha_{\rm CO}$ according to $X_{\rm CO} = 6.3 \times 10^{19}\,\alpha_{\rm CO}$.

The galaxies considered in this work contain environments with metallicities that vary by a factor of 2.5, spanning 0.4--1\,${\rm Z_{\odot}}$ (see Section~\ref{Section:Application}). When considering locales with significantly-varying metallicities, it is important to account for the corresponding variation of $\alpha_{\rm CO}$ with metallicity \citep{Bolatto2013B}. In lower-metallicity environments, there will be reduced abundances of C and O, relative to H. Additionally, there is less dust available in low-metallicity environments to shield the CO --  which is less able to self-shield than ${\rm H_{2}}$ -- from photodisassociation (see \citealp{Wolfire2010A}, \citealp{Clark2015A}, and references therein). Here we opt to use the metallicity-dependent $\alpha_{\rm CO}$ prescription of \citet{Amorin2016A}, described by:
\begin{equation}
\alpha_{\rm CO} =  \alpha_{\rm CO_{\it MW}} \left( \frac{Z}{Z_{\odot}} \right)^{-y_{\rm\, CO}}
\label{Equation:Alpha_CO}
\end{equation}

\noindent where $\frac{Z}{Z_{\odot}}$ is the ISM metallicity in terms of the Solar value, and $y_{\rm\, CO}$ is an empirical power-law index with a value of $1.5 \pm 0.3$.

The \citet{Amorin2016A} rule is calibrated on a sample of galaxies spanning over an order of magnitude in metallicity ($7.69 < 12 + {\rm log}_{10} [ \frac{\rm O}{\rm H} ] < 8.74$), by using the Star Formation Efficiency (SFE) and Star Formation Rate (SFR) to infer the molecular gas supply present. They do this by employing the relation $\frac{\alpha_{\rm CO}}{\alpha_{\rm CO_{\it MW}}} = \tau_{\rm H_{2}} \frac{\it SFR}{M_{\rm H_{2}}}$; effectively inverting the Kennicutt-Schmidt law \citep{Kennicutt1998H} to infer the molecular gas mass present, anchored by the known star formation efficiency of the Milky Way. Resolved studies such as \citet{Bigiel2011A} and \citet{Utomo2019B} find remarkably little variation in SFE within face-on local normal spirals like those studied in this work; this supports the reliability of using a SFE-calibrated method for estimating $\alpha_{\rm CO}$ in a resolved study such as ours. Additionally, the \citet{Amorin2016A} prescription effectively traces the median of the commonly-cited metallicity-dependent literature prescriptions (see Figure~11 of \citealp{Amorin2016A} and Figure~6 of \citealp{Accurso2017B} for comparisons of prescriptions), making it the choice most likely to not conflcit with other works. 

Regarding the Solar metallicity, we use the canonical value for the Solar oxygen abundance of $[12 + {\rm log}_{10} \frac{\rm O}{\rm H} ] _{\odot}= 8.69 \pm 0.05$ \citep{Asplund2009A}, corresponding to a Solar metal mass fraction of $f_{Z_{\odot}} = 0.0134$ (\citealp{Asplund2009A}, uncertainty deemed to be negligible). In common with the literature at large, we assume that oxygen abundance traces total metallicity. Whilst this assumption has its limits, oxygen is the most abundant metal in the Universe, and a dominant constituent of dust \citep{Savage1996D,Jenkins2009B}, making it a useful metallicity tracer for our purposes. Although the ratio of oxygen to carbon (the other main constituent of dust by mass) is known to vary with metallicity \citep{Garnett1995B}, this systematic trend is no more prominent than the intrinsic scatter over the 0.4--1.0\,${\rm Z_{\odot}}$ metallicity range relevant to this work \citep{Pettini2008B,Berg2016A}.

Although a $D^{2}$ term appears in Equation~\ref{Equation:Kappa}, the $M_{\rm HI}$ and $ M_{\rm H_{2}}$ terms are also both proportional to $D^{2}$, which therefore ultimately cancels out. This renders the resulting values of $\kappa_{\lambda}$ independent of distance, removing a potentially large source of uncertainty.

Throughout this work, when employing values from the literature, we take care to only use values that do not themselves rely upon any assumed value of \kappad.

For the value of the dust-to-metals ratio, $\epsilon_{d}$, in Equation~\ref{Equation:Kappa}, we take two approaches. For our fiducial analysis, presented in Section~\ref{Section:Results}, we assume a constant value of $\epsilon_{d} = 0.4 \pm 0.2$. This is smaller than the value of 0.5 assumed in \citet{CJRClark2016A}, as more recent works \citep{McKinnon2016A,DeCia2016A,Wiseman2017B} suggest that for most galaxies with metallicities \textgreater\,0.1\,${\rm Z_{\odot}}$, the dust-to-metals ratio is slightly below the Milky Way's average value of 0.5 \citep{James2002,Jenkins2009B}.

The assumption of a constant dust-to-metals ratio is an approximation that will break down at some point. Therefore, in Section~\ref{Subsubsection:Variable_Dust-To-Metals}, we construct an alternate analysis where \epsilond\ increases as a function of ISM surface density. This is a more physical treatment, as depletion of ISM metals onto dust grains is found to increase in regions of greater ISM column density \citep{Jenkins2009B,Roman-Duval2019A}. This is in agreement with the fact that grain growth in the ISM is required to explain the dust budgets in many galaxies \citep{Galliano2008A,Rowlands2014B,Zhukovska2014A}. As a result, dust grain growth in denser ISM (with the corresponding increase in \epsilond) is a feature of dust evolution models such as The Heterogeneous dust Evolution Model for Interstellar Solids (THEMIS; \citealp{Jones2013C,Jones2017A,Jones2018A}).  Unfortunately, the exact form of the relationship between \epsilond\ and ISM (surface) density is very poorly constrained (the relationship we assume for our analysis is described in detail in Section~\ref{Section:Discussion}). As such, the variable-\epsilond\ model represents a more-physical, but worse-constrained approach; whilst the fixed-\epsilond\ model represents a less-physical, but better-constrained approach. For this reason, whilst the fixed-\epsilond\ approach is our fiducial model, we nonetheless consider both scenarios.

\section{Data} \label{Section:Data}

\begin{figure*}
\centering
\includegraphics[width=0.975\textwidth]{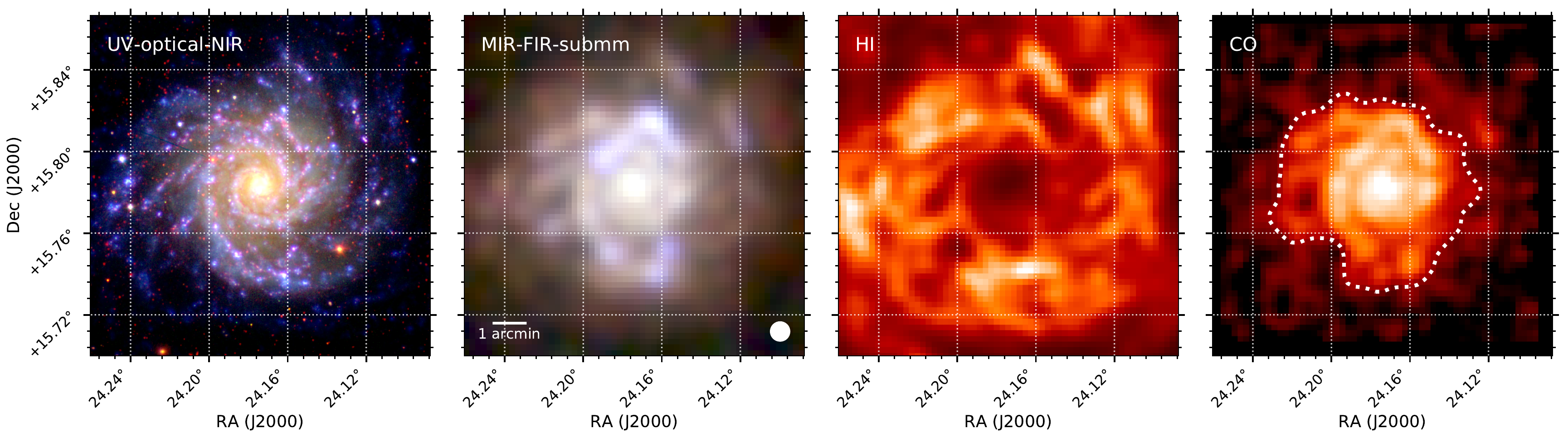}
\caption{Multiwavelength overview of M\,74. {\it 1\textsuperscript{st}:} Three-colour UV--optical--NIR image, composed of GALEX NUV (blue), SDSS {\it g} (green), and \spitz-IRAC 3.6\,\micron\ (red) data. {\it 2\textsuperscript{nd}:} Three-colour MIR--FIR-submm image, composed of WISE 22\,\micron\ (blue), \hersc-PACS 160\,\micron\ (green), and \hersc-SPIRE 350\,\micron\ (red) data. {\it 3\textsuperscript{rd}:} THINGS \HI\ moment-0 map. {\it 4\textsuperscript{th}:} HERACLES CO(2-1) moment-0 map. Except for the UV--optical--NIR image, all maps are convolved to the 36\arcsec\ limiting resolution at which we perform our analysis (beam size indicated in the 2\textsuperscript{nd} panel). The dotted line in the far-right panel marks the SNR\,=\,2 contour of the CO(2-1) map, which is the region within which we mapped \kappad.}
\label{Fig:NGC0628_Multiwavelenth_Grid}
\end{figure*}

\begin{figure*}
\centering
\includegraphics[width=0.975\textwidth]{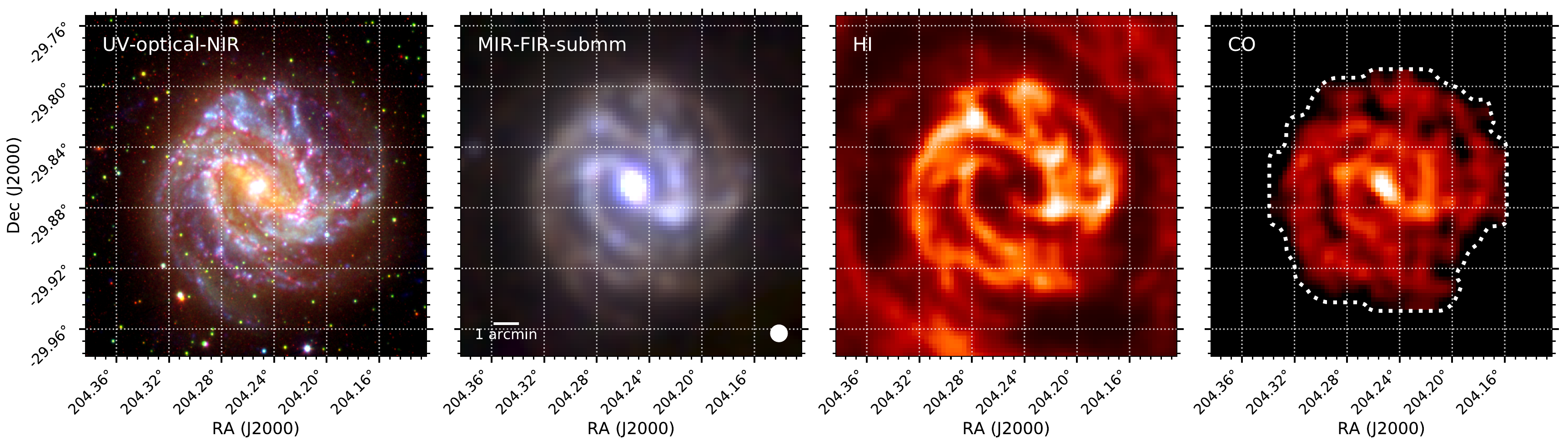}
\caption{Multiwavelength overview of M\,83. Description as per Figure~\ref{Fig:NGC0628_Multiwavelenth_Grid}, with the exceptions that the green channel in the far-left three-colour UV--optical--NIR image corresponds to DSS {\it B}-band, the CO moment-0 map is SEST CO(1-0) data, and that the limiting resolution of our M\,83 data is 42\arcsec\ (images convolved accordingly).}
\label{Fig:NGC5236_Multiwavelenth_Grid}
\end{figure*}

\begin{table}
\centering
\caption{Basic properties of M\,74 and M\,83, the galaxies studied in this work. All values derived from the data presented in \citet{CJRClark2018A}, unless otherwise specified.}
\label{Table:Galaxy_Properties}
\begin{tabular}{lrr}
\toprule \toprule
\multicolumn{1}{c}{} &
\multicolumn{1}{c}{M\,74} &
\multicolumn{1}{c}{M\,83} \\
\cmidrule(lr){2-3}
NGC N\textsuperscript{\underline{o}} & NGC\,628 & NGC\,5236 \\
RA (J2000) & 24.174\degr & 204.254\degr \\
 & (01\textsuperscript{h}\,36\textsuperscript{m}\,41\fs\,8) & (13\textsuperscript{h}\,37\textsuperscript{m}\,01\fs\,0) \\
Dec (J2000) & +15.783\degr &  -29.866\degr \\
 & (+15\degr46\arcmin\,58\farcs\,8) & (-29\degr\,51\arcmin\,57\farcs\,6) \\
Distance (Mpc)\,$^{\rm a}$ & 10.1 & 4.9 \\
Hubble Type & SAc & SBc \\
 & (5.2) & (5.0) \\
${D_{25}}$ (arcmin) & 10.0 & 13.5 \\
${D_{25}}$ (kpc) & 29.4 & 19.2 \\
${A_{25}}$ (${\rm kpc^{2}}$) & 683 & 290 \\
${M_{\ast}}$ (${\rm log_{10}\,M_{\odot}}$)\,$^{\rm b}$ & 10.1 & 10.5 \\
${M_{\rm HI}}$ (${\rm log_{10}\,M_{\odot}}$)\,$^{\rm c}$ & 9.9 & 10.0  \\
${M_{\rm H_{2}}}$ (${\rm log_{10}\,M_{\odot}}$)\,$^{\rm d}$ & 9.4 & 9.5 \\
${M_{d}}$ (${\rm log_{10}\,M_{\odot}}$)\,$^{\rm e}$ & 7.5 & 7.4 \\
SFR (${\rm M_{\odot}\,yr^{-1}}$)\,$^{\rm b}$ & 2.4 & 6.7 \\
FUV$-$$K_{S}$ (mag) & 2.9 & 3.4 \\
NUV$-$$r$ (mag) & 2.5 & 2.8 \\
\bottomrule
\end{tabular}
\footnotesize
\justify
$^{\rm a }$ As a first-order estimate of the uncertainty on the distance, we use the standard deviation of the redshift-independent distances listed in the {\sc Nasa/ipac} Extragalactic Database (NED; \url{https://ned.ipac.caltech.edu/ui/}) for each galaxy. This gives uncertainties of 3.2 and 3.4\,Mpc for M\,74 and M\,83 respectively.\\
$^{\rm b}$ \citet{Nersesian2019A}.\\
$^{\rm c}$ H{\sc i} mass from total single-dish flux in the HI Parkes All Sky Survey (HIPASS; \citealp{Meyer2004,Wong2006}).\\
$^{\rm d}$ This work (see Section \ref{Subsection:Gas_Data}). \\
$^{\rm e}$ This work (using the pixel-by-pixel \kappad\ values calculated in produced \ref{Section:Results}).
\end{table}


An initial attempt by \citet{CJRClark2016A} to detect variation in \kappad\ using the dust-to-metals method was unsuccessful; however, that study only considered the global dust properties of galaxies, and considered a sample of 22 objects, all of which were of similar masses, metallicities, and environments. A promising avenue for finding variation in \kappad\ is to look {\it within} well-resolved nearby galaxies. Many studies have found that dust properties can vary significantly -- and sometimes dramatically -- within galaxies \citep{MWLSmith2012B,Roman-Duval2017B,Relano2018A}. It would be surprising if this variation did not extend to \kappad.

Creating a \kappad\ map of a galaxy using the dust-to-metals method requires resolved data for its dust emission, atomic gas, molecular gas, and metallicity; with the resolution provided by modern observations, it is possible to make many hundreds, or even thousands, of independent \kappad\ determinations within a galaxy. For this proof-of-concept demonstration we map \kappad\ within two nearby face-on spiral galaxies -- M\,74 (NGC\,628) and M\,83 (NGC\,5236). We select these galaxies on account of their particularly extensive metallicity data (see Section~\ref{Subsection:Metallicity_Data}), coupled with their resolution-matched multi-phase ISM observations (see Section ~\ref{Subsection:Gas_Data}).

We obtained the bulk of the necessary data from the DustPedia archive\footnote{\url{https://dustpedia.astro.noa.gr/}}. DustPedia \citep{Davies2017A} is a European Union project working towards a comprehensive understanding of dust in the local Universe, capitalising on the legacy of the \hersc\ Space Observatory \citep{Pilbratt2010D}. A centrepiece of the project is the DustPedia database, which includes every galaxy observed by \hersc\ that has recessional velocity within $3000\,{\rm km\,s^{-1}}$ ($\sim$\,40\,Mpc), has optical angular size in the range 1\arcmin\,\textless\,$D_{25}$\,\textless\,1\degr, and has a detected stellar component\footnote{As defined according to detection by the Wide-Field Infrared Survey Explorer (WISE; \citealp{Wright2010F}), at its all-sky sensitivity, in 3.4\,\micron\ (its most sensitive band).}.

The continuum data we employ is described in Section~\ref{Subsection:Continuum_Data}, the metallicity data (and the process by which we use it to create metallicity maps) is described in Section~\ref{Subsection:Metallicity_Data}, and the atomic \&\ molecular gas data in Section~\ref{Subsection:Gas_Data}.

\subsection{Target Galaxies} \label{Subsection:Target_Galaxes}

We selected M\,74 and M\,83 as the subject galaxies for this work; a summary of their basic characteristics is provided in Table~\ref{Table:Galaxy_Properties}. Both are very nearby, highly extended, and almost perfectly face-on, making them two of the most heavily-studied galaxies in the sky, and ideally suited to serving as our proof-of-concept targets for mapping \kappad.

Both galaxies are classified as `grand design' \mbox{\citep{DMElmegreen1987A}} type Sc spirals, with M\,83 also displaying a prominent bar \citep{deVaucouleurs1991A}. M\,74 has a physical diameter of 29\,kpc -- similar to that of the Milky Way \mbox{\citep{Goodwin1998C,Rix2013A}} -- and about 50\%\ greater than that of M\,83 (diameter defined according to the optical $D_{25}$, being the isophotal major axis at which the optical surface brightness falls beneath 25\,${\rm mag\,arcsec^{2}}$).

Despite being the physically smaller of the two, M\,83 has a stellar mass 2.2 times greater, and a Star Formation Rate (SFR) 2.7 times greater \citep{Nersesian2019A}. M\,83 has a correspondingly higher surface brightness in dust emission, averaging 4.2\,${\rm MJy\,sr^{-1}}$ at 500\,\micron\ within its $D_{25}$, compared to 1.6\,${\rm MJy\,sr^{-1}}$ for M\,74. The nuclear region of M\,83 is currently undergoing a bar-driven starburst, concentrated in the central 250\,pc, accounting for $\sim$10\%\ of the galaxy's total ongoing star-formation (\citealp{Sersic1965A}; \mbox{\citealp{Harris2001B};} \citealp{Fathi2008A}). The optical disc of M\,83 has a minimal systematic metallicity gradient, with oxygen abundances varying by only about 0.1\,dex from place to place; in contrast, M\,74 has a pronounced metallicity gradient, with oxygen abundances in its centre about 0.3\,dex greater than at its $R_{25}$ \citep{DeVis2019B}.

Many of the differences between M\,74 and M\,83 -- such as in their stellar surface densities (and therefore interstellar radiation fields), star formation characteristics, metallicity profiles, ISM distributions, etc -- have the potential to affect dust properties, and thereby provide useful scope for us to contrast how \kappad\ can vary due to a range of factors.

The appearances of both galaxies, in various parts of the spectrum, are illustrated in Figures~\ref{Fig:NGC0628_Multiwavelenth_Grid} and \ref{Fig:NGC5236_Multiwavelenth_Grid}. The stellar masses and SFRs for the DustPedia galaxies, as presented in \citet{Nersesian2019A}, were estimated using the Code Investigating GALaxy Emission (CIGALE; \mbox{\citealp{Burgarella2005F,Noll2009B}}) software, incorporating the THEMIS dust model.

\subsection{Continuum Data} \label{Subsection:Continuum_Data}

Multiwavelength imagery and photometry for the DustPedia galaxies (spanning 42 ultraviolet--millimetre bands), along with distances, morphologies, etc, are presented in \citet{CJRClark2018A}. Our analysis makes use of observations from several of the facilities included in the DustPedia archive.

In the submm, we use observations at 250, 350, and 500\,\micron\ from the Spectral and Photometric Imaging REceiver (SPIRE; \citealp{Griffin2010D}) instrument onboard \hersc. In the FIR, we use observations at 160, and 70\,\micron\ from the  Photodetector Array Camera and Spectrometer (PACS; \citealp{Poglitsch2010B}) instrument, also onboard \hersc\ (PACS did not perform 100\,\micron\ observations for M\,83, so for consistency we make no use of the the PACS 100\,\micron\ data for M\,74). In the MIR, we use observations at 22\,\micron\ from the WISE\footnote{Whilst 24\,\micron\ \spitz\ data does exist for these galaxies, the background is better-behaved in the WISE data, thanks to the superior mosaicing permitted by the larger field of view.}. A compilation of the MIR--FIR--submm data for each galaxy is shown in the centre-left panels of Figures~\ref{Fig:NGC0628_Multiwavelenth_Grid} and \ref{Fig:NGC5236_Multiwavelenth_Grid}.

Although not required for the creation of the \kappad\ maps, we use various additional data for reference and comparison, also drawn from the DustPedia archive. This includes UltraViolet (UV) observations from GALaxy Evolution eXplorer (GALEX; \citealp{Morrissey2007B}); UV, optical, and NIR observations from the Sloan Digital Sky Survey (SDSS; \citealp{York2000B,Eisenstein2011B}); optical observations from the Digitized Sky Survey (DSS); plus NIR observations from the InfraRed Array Camera (IRAC; \citealp{Fazio2004G}) and Multiband Imager for \spitz\ (MIPS; \citealp{Rieke2004K}) instruments onboard the \spitz\ Space Telescope \citep{Werner2004B}. A compilation of the UV--optical--NIR data for each galaxy is shown in the far-left panels of Figures~\ref{Fig:NGC0628_Multiwavelenth_Grid} and \ref{Fig:NGC5236_Multiwavelenth_Grid}.

\subsection{Metallicity Data} \label{Subsection:Metallicity_Data}

\begin{figure*}
\centering
\includegraphics[width=0.475\textwidth]{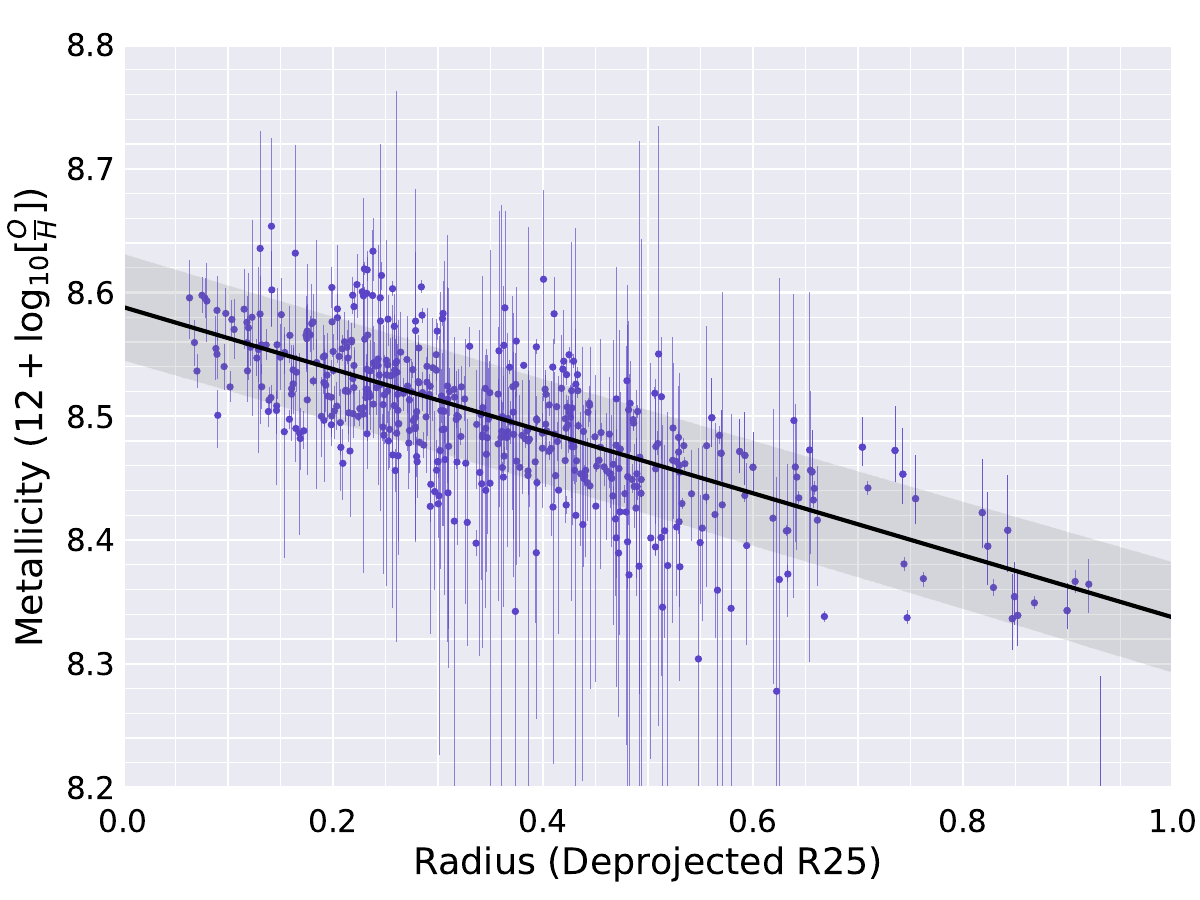}
\includegraphics[width=0.475\textwidth]{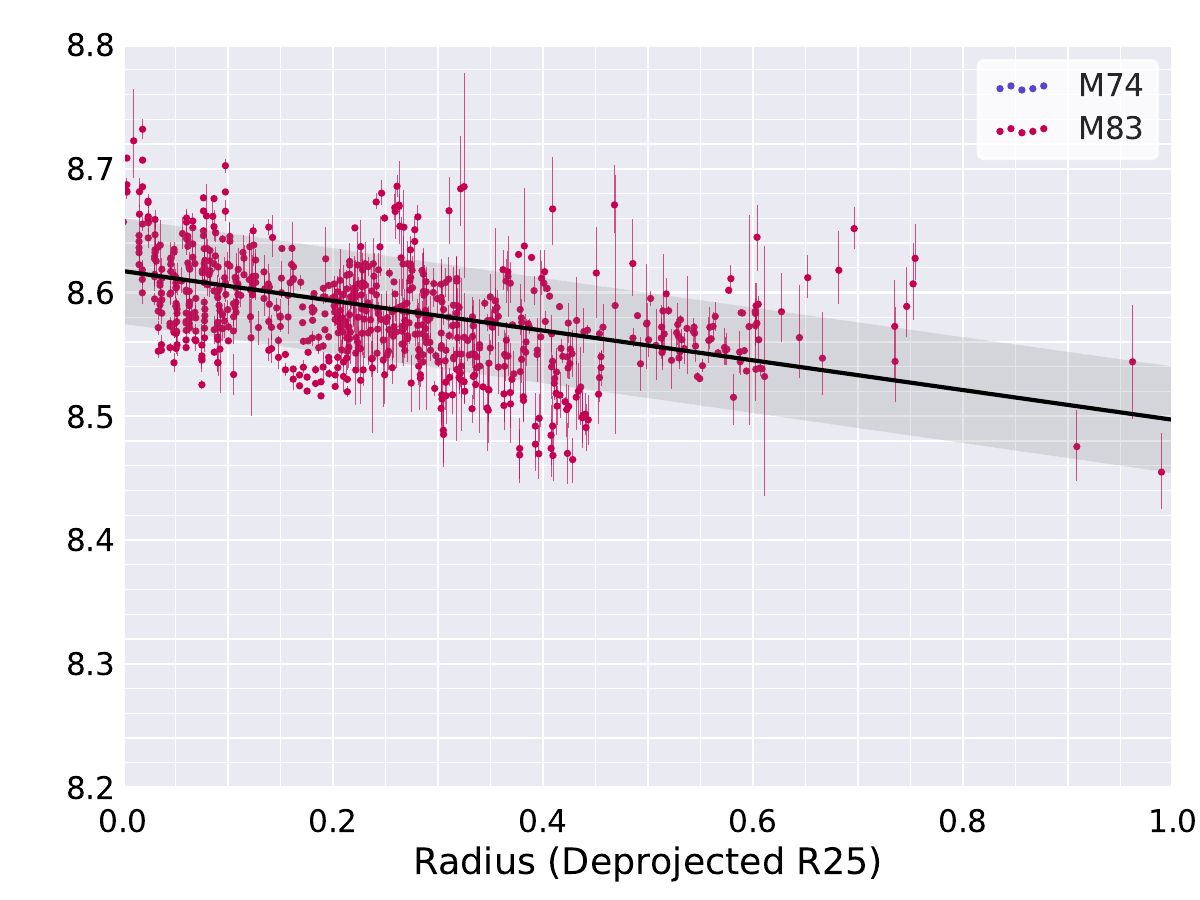}
\caption{The radial metallicity profiles of M\,74 (left) and M\,83 (right). The black lines show the radial metallicity profiles; the shaded grey areas indicate the intrinsic scatter (all based on median posterior values of $m_{Z}$, $c_{Z}$, and $\psi$). For ease of viewing, a handful of points are not shown in these plots (being at \logOH\,\textless\,8.2, and/or radii beyond $R_{25}$); such points are nonetheless included in all modelling.}
\label{Fig:Metallicity_Gradient}
\end{figure*}

Galaxies sufficiently extended to have well-resolved global FIR--submm observations, atomic gas observations, and molecular gas observations, are generally {\it too} extended to have their UV--NIR nebular spectral emission -- and hence metallicities -- fully mapped by Integral Field Unit (IFU) spectrometry. Whilst some large-area IFU surveys of nearby galaxies have now been undertaken, these are still very much the exception rather than the rule, and even the very largest can currently only cover $\sim$\,50\%\ of the area of galaxies as extended as M\,74 and M\,83. \citep{RosalesOrtega2010A,Sanchez2011A,Blanc2013AC}. As such, the few DustPedia galaxies with mostly complete IFU coverage do not have the well-resolved gas and dust data needed for this analysis.

However, extended nearby galaxies are popular targets for spectroscopic observation; most have had large numbers of individual slit and fibre spectra taken, supplementing partial IFU coverage like that described above. For DustPedia, \citet{DeVis2019B} have compiled a sizeable database of emission line fluxes, collated from 42 literature studies plus all available archival Multi Unit Spectroscopic Explorer (MUSE; \citealp{Bacon2010A}) data that covers the DustPedia galaxies. The \citet{DeVis2019B} spectroscopic database contains emission line fluxes from 10,000 spectra, with data for 492 (56\%) of the DustPedia galaxies. \citet{DeVis2019B} also present consistent gas-phase metallicity measurements for all of these spectra, for 5 different strong-line relation prescriptions (all of which yield standard \logOH\ metallicities). Following their tests of the internal consistency of the prescriptions considered, \citet{DeVis2019B} find the \citet{Pilyugin2016A} `S' prescription most reliable; we therefore use these metallicities throughout the rest of this work. A recent study by \citet{Ho2019B} also supports the validity of the \citet{Pilyugin2016A} prescriptions at the metallicities of our target galaxies. As an additional test, we also repeat the entire \kappad-mapping process using metallicity data produced using 4 other strong-line relations; this is presented in Appendix~\ref{AppendixSection:Kappa_Z_Variations}.

M\,74 and M\,83 both have large numbers of metallicities in the \citet{DeVis2019B} database -- 510 and 793 measurements respectively, more than any other DustPedia galaxy (except UGC\,09299, which lacks the resolved gas data we require). These metallicity points sample the entirety of both galaxies' optical discs. The positions of these spectra, and the metallicities derived from them, are plotted in the upper-left panels of Figures~\ref{Fig:NGC0628_Metallicity_Grid} and \ref{Fig:NGC5236_Metallicity_Grid}. Our region of interest for each galaxy\footnote{\label{Footnote:Kappa_Map_Region}The region of interest being the area where we map \kappad; illustrated in Figures~\ref{Fig:NGC0628_Multiwavelenth_Grid} and Figure~\ref{Fig:NGC5236_Multiwavelenth_Grid}, and defined in Section~\ref{Subsection:Data_Preparation}.} extends approximately out to 0.55\,$R_{25}$ for M\,74, and to 0.7\,$R_{25}$ for M\,83. So whilst the bulk of the metallicity points lie within the region of interest of each galaxy, providing dense sampling, there are also sufficient points outside it to constrain the metallicity variations over larger scales.

In order to produce maps of \kappad, it was necessary to first have maps of the metallicity distributions of our target galaxies. The first step towards achieving this was modelling their radial metallicity profiles. The spectra metallicity points for M\,74 and M\,83, plotted as a function of their deprojected galactocentric radius, $r$, are shown in Figure~\ref{Fig:Metallicity_Gradient}. As can be seen, there is significant scatter around the radial trends of both galaxies, far in excess of what would be expected if it were driven solely by the uncertainties on the individual metallicity points. Indeed, if one fits a na\"ive metallictiy profile where the only variables are the gradient and the central metallicity, then the majority of datapoints would count as \textgreater\,$5\,\sigma$ `outliers' in M\,83 (and most would count as \textgreater\,$2\,\sigma$ outliers for M\,74). This scatter represents localised variations in metallicity, which are not azimuthally-symmetric -- and which therefore cannot be captured by a 1-dimensional model. Such variation becomes apparent when sampling the metallicity within galaxies at such high spatial resolution \citep{RosalesOrtega2010A,Moustakas2010B}. For example, note the localised region of significantly depressed metallicity in the western part\footnote{\label{Footnote:M74_Low_Metallicity_Coods}Centred at approximately: $\alpha=204.20^{\circ}$, $\delta=-29.87^{\circ}$.} of the disc of M\,83, visible in the upper-left panel of Figure~\ref{Fig:NGC5236_Metallicity_Grid}.

\begin{table}
\centering
\caption{Results of our modelling of the radial metallicity profiles of M\,74 and M\,83. Stated values are posterior medians, with uncertainties indicating the 68.3\%\ credible interval (all posteriors were symmetric and Gaussian).}
\label{Table:Metalliciy_Profile}
\begin{tabular}{lrr}
\toprule \toprule
\multicolumn{1}{c}{} &
\multicolumn{1}{c}{M\,74} &
\multicolumn{1}{c}{M\,83} \\
\cmidrule(lr){2-3}
$m_{Z}$ (${\rm dex\,r^{-1}_{25}}$) & $-0.27 \pm 0.04$ & $-0.14 \pm 0.02$ \\
$c_{Z}$ (\logOH) & $8.59 \pm 0.02$ & $8.62 \pm 0.01$ \\
$\psi$ (dex) & $0.044 \pm 0.01$ & $ 0.048 \pm 0.01$ \\
\bottomrule
\end{tabular}
\end{table}

We had to take this intrinsic scatter into account when modelling the radial metallicity profiles of our target galaxies; we therefore used a model with 3 parameters: the metallicity gradient $m_{Z}$ (in ${\rm dex\,r^{-1}_{25}}$), the central metallicity $c_{Z}$ (in \logOH), and the intrinsic scatter $\psi$ (in dex). We employed a Bayesian Monte Carlo Markov Chain (MCMC) approach to fit this model, the full details of which are given in Appendix~\ref{AppendixSection:Metallicity_Profile_Fitting}; the resulting parameter estimates, with uncertainties, are listed in Table~\ref{Table:Metalliciy_Profile}.

It would technically be possible to create metallicity maps of our target galaxies using only these fitted radial metallicity profiles. However, using this simple 1-dimensional approach (ie, where metallicity varies only as a function of $r$) leads to very large uncertainties on the metallicity value of each pixel in the resulting maps, thanks to the considerable intrinsic scatter values ($\psi = 0.044\,{\rm dex}$ for M\,74, and $\psi = 0.049\,{\rm dex}$ for M\,83). In contrast, most of the individual spectra metallicity datapoints have uncertainties much smaller than this, with median uncertainties of  0.010 and 0.025\,dex for M\,74 and M\,83 respectively (NB, spectra located in close proximity tend to have metallicities that are in good agreement -- see the densely-sampled area in Figures~\ref{Fig:NGC0628_Metallicity_Grid} and \ref{Fig:NGC5236_Metallicity_Grid}). In other words, there are many areas of these galaxies where the metallicity is known to much greater confidence than is reflected by the global radial metallicity gradient -- therefore, relying upon the global 1-dimensional model alone would mean `throwing away' that information. As such, we opted to model the metallicity distributions of our target galaxies in 2 dimensions. To achieve this, we employed Gaussian process regression.

\subsubsection{Gaussian Process Regression} \label{Subsubsection:GPR}

\begin{figure*}
\centering
\includegraphics[width=0.975\textwidth]{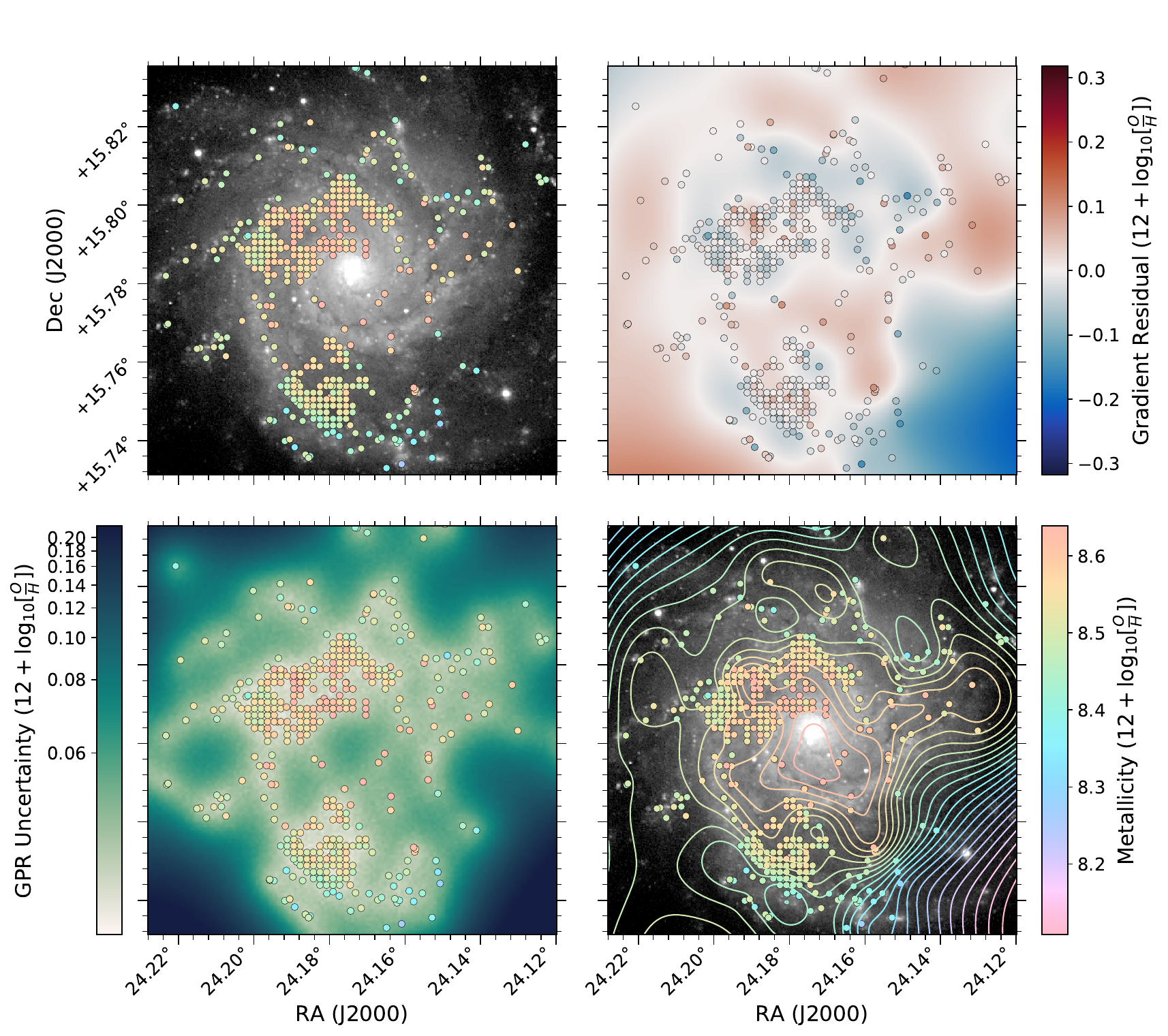}
\caption[]{Illustration of our Gaussian Process Regression (GPR) metallicity mapping procedure, for M\,74. {\it Upper left:} Markers show the positions of spectra, colour-coded to indicate their metallicity (as per the colour bar at the lower right of the figure), plotted on a \spitz\ 3.6\,\micron\ image. {\it Upper right:} Points show the residual between the metallicity of each spectra, and the global radial metallicity profile at that position. Red points have a positive residual, blue points have a negative residual. Background image shows the GPR model to these residuals. {\it Lower left:} Background image shows the uncertainty on the GPR, with positions of spectra plotted on top (again colour coded according their individual metallicities, as per the colour bar at the lower right of the figure). The regression tends to have much lower uncertainty in area more densely sampled with spectra. {\it Lower right:} Same as upper left panel, but now with the final GPR metallicity map traced with colour-coded contours. This final metallicity map was produced by adding the GPR residual model shown, in the upper right panel, to the global radial metallicity profile. The colour scale used to indicate metallicity is red-to-red circular (therefore preserving sequentiality for all kinds of colour blindness) and approximately isoluminant (therefore reverting to a near-constant shade when displayed in greyscale).}
\label{Fig:NGC0628_Metallicity_Grid}
\end{figure*}

\begin{figure*}
\centering
\includegraphics[width=0.975\textwidth]{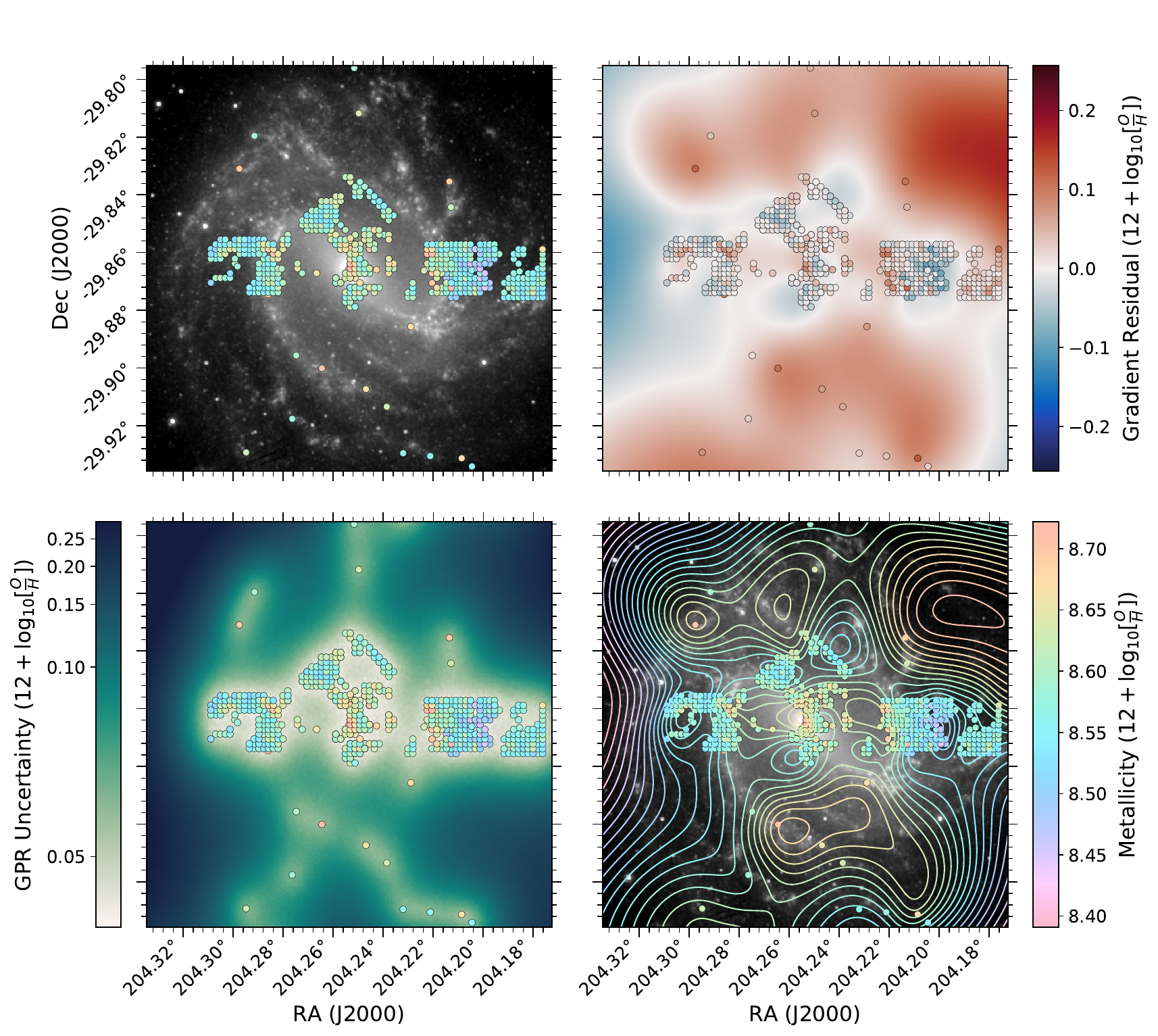}
\caption{Metallicity mapping for M\,83. Description as per Figure~\ref{Fig:NGC0628_Metallicity_Grid}. Localised variations in metallicity are as prominent as the global gradient, as expected given Figure~\ref{Fig:Metallicity_Gradient}. The high-metallicity (and high-uncertainty) region extrapolated by the GPR to the northwest of M\,83 is driven by the fact that the closest spectra to this area have metallicities above what would be predicted from the global gradient.}
\label{Fig:NGC5236_Metallicity_Grid}
\end{figure*}

Gaussian Process Regression (GPR) is a form of probabilistic interpolation, that makes it possible to model a dataset without having to assume any sort of underlying functional form for the model. GPR (and Gaussian process methodology in general) is a commonly-applied tool in the field of machine learning -- and in recent years GPR has seen increasing use in astronomy, to tackle problems where stochastic (and therefore impractical to model directly) processes give rise to complex features in data (for instance, capturing the effect of varying detector noise levels in time-domain data). For a full introduction to Gaussian process methodology, including GPR, see \citet{Rasmussen2006}; for an extensive list of works where Gaussian processes have been successfully applied to problems in astronomy, see Section~1 of \citet{Angus2018A}.

Instead of trying to model the underlying function that gave rise to the observed data, GPR models the {\rm covariance} between the datapoints. The covariance is modelled using a kernel, which describes how the values of datapoints are correlated with one another, as a function of their separation in the parameter space.

This covariance-modelling approach is well-suited to the problem we face with mapping metallicity within our target galaxies. Spectra located very close together (eg, within a few arcseconds) will tend to have very similar metallicities, whilst spectra with greater separations (eg, arcminutes apart) will only be weakly correlated with one another (this is readily apparent from visual inspection of Figures~\ref{Fig:NGC0628_Metallicity_Grid} and \ref{Fig:NGC5236_Metallicity_Grid}).

For the covariance function, we used a M\'atern kernel \citep{Stein1999}. The M\'atern function is a standard choice for modelling the spatial correlation of 2-dimensional data (\mbox{\citealp{Minasny2005C};}\citealp{Rasmussen2006,Cressie2011}) -- especially physical data \citep{Schon2018}. In practice, a M\'atern kernel is similar to a Gaussian kernel, but has a narrower peak (allowing it to be sensitive to variations over short distances) whilst also having thicker tails (letting it maintain sensitivity to the covariance over large distances). Like a Gaussian, the tails extend to infinity. The M\'atern kernel has two hyperparameters: kernel scale, and kernel smoothness (essentially how `sharp' the peak of the kernel is). 

Once the covariance has been modelled, it is used in combination with the observed data to trace the underlying distribution. The result is a full posterior Probability Distribution Function (PDF) for the likely value of the underlying function at that location. The uncertainties in each input datapoint are fully considered by GPR. In regions where the input datapoints have large uncertainties, or where datapoints in close proximity disagree with one another, the output PDF will be less well constrained, reflecting the greater uncertainty on the underlying value at that location.


\subsubsection{Metallicity Maps Via Gaussian Process Regression} \label{subsubsubsection:GPR_Maps}

We opted to apply the GPR to the {\it residuals} between the individual spectra metallicity points and the global radial metallicity profile (ie, Figure~\ref{Fig:Metallicity_Gradient}). By fitting to the residuals, the global radial metallicity profile effectively serves as the prior for the regression. The regression then traces the structure of the local deviations from the global radial metallicity profile. In regions where there are no data points, the GPR therefore tends to revert to the metallicity implied by the global radial profile.

This process is illustrated in the upper-right panels of Figures~\ref{Fig:NGC0628_Metallicity_Grid} and \ref{Fig:NGC5236_Metallicity_Grid} for M\,74 and M\,83 respectively. The circular points mark the positions of the individual spectra metallicities, colour-coded to show the residual of each (the median absolute residual is 0.026\,dex for both galaxies). The coloured background shows the Gaussian process regression to these residuals, similarly colour-coded. We used {\tt GaussianProcessRegressor}, the GPR implementation of the Scikit-Learn machine learning package for {\sc python} \citep{Scikit-Learn2011}. The hyperprior for the kernel scale was flat, but limited to a range of 0.05--0.5\,$D_{25}$, to prevent the modelled regression being either featurelessly smooth, or unrealistically granular. The kernel smoothness hyperprior was set to 1.5, which is a standard choice due to being computationally efficient, differentiable, and often found to be effective in practice \citep{Rasmussen2006,Gatti2015A}.

The final metallicity map for each galaxy was produced by adding the residual distribution traced by the GPR to the global radial metallicity profile, for each pixel. The resulting metallicity maps are plotted as contours in the lower-right panels of Figures~\ref{Fig:NGC0628_Metallicity_Grid} and \ref{Fig:NGC5236_Metallicity_Grid}, for M\,74 and M\,83 respectively. Visual inspection indicates that the GPR does a good job of tracing the metallicity distribution as sampled by the spectra metallicity points (ie, the contours consistently have the same levels as the points they pass through).

Our full procedure for calculating the uncertainty on the GPR metallicity in each pixel is presented in Appendix~\ref{AppendixSection:GPR_Uncertainties}. The resulting metallicity uncertainty maps are shown in the lower-left panels of Figures~\ref{Fig:NGC0628_Metallicity_Grid} and \ref{Fig:NGC5236_Metallicity_Grid}.

We validated the reliability of the metallicities predicted by GPR by performing a jackknife cross-validation analysis, which is described in detail in Appendix~\ref{AppendixSection:GPR_Validation}. This analysis found that the predicted values exhibit no significant bias, and the associated uncertainties are reliable.

There are areas in both galaxies where the datapoints suggest a steadily-increasing residual in a certain direction; the GPR then extrapolates that this increase continues for some distance (defined by the modelled kernel scale) into regions where there are no datapoints. For instance, in the south-western part of M\,74, the datapoints suggest that the metallicity gradient is steeper than for the rest of the galaxy (ie, a trend of increasingly negative residuals) -- the GPR extrapolates that this increased steepness will continue for a certain distance into an area where there are no metallicity points. A similar situation occurs in the north-west portion of M\,83 (but instead with a positive residual). Naturally, extrapolations such as these are highly uncertain; but this is quantified by the uncertainty on the regression at these locations. This is illustrated in the lower-left panels of Figures~\ref{Fig:NGC0628_Metallicity_Grid} and \ref{Fig:NGC5236_Metallicity_Grid}, which show the uncertainty for each pixel's predicted metallicity.

Utilising GPR provides a marked reduction in the uncertainty of our metallicity maps, relative to using the global radial metallicity profiles alone. If we were to use that simple global approach, every pixel in our metallicity map for M\,74 would have an uncertainty at least as large as the intrinsic scatter of 0.044\,dex (Table~\ref{Table:Metalliciy_Profile}). In contrast, with our GPR metallicity map of M\,74, 91\% of the pixels {within the region of interest\footnoteref{Footnote:Kappa_Map_Region} have uncertainties \textless\,0.044\,dex; the median GPR uncertainty within this region is only 0.016\,dex. Similarly, whereas the intrinsic scatter on the global radial profile of M\,83 is 0.048\,dex, the median error on the GPR metallicity map is only 0.037\,dex within the region of interest; the GPR uncertainty is less than the global intrinsic scatter for 66\% of the pixels within this region. 

There exist `direct' electron temperature metallicity measurements for M\,74, produced by the CHemical Abundances Of Spirals (CHAOS; \mbox{\citealp{Berg2015A}}). Electron temperature metallicities are at reduced risk of systematic errors, compared to strong-line values like those provided by \citet{DeVis2019B}. However, the CHAOS data for M\,74 only consists of 45 measurements. Whilst we trialled producing metallicity maps with this data, the sparse sampling meant that the uncertainty on the metallicity at any given point was extremely large. Maps of \kappad\ produced with these metallicity maps (as per the procedure described in Section~\ref{Section:Application}) were so dominated by the resulting noise that they were not informative.

\subsection{Atomic \& Molecular Gas Data} \label{Subsection:Gas_Data}

Atomic and molecular gas data for a sample of extended, face-on spiral galaxies in DustPedia -- including those studied in this work -- is presented in \citet{Casasola2017A}. For both of our target galaxies, we followed \citet{Casasola2017A} and use \HI\ data from The HI Nearby Galaxy Survey (THINGS, \citealp{Walter2008O}), which conducted 21\,cm observations of 34 nearby galaxies with the Very Large Array, at 6--16\arcsec\ resolution. We retrieved the naturally-weighted moment 0 maps for M\,74 and M\,83 from the THINGS website\footnote{\url{https://www.mpia.de/THINGS/Overview.html}}. The \HI\ maps for both galaxies are shown in the 3\textsuperscript{rd} panels of Figures~\ref{Fig:NGC0628_Multiwavelenth_Grid} and \ref{Fig:NGC5236_Multiwavelenth_Grid}.

To obtain CO observations for M\,74 we again followed \citet{Casasola2017A}, and used data from the HERA {\sc Co} Line Extragalactic Survey (HERACLES; \citealp{Leroy2009B}), which performed CO(2-1) observations of 18 nearby galaxies using the IRAM 30\,m telescope, at 13\arcsec\ resolution. We retrieved the moment 0 maps, as associated uncertainty maps, from IRAM's official HERACLES data repository\footnote{\url{https://www.iram-institute.org/EN/content-page-242-7-158-240-242-0.html}}. The CO(2-1) map for M\,74 is shown in the 4\textsuperscript{th} panel of Figure~\ref{Fig:NGC0628_Multiwavelenth_Grid}.

Although M\,74 has been observed in CO(1-0) by various authors \citep{Young1995A,Regan2001A}, these observations are all lacking in either resolution, sensitivity, and/or coverage, in comparison to the HERACLES data. We therefore found it preferable to use the CO(2-1) data of HERACLES, despite the fact this requires applying a line ratio, $r_{2:1} = I_{\rm CO(2-1)} / I_{\rm CO(1-0)}$, in order to find $I_{\rm CO(1-0)}$, and hence calculate ${\rm H_{2}}$ mass as per Equation~\ref{Equation:H2_Mass}.

\begin{figure}
\centering
\includegraphics[width=0.475\textwidth]{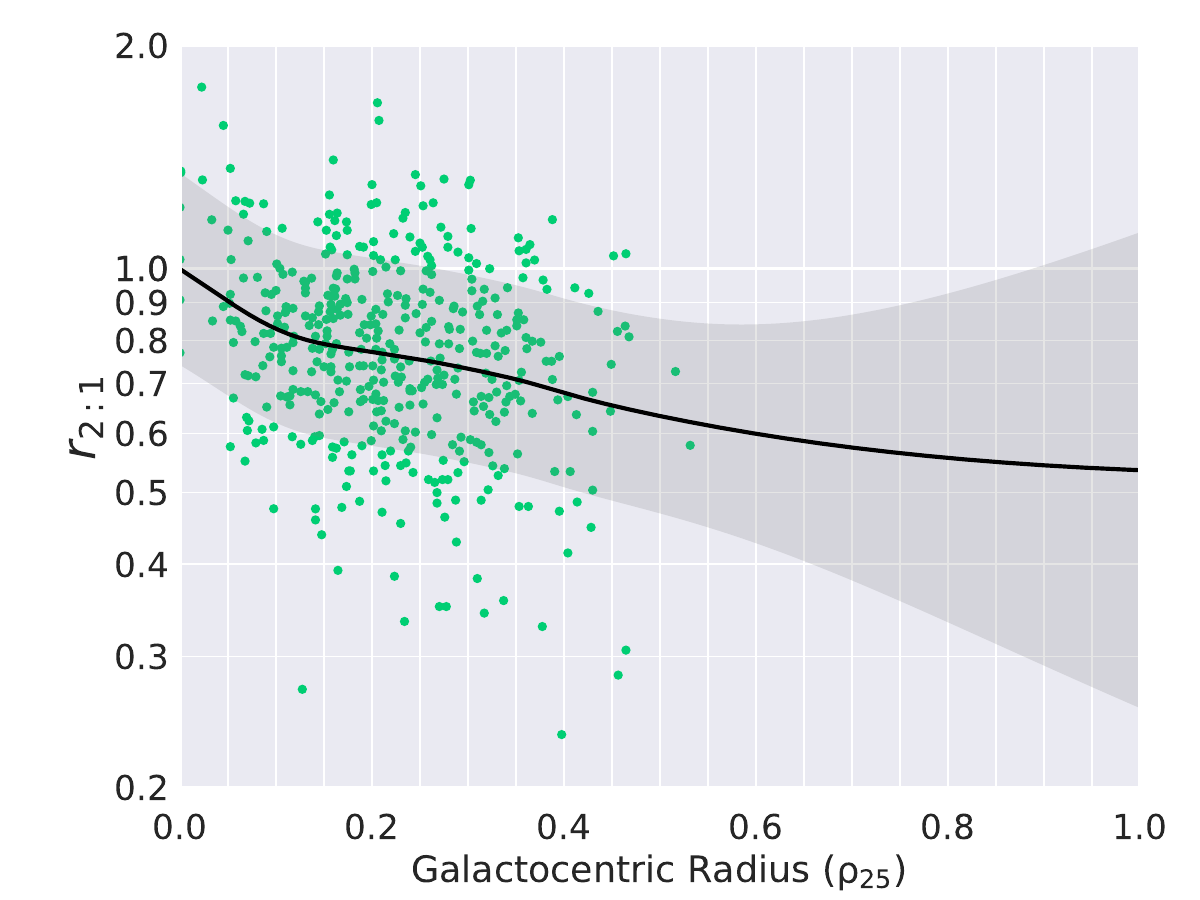}
\caption{$r_{2:1}$ values from Figure~34 (lower-right panel) of \citet{Leroy2009B}, plotted against galactocentric radius in terms of $R_{25}$. The black line shows our Gaussian process regression to this data, with the grey shaded area indicating the 1$\sigma$ uncertainty.}
\label{Fig:Leroy+2009_r21_GPR}
\end{figure}

In nearby late-type galaxies, $r_{2:1}$ has an average value of $\sim$\,0.7 \citep{Leroy2013B,Casasola2015B,Saintonge2017A}. However, it is also known that $r_{2:1}$ varies significantly with galactocentric radius \citep{Casoli1991D,Sawada2001B,Leroy2009B}. As such, accurately inferring the CO(1-0) distribution in M\,74 using the HERACLES CO(2-1) map required a radially-dependent $r_{2:1}$. To produce this, we used the data presented in Figure~34 (lower-right panel) of \citet{Leroy2009B}, where they compare the HERACLES $I_{\rm CO(2-1)}$ maps to literature $I_{\rm CO(1-0)}$ maps of the same galaxies produced by several other telescopes (with appropriate corrections applied to account for differences in spatial and velocity resolution). This yielded $\approx$\,450 directly-measured $r_{2:1}$ values, spanning radii from 0--0.55\,$R_{25}$, for 9 of the HERACLES galaxies. \citet{Leroy2009B} simply binned these points to trace the radial variation in $r_{2:1}$; however, we chose to take a fully probabilistic approach, and use GPR to infer the underlying radial trend in $r_{2:1}$. In Figure~\ref{Fig:Leroy+2009_r21_GPR}, we plot all of the $r_{2:1}$ points from Figure~34 (lower-right panel) of \citet{Leroy2009B}. We applied a GPR to this data, using a M\'atern covariance kernel. Because $r_{2:1}$ is a ratio, we constructed the regression so that the output uncertainties are symmetric in logarithmic space; otherwise, output uncertainties symmetric in linear space would extend to unphysical values of $r_{2:1} < 0$ at larger radii. The resulting regression is shown in black in Figure~\ref{Fig:Leroy+2009_r21_GPR}. It is in excellent agreement with the radial trend that \citet{Leroy2009B} traced by binning the data, with $r_{2:1}$ elevated to $\sim$\,1 in the galaxies' centres, falling to 0.7--0.8 over the rest of the sampled region -- but our approach has the added benefit over binning of providing well-constrained uncertainties on $r_{2:1}$ values produced using the regression. The uncertainty associated with the regression is a factor of $\approx$\,1.3 over the $0 < R/R_{25} < 0.55$ range in radius sampled by the HERACLES measurements, reflecting the intrinsic scatter present in the datapoints; beyond this, the uncertainty steadily increases, reaching a factor of $\approx$\,2 at $R = R_{25}$. Given the uncertainty on $\alpha_{\rm CO}$, this does not represent a large addition to the total uncertainty on the molecular gas masses we calculated.

M\,83 was not observed by HERACLES. So we instead used the CO(1-0) observations presented in \citet{Lundgren2004A}, which were made using the Swedish--{\sc Eso} Submillimetre Telescope (SEST) at a resolution of 42\arcsec, to a uniform depth of 74\,mK ($T_{\it mb}$). The CO(1-0) map for M\,83 is shown in the far-right panel of Figure~\ref{Fig:NGC5236_Multiwavelenth_Grid}.

We determined $\alpha_{\rm CO}$ pixel-by-pixel using our metallicity maps according to Equation~\ref{Equation:Alpha_CO}, and thereby produced ${\rm H_{2}}$ maps of our target galaxies. The total ${\rm H_{2}}$ masses contained in these maps are the ${\rm H_{2}}$ masses listed in Table~\ref{Table:Galaxy_Properties}.

\section{Application} \label{Section:Application}

\subsection{Data Preparation} \label{Subsection:Data_Preparation}

We background-subtracted all continuum maps following the procedure described in \citet{CJRClark2018A}, using the background annuli they specify for our target galaxies.

All data (continuum  observations, gas observations, and metallicity maps) were smoothed to the resolution of the most poorly-resolved observations for each galaxy. This was done by convolving each image with an Airy disc kernel of Full-Width Half-Maximum (FWHM) given by $\smash{\theta_{\it kernel} = ( \theta^{2}_{\it worst} - \theta^{2}_{\it data}} )^{\frac{1}{2}}$. We therefore convolve all of our M\,74 data to the 36\arcsec\ resolution of the \hersc-SPIRE 500\,\micron\ observations. Likewise, we convolved all of our M\,83 data to the 42\arcsec\ resolution of the SEST \HI\ observations.

We reprojected all of our data to a common pixel grid for each galaxy, on an east--north gnomic tan projection. We wished to preserve angular resolution, ensuring that our data remain Nyquist sampled, to maximise our ability to identify any spatial features or trends in our final \kappad\ maps. We therefore used projections with 3 pixels per convolved FWHM. This corresponds to 12\arcsec\ pixels for M\,74, and 14\arcsec\ pixels for M\,83.

For each galaxy, we defined a region of interest, within which all required data is of sufficient quality to effectively map \kappad. We defined this as being the region within which all pixels in the smoothed \& reprojected versions of the \HI\ map, CO map, and 22--500\,\micron\ continuum maps, have SNR\,\textgreater\,2 (as defined by comparison to their respective uncertainty maps). For both M\,74 and M\,83, the data with the limiting sensitivity are the CO observations. The borders of our regions of interest for both galaxies are shown in the far-right panels of Figures~\ref{Fig:NGC0628_Multiwavelenth_Grid} and \ref{Fig:NGC5236_Multiwavelenth_Grid}.


\subsection{SED Fitting} \label{Subsection:SED_Fitting}

\begin{figure}
\centering
\includegraphics[width=0.475\textwidth]{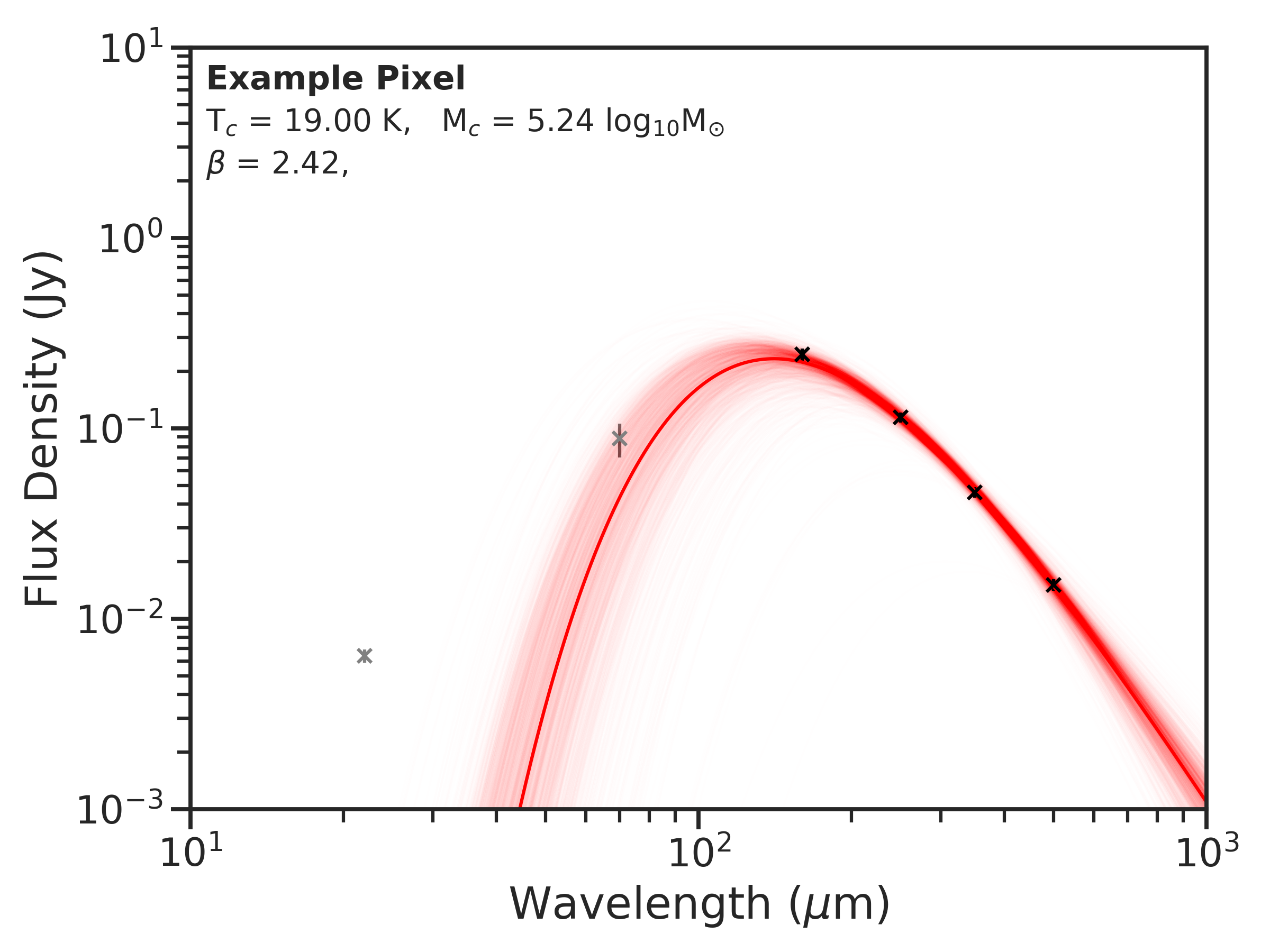}
\caption[]{The posterior SED modelled by our SED fitting for an example pixel in M\,74 (specifically, at $\alpha = 24.1820^{\circ}, \delta = 15.7755^{\circ}$). The black crosses show standard fluxes, whilst the grey crosses are fluxes that serve as upper limits; error bars are drawn for both. The pale red translucent lines show the SEDs corresponding to 500 samples from the posterior distribution. The solid red line shows the data space median\ posterior SED (being the posterior sample for which half of all other samples are brighter, and half fainter, averaged over the wavelength range for which data is present), and the text in the figure give its parameters. The corresponding posterior parameter distributions are shown in Figure~\ref{Fig:SED_Corner}.} 
\label{Fig:SED_Example}
\end{figure}

As described in Section~\ref{Section:Theory}, the dust-to-metals method lets us establish dust masses {\it a priori}; then, by comparing this {\it a priori} dust mass to observed FIR--submm dust emission, we can calibrate the value of \kappad. This necessitates having a model that describes that FIR--submm dust emission. We wished to minimise the scope for potentially-incorrect model assumptions to corrupt our resulting \kappad\ values. We therefore modelled the dust emission with the simplest model that is able to fit FIR--submm fluxes -- a one-component MBB (ie, Equation~\ref{Equation:Greybody}, with $n = 1$).  A one-component MBB model has been shown by many authors to break down in various circumstances (eg: \citealp{Jones2013A,CJRClark2015A,Chastenet2017A}; Lamperti  et al. {\it accepted}). However, these primarily concern either submillimetre excess in low-metallicity and/or low-density environments (which are not present in the regions of interest within our target galaxies), the emission from hotter dust components at short wavelengths (which we do not attempt to model; see below), or features only discernable in spectroscopy (which we are not employing). In `normal' galaxies, a one-component MBB can be expected to fit FIR--submm fluxes successfully \citep{Nersesian2019A}. 

Note that, as a test, we also repeated the entire SED fitting process described in this section with a two-component MBB model (ie, Equation~\ref{Equation:Greybody}, with $n = 2$, giving dust components at two temperatures). However, when comparing the $\chi^{2}$ values of both sets of fits, we found that adopting the two-component MBB approach adds little benefit to the quality of the fits. The median reduced $\chi^{2}$ values (of all posterior samples, from all pixels) for the one-component MBB fits were 0.61 for M\,74 and 0.94 for M\,83 -- compared to 0.59 and 0.65 respectively for the two-component fits. This indicates that the two-component MBB fits offer minimal improvement over the one-component fits (and, indeed, may be straying into the realm of over-fitting). Given our desire to employ the simplest applicable model, we therefore opt to proceed with the one-component MBB approach for this work. Nonetheless, in Appendix~\ref{AppendixSection:Kappa_2MBB}, we verify that the choice of one- or two-component SED fitting does not result in considerable changes to our overall results.

\begin{figure}
\centering
\includegraphics[width=0.475\textwidth]{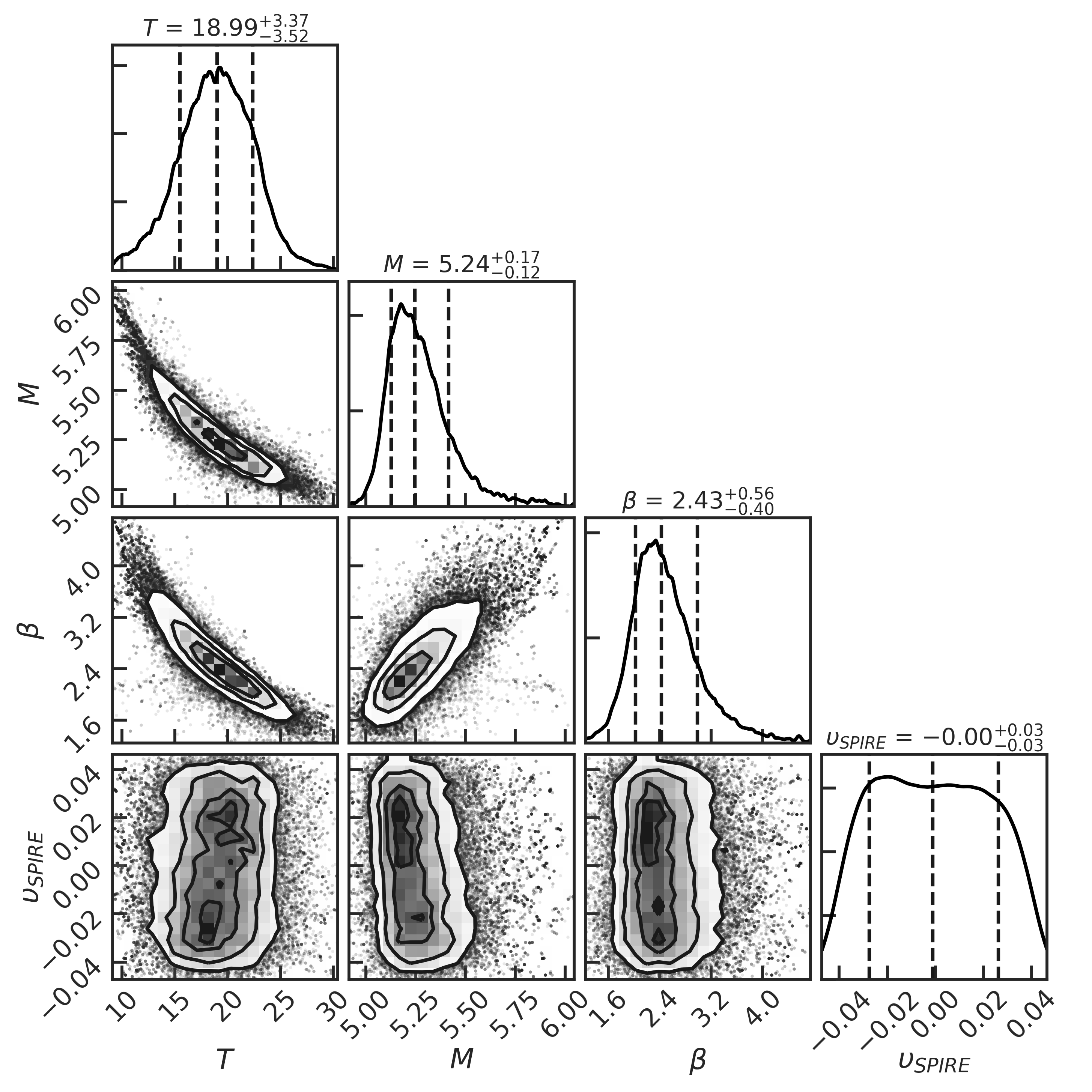}
\caption[]{Corner plot showing the covariances of the posterior distributions of the free parameters modelled in our SED fitting, for an example pixel in M\,74 (specifically, at $\alpha = 24.1820^{\circ}, \delta = 15.7755^{\circ}$). The two-parameter distributions have contours indicating the regions containing 68.3\%, 95.5\%, 99.7\%, and 99.9\% of the posterior samples; probability density is indicated as a shaded density histogram within the contoured region, whilst outside of the contoured region the samples are plotted as individual points. The individual parameter distributions, plotted at the top of each column as KDEs, are annotated with the median values, along with the boundaries of the 68.3\% credible interval as $\pm$ values (with masses given in units of ${\rm log_{10}\,M_{\odot}}$). The corresponding posterior SEDs, plotted in data space, are shown in Figure~\ref{Fig:SED_Example}.}
\label{Fig:SED_Corner}
\end{figure}

By performing our SED fitting pixel-by-pixel, we are reducing the degree to which there will be contributions from multiple dust components at different temperatures. Nonetheless, there will inevitably be some degree of line-of-sight mixing of dust populations. This risk will be greatest in the densest regions, where fainter emission from colder, but potentially more massive, dust components can be dominated by brighter emission from warmer, but less massive, components heated by star-formation \citep{Malinen2011A,Juvela2012C}. If this does occur, then the resulting \kappad\ values will, in effect, factor in the mass of any cold dust component too faint to affect the SED (assuming the {\it a priori} dust masses calculated by the dust-to-metals method are accurate). In this scenario, the \kappad\ values we calculate may not be valid if applied to observations with good enough spatial resolution that line-of-sight mixing becomes negligible.

Although we use Equation~\ref{Equation:Greybody} to model SEDs, we assign an arbitrary value of $\kappa_{\lambda}$ during the fitting process (as, of course, the SED fitting is being performed in order to allow us to {\it find} a value of $\kappa_{\lambda}$ using the results). This means that the `mass' parameter yielded by our SED fitting merely serves as a normalisation term for the SED amplitude. This is not a problem, as the only output values actually required is the temperature of the dust, and its flux at the reference wavelength; these are needed in Equation~\ref{Equation:Kappa} to calculate values of \kappad.


\begin{figure*}
\centering
\includegraphics[width=0.975\textwidth]{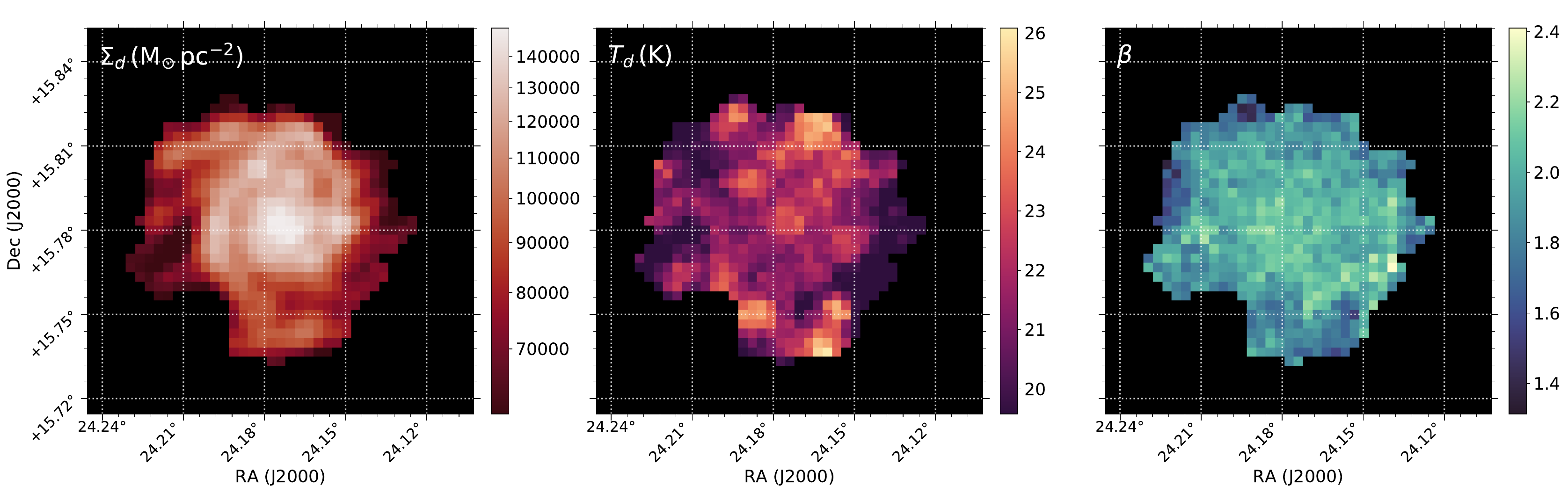}
\caption{Maps showing the results of our SED fitting of M\,74. {\it Left:} Map of dust mass surface density ($\Sigma_{d}$, in ${\rm M_{\odot}\,pc^{2}}$). {\it Centre:} Map of dust temperature ($T_{d}$, in K). {\it Right:} Map of dust emissivity spectral index ($\beta$).}
\label{Fig:NGC0628_SED_Grid}
\end{figure*}

\begin{figure*}
\centering
\includegraphics[width=0.975\textwidth]{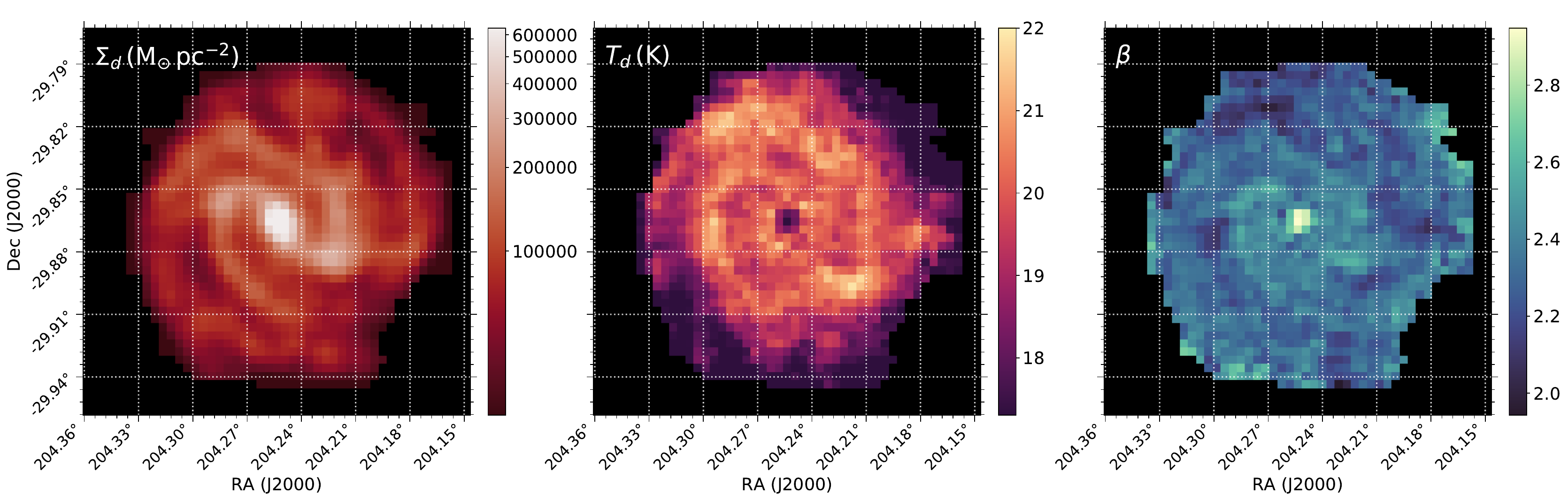}
\caption{Maps showing the results of our SED fitting of M\,83. Description as per Figure~\ref{Fig:NGC0628_SED_Grid}.}
\label{Fig:NGC5236_SED_Grid}
\end{figure*}

We also incorporate a correlated photometric error parameter, $\upsilon_{\rm SPIRE}$, into our SED-fitting. The photometric calibration uncertainty of the \hersc-SPIRE instrument contains a systematic error component that is correlated between bands \citep{Griffin2010D,Bendo2013A,Griffin2013A}. This arises from the fact that \hersc-SPIRE was calibrated using observations of Neptune; however, the reference model of Neptune's emission has a $\pm 4 \%$ uncertainty. We account for this by parameterising the correlated \hersc-SPIRE error as $\upsilon_{\rm SPIRE}$. The $\pm 4 \%$ scale of $\upsilon_{\rm SPIRE}$ accounts for the majority of the combined 5.5\% calibration uncertainty of \hersc-SPIRE\footnote{\label{Footnote:SPIRE_Wiki}SPIRE Instrument \& Calibration Wiki: \url{https://herschel.esac.esa.int/twiki/bin/view/Public/SpireCalibrationWeb}}. As such, for high-SNR sources (such as bright pixels within our target galaxies), where the photometric noise is minimal, the correlated calibration error can actually dominate the entire uncertainty budget. Moreover, the $\pm 4 \%$ error on $\upsilon_{\rm SPIRE}$ does not follow the Gaussian or Student's $t$ distribution typically assumed for photometric uncertainties -- rather, it is essentially flat, with the true value of the correlated systematic error almost certainly lying somewhere within the $\pm 4 \%$ range (\citealp{Bendo2013A}; A.\,Papageorgiou, {\it priv. comm.}; C.\,North, {\it priv. comm.}). Explicitly handling $\upsilon_{\rm SPIRE}$ as a nuisance parameter allows us to properly account for this with a matching prior. \citet{Gordon2014B} highlight the significant differences that can be found in dust SED fitting when the correlated photometric uncertainties are considered, compared to when they are not.

The \hersc-PACS instrument also has a systematic calibration error, of $\pm 5 \%$, arising from uncertainty on the emission models of its calibrator sources, a set of 5 late type giant stars \citep{Balog2014B}. However, the error budget on the emission models is dominated by the $\pm 3 \%$ uncertainty on the line features in the atmospheres of the calibrator stars (see Table~2 of \citealp{Decin2007B}), which are different in each band, and hence not correlated. Only the uncertainty on the continuum component of the emission model, of $\pm 1$--$2 \%$, will be correlated between bands. Given the small scale of this correlated error component, and given that systematic error makes up a smaller fraction of the total \hersc-PACS calibration uncertainty than it does for \hersc-SPIRE, and given that the greater instrumental noise for \hersc-PACS means that calibration uncertainty makes up a small fraction of the total photometric uncertainty budget than it does for \hersc-SPIRE, we opt to not model the correlated uncertainty for \hersc-PACS as we do with $\upsilon_{\rm SPIRE}$.

Our one-component MBB SED model therefore has 4 variables: the dust temperature, $T_{d}$;  the dust `mass' normalisation, $M_{d}^{\rm (norm)}$; the emissivity slope, $\beta$; and the correlated photometric error in the \hersc-SPIRE bands, $\upsilon_{\rm SPIRE}$.

The resulting likelihood function, for a set of fluxes $S$ (in Jy), observed at a set of wavelengths $\lambda$ (in m), with a corresponding set of uncertainties $\sigma$ (in Jy), for a set of size $n_{\lambda}$, takes the form:
\begin{multline}
\mathcal{L}(S |\, \lambda, \sigma, T_{d}, M_{d}^{\rm (norm)}, \beta, \upsilon_{\rm SPIRE}) =\\
\prod^{n_{\lambda}}_{i} \left( t(d,S_{d_{i}},\sigma_{i}) + S_{d_{i}}\upsilon_{\rm SPIRE} \right)
\label{Equation:SED_Likelihood}
\end{multline}

\noindent where, for the $i$\textsuperscript{th} wavelength in the set, $S_{d_{i}}$ is the flux arising from dust emission given the SED model parameters, and $\sigma_{i}$ is the corresponding uncertainty; $t(d,S_{d_{i}},\sigma_{i})$ is a $d$\textsuperscript{th}-order Student $t$ distribution\footnote{Standardised to allow modes and widths other than zero, as per the SciPy \citep{SciPy2001} implementation: \url{https://docs.scipy.org/doc/scipy/reference/generated/scipy.stats.t.html}}, centred at a mode of $S_{d_{i}}$, with a width of $\sigma_{i}$. The expected dust emission $S_{d_{i}}$ is given by:
\begin{equation}
S_{d_{i}} = \frac{1}{D^{2}}\, \kappa_{0} \left( \frac{\lambda_{0}}{\lambda_{i}}\right)^{\beta} M_{d}^{\rm (norm)} B(\lambda_{i},T_{d})
\label{Equation:SED_Dust_Flux}
\end{equation}

We treat photometric uncertainties as being described by a 1\textsuperscript{st}-order (ie, one degree of freedom) Student $t$ distribution. The Student $t$ distribution has more weight in the tails than a Gaussian distribution, allowing it to better account for outliers. This makes the Student $t$ distribution a standard choice for Bayesian SED fitting \citep{DaCunha2008A,Kelly2012B,Galliano2018B}.

For the photometric uncertainty in each pixel, we used the values provided by the uncertainty maps, added in quadrature to the calibration uncertainty of each band: 5.6\% for WISE 22\,\micron\footnote{WISE All-Sky Release Explanatory Supplement \citep{Cutri2012E}: \url{https://wise2.ipac.caltech.edu/docs/release/allsky/expsup/sec4_4h.html}}, 7\% for \hersc-PACS 70--160\,\micron\footnote{PACS Instrument \& Calibration Wiki: \url{https://herschel.esac.esa.int/twiki/bin/view/Public/PacsCalibrationWeb}}, and 2.3\%\footnote{2.3\% being the non-correlated component of the \hersc-SPIRE calibration uncertainty, separate from $\upsilon_{\rm SPIRE}$.} for \hersc-SPIRE 250--500\,\micron\footnoteref{Footnote:SPIRE_Wiki}. Both of our target galaxies lie in regions with negligible contamination from Galactic cirrus. The WISE and \hersc-PACS backgrounds are dominated by instrumental noise, whilst the \hersc-SPIRE background has a significant contribution from the confused extragalactic background. Therefore, for the \hersc-SPIRE data, we also add in quadrature the contribution of confusion noise; for this we use the values given in \citet{MWLSmith2017A}, of 0.282, 0.211, 0.105\,${\rm MJy\,sr^{-1}}$ at 250, 350, and 500\,\micron\ respectively, derived from the \hersc-ATLAS fields (although the instrumental noise level still dominates over this in all of our \hersc-SPIRE data).

We treat fluxes at wavelengths \textless\,100\,\micron\ as upper limits, as emission in this regime will include contributions from hot dust and stochastically heated small grains \citep{Boulanger1988,Desert1990,Jones2013C} that will not be accounted for by our MBB model. Therefore at these wavelengths, any proposed model flux that falls below the observed flux will be deemed as likely as the observed flux itself (ie, no proposed model will be penalised for under-predicting the flux in these bands). Only for proposed model fluxes greater than the observed flux will the likelihood decrease according to the Student $t$ distribution, as per usual.

We sample the posterior probability distribution of the SED model parameters in each pixel using the {\sc emcee} \citep{ForemanMackey2013B} MCMC package for {\sc python}. We perform 750 steps with 500 chains (`walkers'); the first 500 steps from each chain were discarded as burn-in, and non-convergence was checked for using the Geweke diagnostic\footnote{Comparing the means of the last 90--100\% quantile of the combined chains to the 50--60\% quantile.} \citep{Geweke1992}. Our priors are detailed in Appendix~\ref{AppendixSection:SED_Priors}. 

Our SED fitting routine incorporates colour-corrections to account for the effects of the instrumental filter response functions and beam areas\footnote{WISE colour corrections from \citet{Wright2010F}.}\textsuperscript{,}\footnote{\spitz-MIPS colour corrections from the MIPS Instrument Handbook, version 3 \citep{Spitzer-MIPS2011}: \url{https://irsa.ipac.caltech.edu/data/SPITZER/docs/mips/mipsinstrumenthandbook/51/\#_Toc288032329}}\textsuperscript{,}\footnote{\hersc-PACS colour corrections from the PACS Handbook, version 4.0.1 \citep{Herschel-PACAS2019}: \url{https://www.cosmos.esa.int/documents/12133/996891/PACS+Explanatory+Supplement}}\textsuperscript{,}\footnote{\hersc-SPIRE colour corrections from the SPIRE Handbook, version 3.1 \citep{Herschel-SPIRE2017}: \url{https://herschel.esac.esa.int/Docs/SPIRE/spire_handbook.pdf}}. An example posterior SED, along with the corresponding parameter distributions, are shown in Figures~\ref{Fig:SED_Example} and \ref{Fig:SED_Corner}. 

Figures~\ref{Fig:NGC0628_SED_Grid} and \ref{Fig:NGC5236_SED_Grid} show maps of the median values of dust mass surface density, temperature, and $\beta$ values for each pixel. We assume that the low temperatures and large $\beta$ values found in the centre of M\,83 are non-physical, and instead are due to non-thermal emission from the nuclear starburst affecting the SED-fitting. This is limited to a beam-sized area, consisting of 9 pixels - we therefore exclude these pixels from analysis in later sections, where noted.

Unsurprisingly, the maps of dust mass surface density closely match the morphology of the dust emission (see Figures~\ref{Fig:NGC0628_Multiwavelenth_Grid} and \ref{Fig:NGC5236_Multiwavelenth_Grid}). The temperature map for M\,74 is `blotchy', with warmer dust being located around areas of particularly active star formation (compare to the regions of bright MIR emission in Figure~\ref{Fig:NGC0628_Multiwavelenth_Grid} in the northern and southern parts of the disc). The temperature map for M\,83 more visibly traces the overall spiral structure; in particular, elevated temperatures are found on the exterior edges of the spiral arms. The $\beta$ maps for both galaxies show correlations with the dust mass surface density; in M\,74 this manifests as a broad global trend of beta decreasing with radius, whilst in M\,83 beta again more obviously traces the spiral structure.

There is a well-known anticorrelation between temperature and $\beta$ when performing MBB SED fits \citep{Shetty2009A,Kelly2012B,Galliano2018C}. This is clearly in evidence in Figure~\ref{Fig:SED_Corner}. However, as demonstrated by \citet{MWLSmith2012B}, this does not introduce {\it systematic} errors into the results of such fits. And given this lack of systematic bias, the anticorrelation will not introduce spurious trends into resolved SED fits -- because fits separated by more than one beam-width will be independent, and will be no more likely to be biased one way than the other. Combined with the fact that we sample the full posterior in our SED fits, and propagate this into the final calculation of our \kappad\ maps (see Section~\ref{Section:Results}), we do not believe that the temperature-$\beta$ anticorrelation will compromise the validity of our final results.

Our SED fitting code has been made freely available online as a {\sc python} 3 package\footnote{\label{Footnote:ChrisFit_URL}\url{https://github.com/Stargrazer82301/ChrisFit}}.

\section{Results} \label{Section:Results}

\begin{figure*}
\centering
\includegraphics[width=0.85\textwidth]{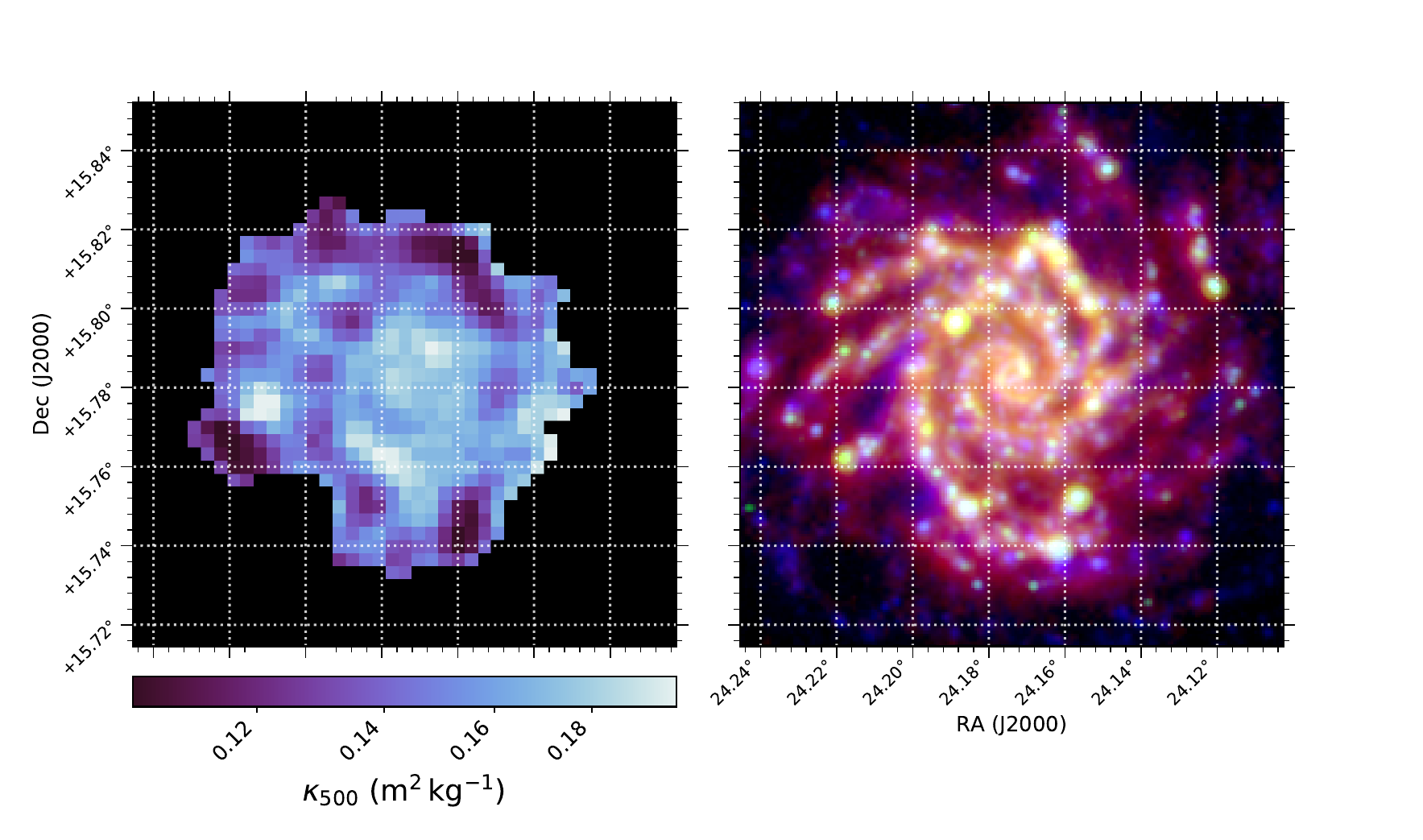}
\caption[]{{\it Left:} Map of \kappamu\ within M\,74. {\it Right:} UV--NIR--FIR three-colour image of M\,74, shown for comparison.}
\label{Fig:NGC0628_Kappa_Map}
\includegraphics[width=0.85\textwidth]{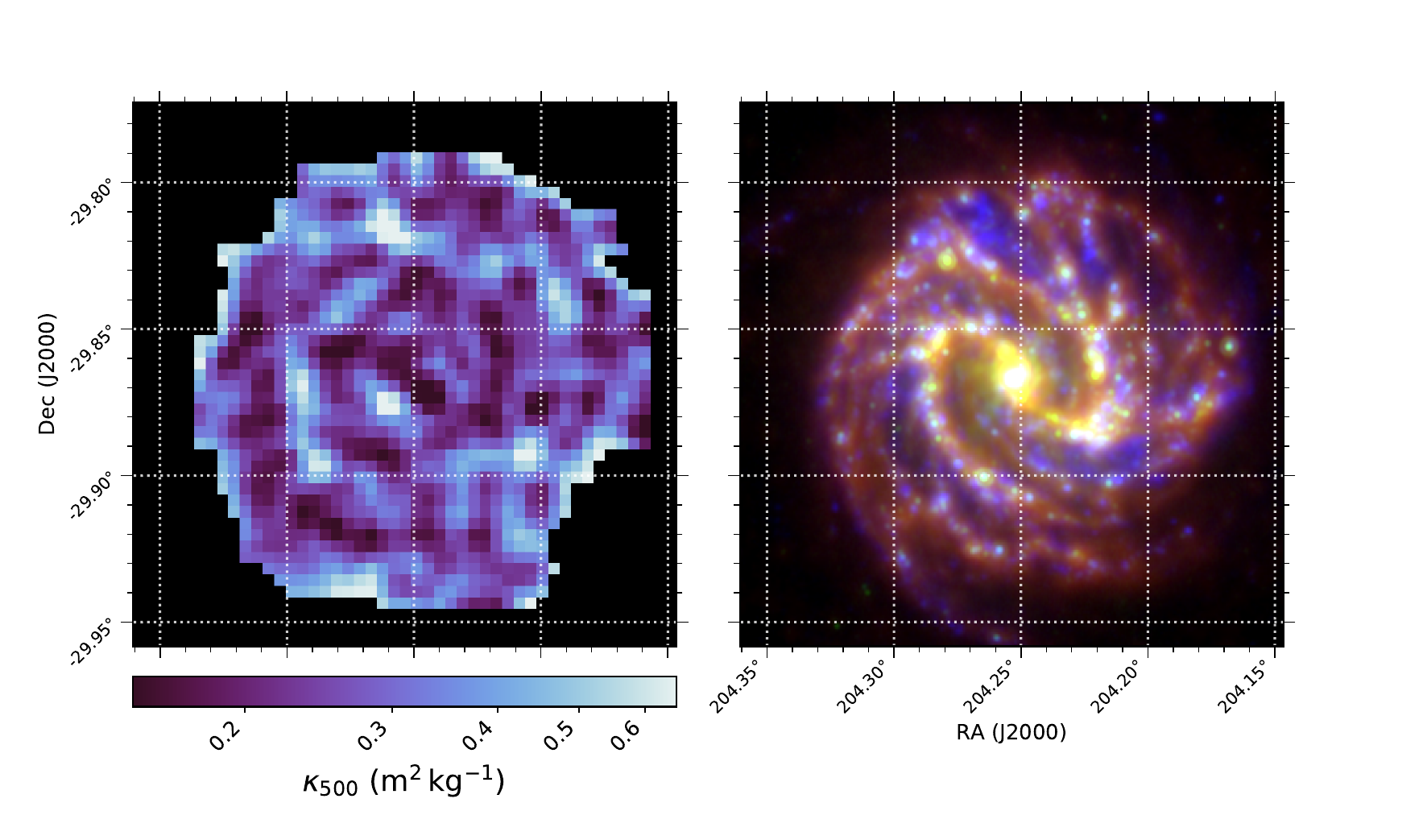}
\caption[]{{\it Left:} Map of \kappamu\ within M\,83. {\it Right:} UV--NIR--FIR three-colour image of M\,83, shown for comparison}
\label{Fig:NGC5236_Kappa_Map}
\end{figure*}

We now have the atomic gas, molecular gas, metallicity, and dust emission data necessary for every pixel in order to create maps of \kappad\ for our target galaxies.

For every pixel within the region of interest for each galaxy, we produced a full posterior probability distribution for \kappad. We did this by drawing random samples from the posterior distributions provided by our SED and metallicity maps (which are independent of one another), and inputting them into Equation~\ref{Equation:Kappa} (with number of MBB SED components $i=1$, as per Section~\ref{Subsection:SED_Fitting}). For all other input values ($S_{\rm HI}$, $I_{\rm CO}$, $\alpha_{\rm CO}$, $\alpha_{\rm CO_{\it MW}}$, $y_{\rm\, CO}$, $r_{2:1}$, $\delta_{O}$, $f_{Z_{\odot}}$, $[12 + {\rm log}_{10} \frac{\rm O}{\rm H}]_{\odot}$, $f_{\it He_{p}}$, $[\frac{\Delta f_{\it He}}{\Delta f_{Z}}]$, and \epsilond) we drew random samples from the Gaussian distributions described by their adopted values and associated uncertainties (effectively assuming flat priors, so that these can be treated as posterior probabilities).

We calculated \kappad\ for a reference wavelength of 500\,\micron, as this is the longest wavelength for which we have data, and therefore the wavelength where emission is least sensitive to dust temperature; this minimises the degree to which uncertainty in temperature is propagated to \kappad. Our resulting maps of \kappamu, produced by taking the posterior median in each pixel, are shown in Figures~\ref{Fig:NGC0628_Kappa_Map}, and \ref{Fig:NGC5236_Kappa_Map}. These maps contain 585 and 1269 pixels for M\,74 and M\,83 respectively. Throughout the rest of this work, quoted \kappamu\ values are pixel medians. The overall median across M\,74 is \kappamu\ = 0.15\,${\rm m^{2}\,kg^{-1}}$, whilst the overall median across M\,83 is \kappamu\ = 0.26\,${\rm m^{2}\,kg^{-1}}$.


\begin{figure}
\centering
\includegraphics[width=0.475\textwidth]{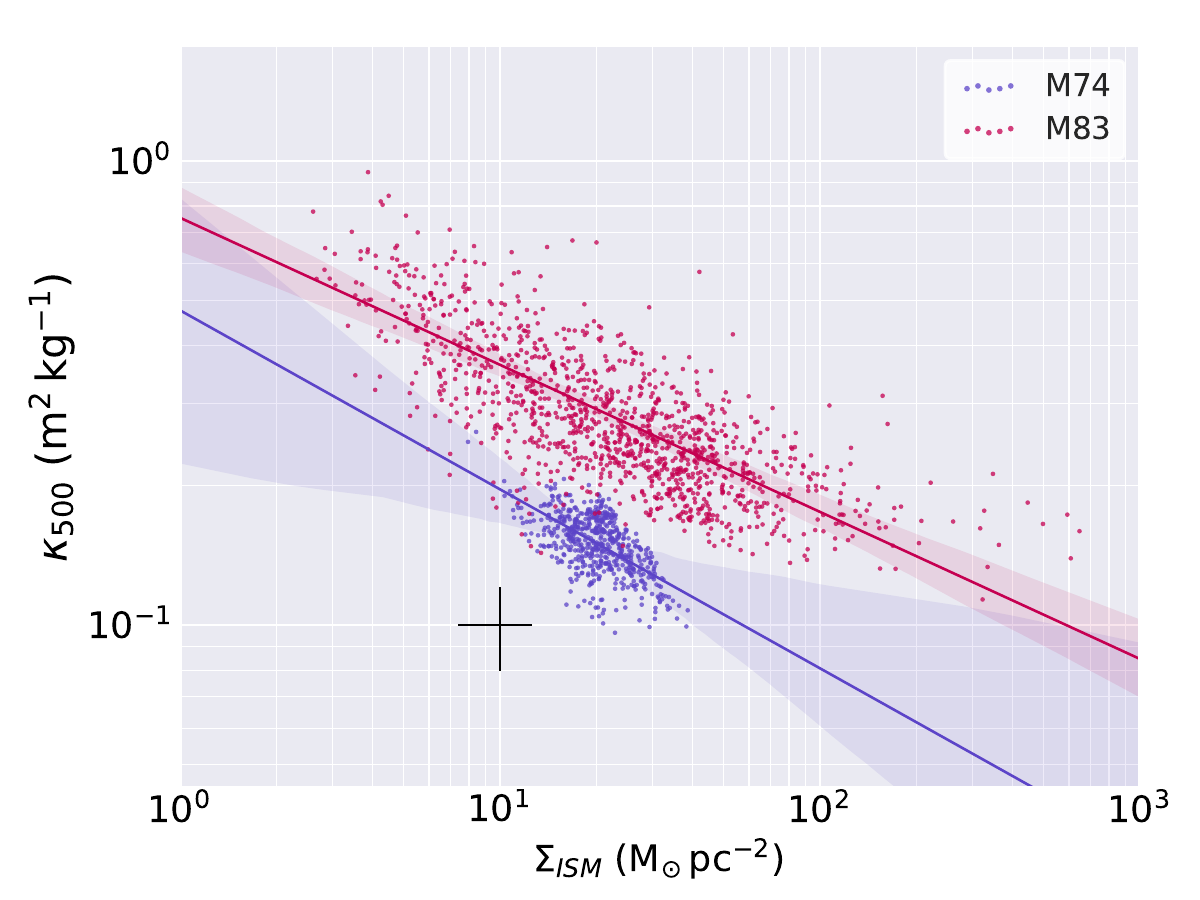}
\caption{Plot of \kappamu\ against ISM surface density (as traced by molecular and atomic gas) for M\,74 and M\,83. The best fit power laws for both galaxies are shown, with shaded regions indicating the 68.3\% credible intervals. The black cross indicates the median 1$\sigma$ error bars (indicating only the statistical uncertainty, omitting systematics uncertainties, as discussed in the text).}
\label{Fig:Kappa_vs_Sigma-ISM}
\end{figure}


\begin{figure*}
\centering
\includegraphics[width=0.975\textwidth]{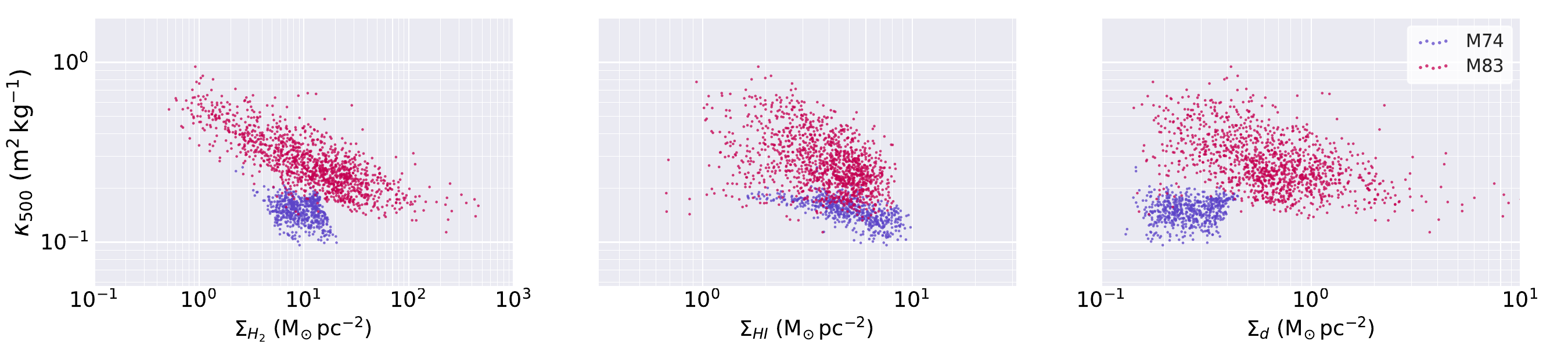}
\caption{Plots of \kappamu\ against the surface density of molecular gas ({\it left}), atomic gas ({\it centre}), and dust ({\it right}), for M\,74 and M\,83.}
\label{Fig:Kappa_vs_Sigma-ISM-All}
\end{figure*}

\begin{figure}
\centering
\includegraphics[width=0.475\textwidth]{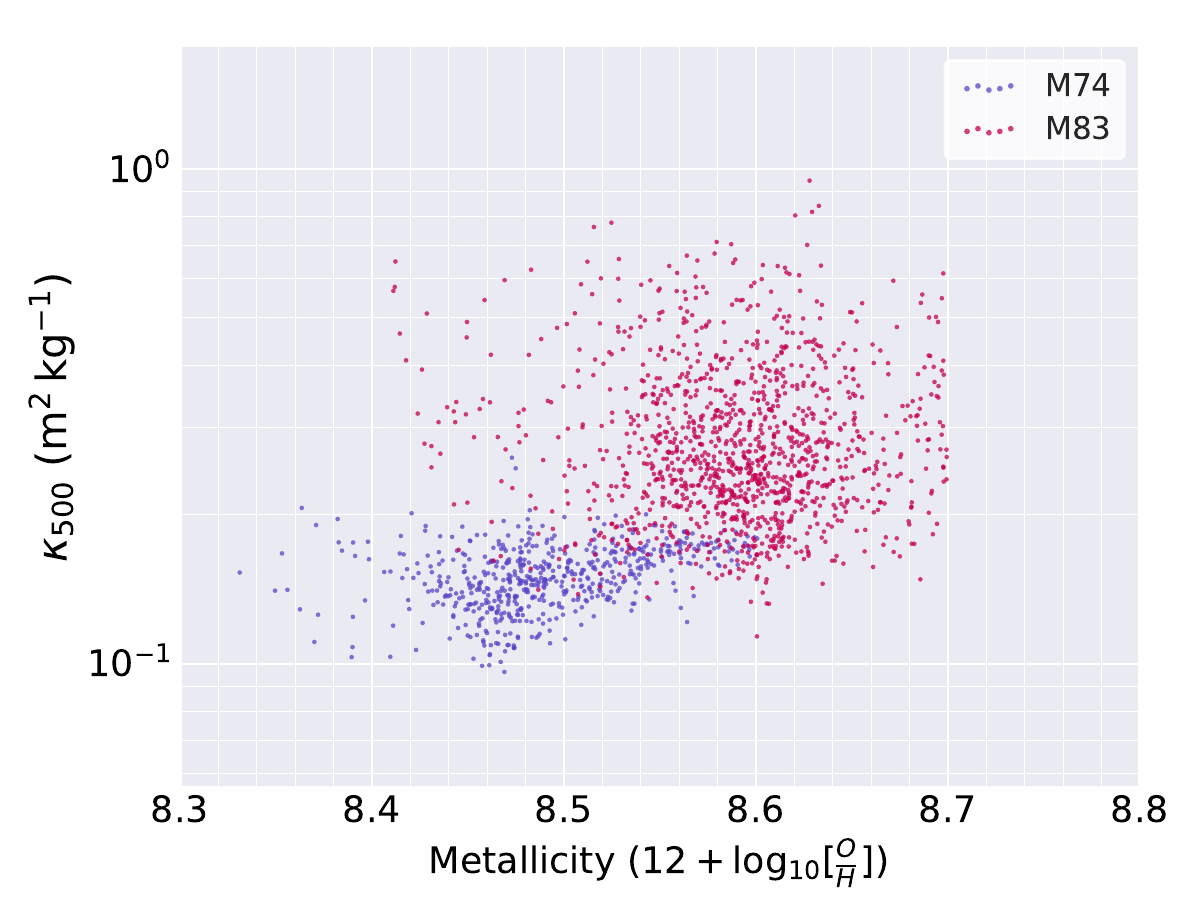}
\caption{Plot of \kappamu\ against gas-phase metallicity, expressed in terms of oxygen abundance, for M\,74 and M\,83.}
\label{Fig:Kappa_vs_Z}
\end{figure}

\begin{figure*}
\centering
\includegraphics[width=0.975\textwidth]{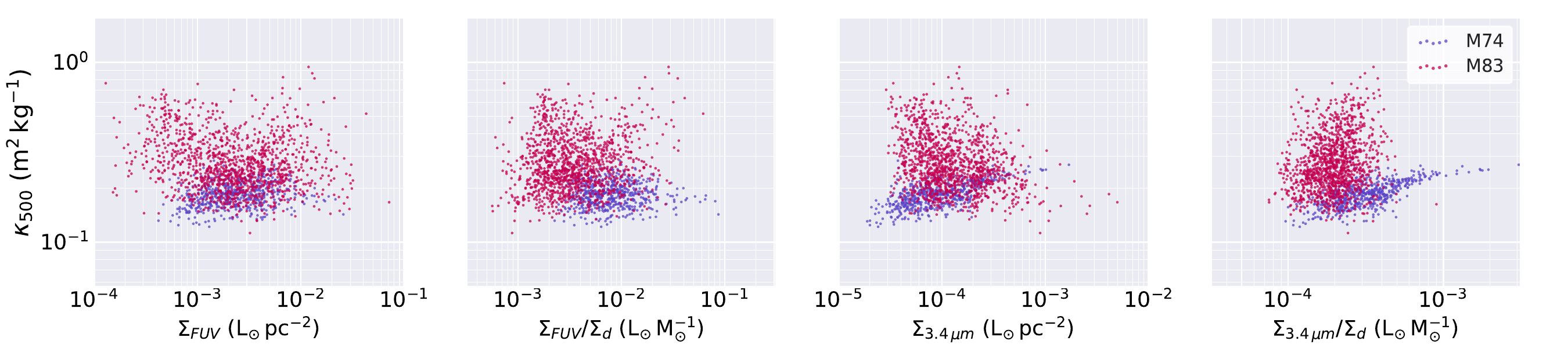}
\caption{Plots of \kappamu\ against the surface density of FUV luminosity surface density ({\it 1\textsuperscript{st}}), FUV luminosity per dust mass ({\it 2\textsuperscript{nd}}), 3.4\,\micron\ luminosity surface density ({\it 3\textsuperscript{rd}}), and 3.4\,\micron\ luminosity per dust mass ({\it 4\textsuperscript{th}}), for M\,74 and M\,83.}
\label{Fig:Kappa_vs_Sigma-Stellar-All}
\end{figure*}

The uncertainties on these \kappamu\ values (defined by the 68.3\% quantile in absolute deviation away from the median along the posterior distribution) span the range 0.21--0.28\,dex, with a mean uncertainty of 0.25\,dex for both galaxies. Note that a large degree of this uncertainty is shared across all pixels, due to the contributions of systematics (such as the uncertainties on \epsilond, $\alpha_{\rm CO_{\it MW}}$, etc), which is why the 0.25\,dex average uncertainty is large relative to the scatter in \kappamu\ values. We determined the contribution of the systematic components to the overall uncertainty via a Monte Carlo simulation, in which \kappamu\ values were generated according to Equation~\ref{Equation:Kappa}, but where only input parameters with systematic uncertainties were allowed to vary. The scatter on the output dummy values of \kappamu\ was taken to represent the total systematic uncertainty. On average, we found that the systematic components contribute 0.20\,dex to the uncertainty. Taking the quadrature difference between this and our average total uncertainty gives an average statistical uncertainty of 0.15\,dex in \kappamu.

The values in our \kappamu\ maps are not fully independent, as they have a pixel width of 3 pixels per FWHM; this will render adjacent pixels correlated. Therefore we also produced a version of the \kappamu\ maps with pixels large enough to be independent (ie, 1 pixel per FWHM). These maps contained 65 and 141 independent \kappamu\ measurements for M\,74 and M\,83 respectively. When performing statistical analyses throughout the rest of this work, we used these maps in order to ensure the validity of the results. However, the use of larger pixels for these maps does involve throwing away spatial information. We therefore present the standard, Nyquist-sampled maps in Figures~\ref{Fig:NGC0628_Kappa_Map}, \ref{Fig:NGC5236_Kappa_Map}, and elsewhere, in order to display all of the spatial information our data is able to resolve. Similarly, individual points plotted in Figure~\ref{Fig:Kappa_vs_Sigma-ISM} and elsewhere represent the pixels from the Nyquist-sampled maps, although the trend lines shown on these plots are derived from the independent-pixel data.

In order to calculate a robust estimate of the underlying range of \kappamu\ values, we performed a non-parametric bootstrap resampling of the pixel medians. This non-parametric bootstrap approach will account for the statistical scatter, and not encompass the systematics. This gives a median underlying range for 0.11--0.25\,${\rm m^{2}\,kg^{-1}}$ for M\,74 (a factor of 2.3 variation), and 0.15--0.80\,${\rm m^{2}\,kg^{-1}}$ for M\,83 (a factor of 5.3 variation). 


There is a strong relationship between \kappamu\ and $\Sigma_{\it ISM}$ (the ISM mass surface density, where $\Sigma_{\it ISM} = \Sigma_{\it H_{\sc i}} + \Sigma_{\rm H_{2}} + \Sigma_{\it d}$) as shown in Figure~\ref{Fig:Kappa_vs_Sigma-ISM}. Both galaxies exhibit this relation, but are curiously separated, with the relation for M\,74 lying $\sim$\,0.3\,dex beneath that of M\,83. We are able to trace this behaviour over a much larger range of $\Sigma_{\it ISM}$ for M\,83 than for M\,74 -- the densest regions of M\,83 are much denser than those of M\,74, whilst the deeper CO data for M\,83 allows us to probe to regions of lower density. This neatly accounts for the fact that we find a narrower range of \kappamu\ values for M\,74 than M\,83 -- whilst we probe a 1.7\,dex range in density in the latter, we only probe 0.7\,dex in the former. We estimated \kappamu\ vs $\Sigma_{\it ISM}$ power laws for each galaxy by performing a Theil-Sen regression \citep{Theil1992} to each set of posterior samples in our \kappamu\ and $\Sigma_{\it ISM}$ maps (specifically, the independent-pixel version of the maps, as discussed above). The the resulting power law slopes for both galaxies are in good agreement, with their indices being $-0.35^{+0.26}_{-0.21}$ for M\,74 and $-0.36^{+0.04}_{-0.05}$ for M\,83. As discussed in Section~\ref{Subsection:Implications_Discussion}, this behaviour is in contradiction to positive correlation between \kappad\ and ISM density predicted by standard dust models. The median statistical uncertainty on pixel values of $\Sigma_{\it ISM}$ is 0.13\,dex; given the similarly-small 0.15\,dex average statistical uncertainty on \kappamu, we can be confident that the trend in Figure~\ref{Fig:Kappa_vs_Sigma-ISM}, which spans 1.7\,dex for M\,83, isn't merely a spurious noise induced correlation. The rank correlation coefficient of the relationship is $\tau = -0.36$ for M\,74, and $\tau = -0.57$ for M\,83 (from a Kendall tau rank correlation test; \citealp{Kendall1990}).

In Figure~\ref{Fig:Kappa_vs_Sigma-ISM-All}, we see that it is the {\it overall} ISM density that is driving this trend, rather than the density of either the molecular gas, atomic gas, or dust dust components of the ISM alone, as all three have much weaker relationships with \kappamu\ than is the case for the combined $\Sigma_{\it ISM}$. For $\Sigma_{\rm H_{2}}$, $\tau_{\rm M74} = -0.18$ and $\tau_{\rm M83} = -0.55$; for $\Sigma_{d}$, $\tau_{\rm M74} = -0.28$ and $\tau_{\rm M83} = -0.42$; for $\Sigma_{d}$, $\tau_{\rm M74} = 0.10$ and $\tau_{\rm M83} = -0.34$.


The relationship between \kappamu\ and gas-phase metallicity is plotted in Figure~\ref{Fig:Kappa_vs_Z}. Once again, whilst M\,83 shows no correlation, there does appear to be a trend for M\,74, with larger values of \kappamu\ being associated with higher metallicities ($\mathcal{P}_{\it null} = 10^{-3.5}$ from a Kendall rank correlation test). On the one hand, metallicity is a parameter in Equation~\ref{Equation:Kappa}, so once again there is a definite risk of spurious correlations arising. However, if all other parameters in Equation~\ref{Equation:Kappa} are held fixed, higher metallicity (therefore higher $f_{Z}$) leads to {\it lower} values of \kappamu, meaning the trend for M\,74 in Figure~\ref{Fig:Kappa_vs_Z} is being driven by the data in spite of this. Greater ISM metallicity will lead to increased grain growth \citep{Dwek1998B,Zhukovska2014A,Galliano2018C}, and larger grains should give rise to larger values of \kappad\ \citep{Li2005D,Kohler2015A,Ysard2018A}.

We wished to assess whether local star formation has an effect on our calculated values of \kappamu. There are several mechanisms by which recent star formation can process dust grains in its vicinity (see review in \citealp{Galliano2018C}). For instance, photo-destruction by high-energy photons from massive (therefore young) stars can directly break down dust grains \citep{Boulanger1998A,Beirao2006A}, whilst the shocks produced by the supernov\ae\ of massive stars will sputter dust grains \citep{Bocchio2014B,Slavin2015C}. FUV emission should be a good proxy of these two environmental conditions; unobscured FUV emission is indicative of massive stars that are old enough to cleared their birth clouds, and hence represent the regions where supernov\ae\ will be occurring. And of course, regions with greater amounts of unobscured FUV emission demonstrably have an InterStellar Radiation Field (ISRF) with greater amounts of high-energy photons. If the environmental effects of recent star formation were impacting \kappamu, this could manifest as a correlation with the total UV energy density, or with the UV energy density per dust mass (similar to the `heating parameter' of \citealp{Foyle2013}), as the dust will be better shielded in areas with greater dust density. Therefore, in the two leftmost panels of Figure~\ref{Fig:Kappa_vs_Sigma-Stellar-All}, we plot \kappamu\ against both the GALEX Far-UltraViolet (FUV) luminosity surface density\footnote{\label{Footnote:FUV_NIR_Background_Subtraction}Maps were reprojected to the same pixel grid as the \kappamu\ maps, then background-subtracted in the same manner as the continuum maps in Section~\ref{Subsection:Data_Preparation}. We manually masked pixels containing obvious foreground Milky Way stars.}  ($\Sigma_{\it FUV}$), and against the FUV luminosity per dust mass surface density ($\Sigma_{\it FUV}/\Sigma_{\it d}$).  No trend is apparent in either plot; M\,74, with its generally lower values of \kappamu, has a higher average value of $\Sigma_{\it FUV}/\Sigma_{\it d}$, but this is to be expected given its bluer colours and lower submm surface brightness (see Table~\ref{Table:Galaxy_Properties}).

We also wished to assess whether the ISRF arising from evolved stars could be influencing \kappamu, given that radiation from evolved stars can be the dominant source of energy received by dust in certain environments (\citealp{Boquien2011C,Bendo2012A}; Nersesian {\it accepted}). Observations in the NIR provide a good tracer of the evolved stellar population, and the ISRF it produces. Therefore, as with FUV, we plot \kappamu\ against the WISE 3.4\,\micron\ luminosity surface density\footnoteref{Footnote:FUV_NIR_Background_Subtraction}\ ($\Sigma_{\it 3.4\,\mu m}$), and against the 3.4\,\micron\ luminosity per dust mass surface density ($\Sigma_{\it 3.4\,\mu m}/\Sigma_{\it d}$), shown in the two rightmost panels of Figure~\ref{Fig:Kappa_vs_Sigma-Stellar-All}.  In M\,74, it seems that the pixels with $\frac{\Sigma_{\it 3.4\,\mu m}}{\Sigma_{\it d}} > 6\,\times\,10^{-4}\,{\rm L_{\odot}\,M_{\odot}^{-1}}$ are exclusively associated with higher values of \kappamu. And most interestingly, there is for both galaxies a positive correlation between \kappamu\ and $\Sigma_{\it 3.4\,\mu m}/\Sigma_{\it d}$). Whilst there is appreciable scatter, a Kendall rank correlation test gives $\mathcal{P}_{\it null} > 0.023$ for both -- so it seems that this relationship, whilst broad, has probably not arisen by chance\footnote{Spearman and Pearson rank correlation tests similarly both give $\mathcal{P}_{\it null} < 0.025$, with correlation coefficients \textgreater\,0.2.}. Plus, the WISE 3.4\,\micron\ data played no part in our \kappamu\ calculations, making it hard to see how this relation could have arisen  spuriously from our methodology.

A downside to using 500\,\micron\ as the reference wavelength is that carbonaceous species are expected to have considerably larger \kappamu\ values than silicate species at these longer wavelengths (due to the steeper $\beta$ for silicates; \citealp{Ysard2018A}). Whereas at shorter wavelengths, the difference in \kappad\ between carbonaceous and silicate dust is smaller. Thus the choice of the longer reference wavelength might be limiting our ability to use the \kappad\ maps to trace such compositional variation. We therefore also produced versions of our \kappad\ maps at a reference wavelength of 160\,\micron. These $\kappa_{160}$ maps are presented in Appendix~\ref{AppendixSection:Kappa_160_Maps}; however, they exhibit no difference in structure to the \kappamu\ maps.

\section{Discussion} \label{Section:Discussion}

\subsection{Robustness of Findings} \label{Subsection:Robustness_Discussion}

Within M\,74 and M\,83, we find values of \kappamu\ that vary by factors of 2.3 and 5.3 respectively. This is, to our knowledge, the first observational mapping of variation in \kappad\ within other galaxies. However, it is important to critically evaluate how much of this apparent variation could simply be an artefact of our method. 

In a companion study to this work, Bianchi et al. ({\it in prep.}) use the dust-to-metals method to calculate global \kappad\ values for 204 DustPedia galaxies. As that study uses integrated gas measurements, they are unable to directly constrain ISM density. However, they do find that galaxies with higher ${\rm H}_{2}$/H{\sc i} ratios (typically associated with denser ISM) tend to have lower values of \kappad. This is what would be expected if the anticorrelation we find between \kappad\ and $\Sigma_{\it ISM}$ continues on global scales, between galaxies. They also find large (a factor of several) scatter in their \kappad\ values between galaxies; in this context, the differences between the values we find for M\,73 and M\,83 are not conspicuous.

Our key assumption of a fixed dust-to-metals ratio, \epsilond, deserves particular scrutiny. As mentioned in Section~\ref{Section:Theory}, the vast majority of directly-measured\footnote{By `direct', we refer to those measurements where \epsilond\ is determined from observing the mass fraction of metals depleted from the gas phase.} values of \epsilond\ lie in the range 0.2--0.6. Whilst this factor of 3 variation could notionally, in the worse-case-scenario, be sufficient to nullify the factor 2.3 variation in \kappad\ we find in M\,74, it could not nullify the factor 5.3 variation in M\,83. Moreover, as we show in Section~\ref{Subsubsection:Variable_Dust-To-Metals}, in the physically most likely scenario where \epsilond\ scales with density, the variation in \kappamu\ actually increases. Nonetheless, it is undoubtably worth considering how, precisely, different kinds of systematic variations in \epsilond\ within our target galaxies could be influencing our results.

There is evidence that \epsilond\ is significantly reduced at low metallicities \citep{Galliano2005B,DeCia2016A,Wiseman2017B}. However, there appears to be reduced variation in \epsilond\ at intermediate-to-high metallicity. \citet{DeCia2016A} and \citet{Wiseman2017B} use depletions in damped Lyman-$\alpha$ absorbers to find only a factor of $\sim$\,2 variation in \epsilond\ at metallicities above 0.1\,$Z_{\odot}$, with at most a weak dependence on metallicity in that regime. Given that our analysis is concerned only with environments at $\gg 0.1\,Z_{\odot}$, our results should be minimally susceptible to this scale of metallicity effect. Additionally, it should be noted that a number of studies have used visual extinction per column density of metals as a proxy for \epsilond, and found it to be constant down to metallicities of 0.01\,$Z_{\odot}$, over a redshift range of $0.1<z<6.3$ \citep{Watson2011A,Zafar2013D,Sparre2014A}.

A number of simulations have addressed the question of how \epsilond\ varies. \citet{McKinnon2016A} trace \epsilond\ in cosmological zoom-in simulations, finding it varies by up to a factor of $\sim$\,3.5 in the modern universe; however, they find minimal systematic variation within galaxies, except for enhanced values in galactic centres (see their Figures 1, 2, and 14). \citet{Popping2017C} trace \epsilond\ in semi-analytic models, and find that it can vary with metallicity by up to a factor of $\sim$\,2 at metallicities \textgreater\,0.5\,$Z_{\odot}$ (with the degree and nature of this variation depending considerably upon the specific model).

However, if \epsilond\ does indeed vary significantly with metallicity within our target galaxies, that will actually {\it increase} the amount of variation in \kappamu\ in M\,83. The highest metallicities are at the inner regions of the disc, where \kappamu\ is already lowest; if increasing $f_{Z}$ in Equation~\ref{Equation:Kappa} also increases \epsilond, then this will drive down \kappad\ still further. On the other hand, because the lowest values of \kappad\ in M\,74 are found in the spiral arms, away from the centre, a correlation of \epsilond\ with metallicity could indeed suppress some variation in \kappamu\ -- although M\,74 already exhibits a much smaller range in \kappamu\ than M\,83.

Theoretical dust models can make specific predictions about how \epsilond\ is expected to vary in different conditions. For instance, the THEMIS model traces how dust populations are expected to change in different interstellar environments, predicting that \epsilond\ will increase monotonically with ISM density by a factor of $\sim$\,3.5, from 0.27 in the diffuse ISM ($n_{H} = 10^{3}\,{\rm cm^{3}}$) to 0.88 in the dense ISM ($n_{H} = 10^{6}\,{\rm cm^{3}}$), driven by the accretion of gas-phase metals onto grains \citep{Jones2018A}. We explore the potential effects of this in detail in Section~\ref{Subsubsection:Variable_Dust-To-Metals}, where we find that it would further increase the variation in \kappamu.

There are several observational studies that report variation of \epsilond\ between and within galaxies, inferred from the fact that the gas-to-dust ratio is found to vary with metallicity \citep{Remy-Ruyer2014A,Chiang2018A,DeVis2019B}. However, these studies all rely upon an assumed value of \kappad\ to infer dust masses, and hence \epsilond. Given that we, conversely, use an assumed \epsilond\ to infer \kappad, it is not really possible to compare such results with ours in a valid way. However, we note with interest that these studies tend to find much larger ranges of \epsilond\ than are suggested by either depletions, simulations, or theoretical dust models -- up to 1\,dex of scatter at a given metallicity, with up to 3\,dex total range over all metallicities. One way to explain this discrepancy would be if \kappad\ is depressed at lower metallicity (which is potentially hinted at for M\,74 in Figure~\ref{Fig:Kappa_vs_Z}).

Beside a breakdown in our assumption of a fixed \epsilond, it is possible that our method is being corrupted by the presence of `dark gas' -- H$_{2}$ at intermediate densities that CO fails to trace \citep{Reach1994C,Grenier2005D,Wolfire2010A}. The presence of dark gas would have the effect of causing us to underestimate the value of $M_{\rm H_{2}}$ in Equation~\ref{Equation:Kappa}, thereby artificially driving up \kappamu. The elevated areas of \kappamu\ in our maps are indeed mainly associated with the inter-arm regions, where the fraction of dark gas is expected to be greatest \citep{Langer2014A,RJSmith2014A}. Estimates of the fraction of galactic gas mass that is dark range from 0\% from dust and gas observations in M\,31 \citep{MWLSmith2012B}, to 30\% in theoretical models \citep{Wolfire2010A}, to 42\% in hydrodynamical simulations of galactic discs \citep{RJSmith2014A}, to 10--60\% from \planck\ observations of the Milky Way \citep{Planck2011XXI}, to 6--60\% from Milky Way $\gamma$-ray absorption studies \citep{Grenier2005D}. Even assuming a worst-case scenario of a 60\% dark gas fraction for the inter-arm regions of our target galaxies (an extreme scenario, given that the 60\% represents the single largest fraction amongst the wide range of values reported within the Milky Way), dark gas could only reduce the variation in \kappamu\ we find by a factor of 1.7. 

In a similar vein, another potential confounder would be systematic variation in $\alpha_{\rm CO}$. If $\alpha_{\rm CO}$ increases in denser ISM ({\it independent of metallicity}, which we account for), then this could counteract the variation in \kappad\ we find. However, evidence to date does not indicate that $\alpha_{\rm CO}$ varies systematically in this way \citep{Sandstrom2013B}. This of course could be due to the fact that the uncertainty on $\alpha_{\rm CO}$ (and the scatter on the relations used to derive it) is large -- however this uncertainty is propagated through our calculations.

In the course of determining \kappamu\ for each pixel, values for the gas-to-dust ratio, $G/D$, are also generated. We find  $176 < G/D < 277$ for M\,74, and $140 < G/D < 275$ for M\,83. Note that these are the ratios of {\it total} gas mass to dust mass. In the literature, quoted $G/D$ values are often {\it hydrogen} to dust ratios (ie, no factor of $\xi$ is applied to account for the masses of helium and metals); our hydrogen-to-dust ratios, $G_{H}/D$, are $127 < G_{H}/D <201$ for M\,74, and $100 < G_{H}/D < 196$. For high-metallicity systems such as of our target galaxies, these are normal values when compared to the literature \citep{Sandstrom2013B,Remy-Ruyer2015B,DeVis2017B}. Indeed, we neatly reproduce the factor of 2--3 radial variation in $G_{H}/D$ in M\,74 reported by \citet{Vilchez2019A} and \citet{Chiang2018A} over the 8.35--8.60 (\logOH) metallicity range we sample; although their adoption of fixed \kappad\ limits the scope for detailed comparison. Nonetheless we can say that our inferred dust masses yield sensible $G/D$ values, following expected trends.

Overall, we are confident that our finding of an inverse correlation of \kappamu\ is indeed robust against a wide range of changes to the initial assumptions of our method.

\subsection{Alternate Models} \label{Subsection:Alternate_Models}

\begin{figure}

\centering
\includegraphics[width=0.23\textwidth]{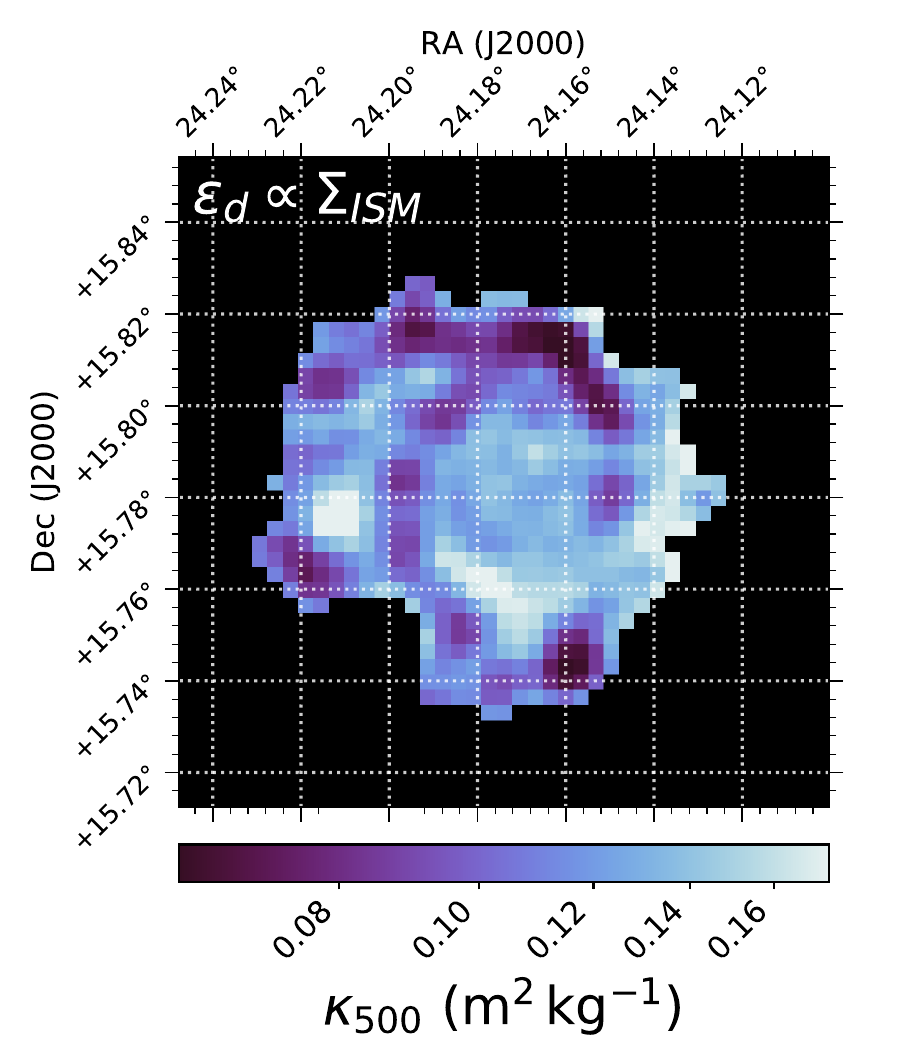}
\includegraphics[width=0.23\textwidth]{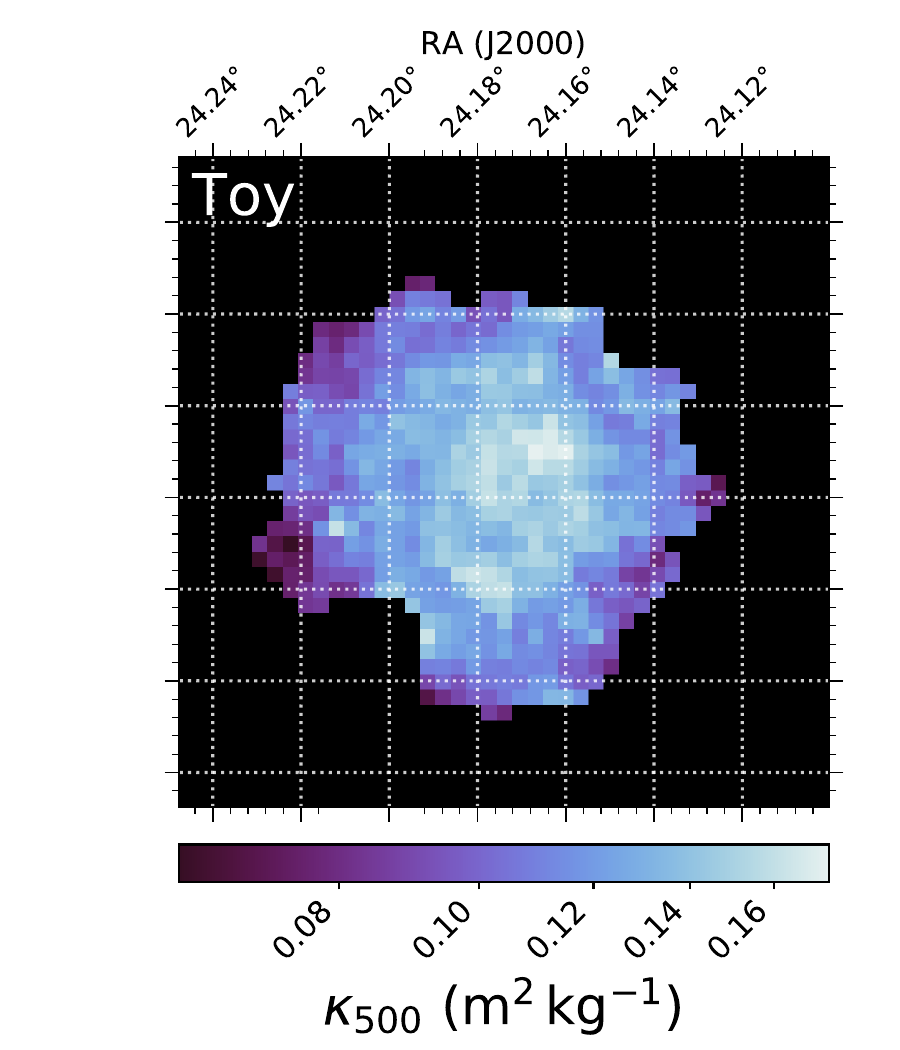}
\caption{Maps of \kappamu\ within M\,74, calculated using different model assumptions than for our fiducial map in Figure~\ref{Fig:NGC0628_Kappa_Map}. {\it Left:} With the dust-to-metals ratio,  \epsilond, set to vary linearly as a function of $\Sigma_{\it ISM}$. {\it Right:} With a toy model where $\alpha_{\rm CO}$, $r_{2:1}$, $T_{d}$, $\beta$, \epsilond, and $[12 + {\rm log}_{10} \frac{\rm O}{\rm H}]$ are kept constant.}
\label{Fig:NGC0628_Kappa_Variations}

\includegraphics[width=0.23\textwidth]{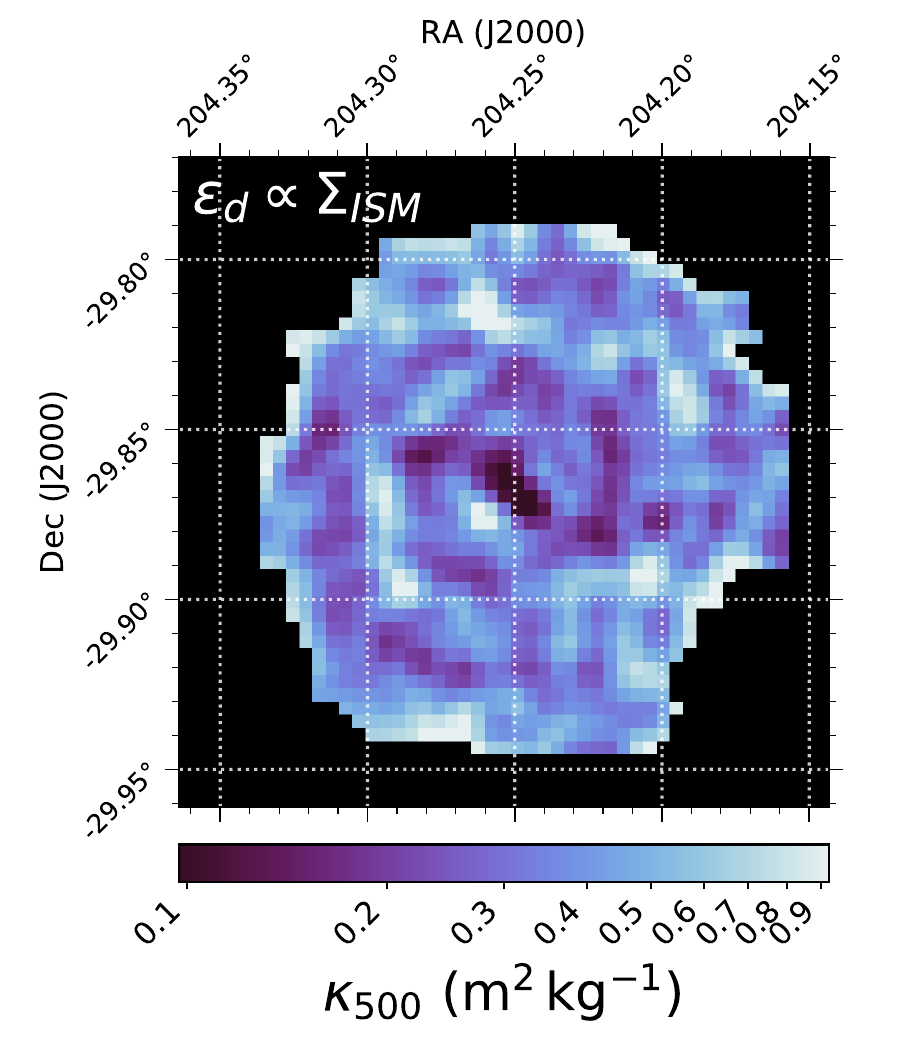}
\includegraphics[width=0.23\textwidth]{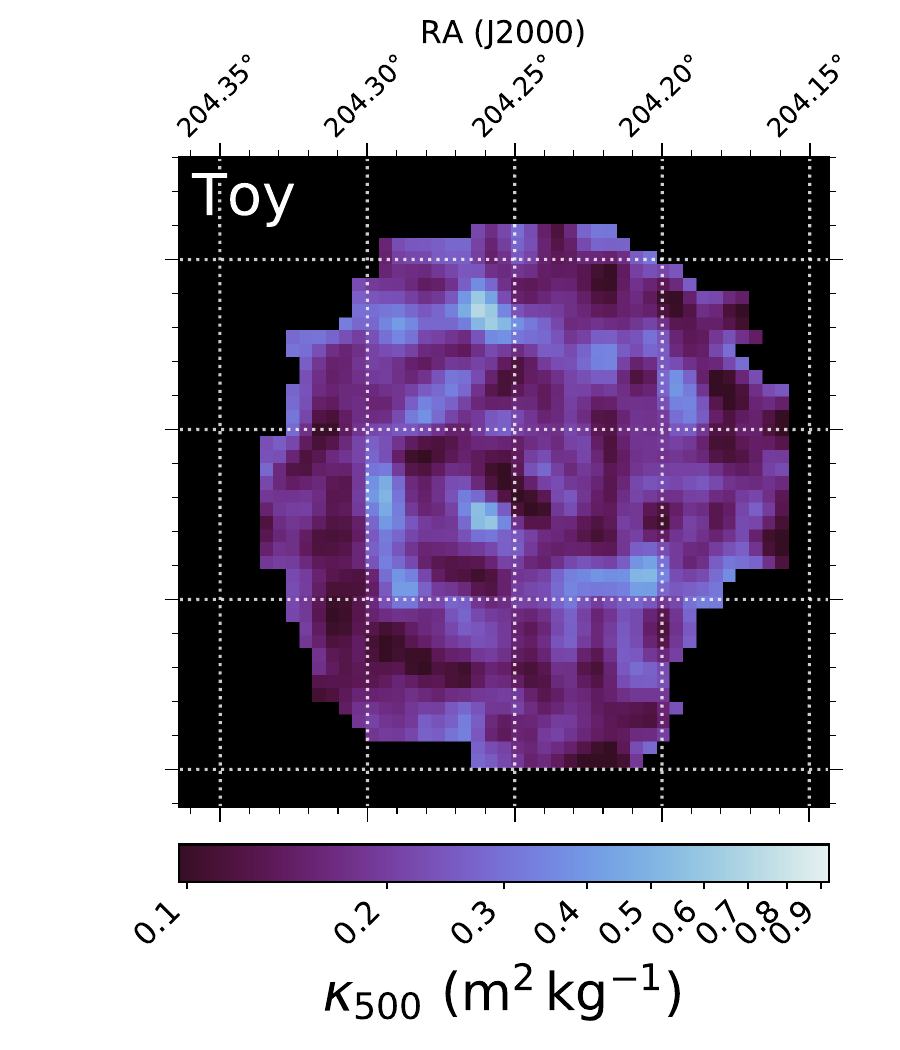}
\caption{Maps of \kappamu\ within M\,83, each calculated using different model assumptions than for our fiducial map in Figure~\ref{Fig:NGC5236_Kappa_Map}. Model descriptions the same as for Figure~\ref{Fig:NGC0628_Kappa_Variations}.}
\label{Fig:NGC5236_Kappa_Variations}

\end{figure}

\subsubsection{Variable Dust-To-Metals Ratio} \label{Subsubsection:Variable_Dust-To-Metals}

As discussed in Section~\ref{Section:Theory}, the assumption of a fixed \epsilond\ is a simplification. Observed depletions in nearby portions of the Milky Way's diffuse ISM indicate that in reality, \epsilond\ increases with column density \citep{Jenkins2009B,Draine2014A,Roman-Duval2019A}. However, the form of this relation in extragalactic systems, where only integrated column density data is available, is not well constrained. Nonetheless, we can still explore, in general terms, how such a model would affect the manner in which \kappad\ scales. Even if this approach requires more assumptions, it may be more physical than our fiducial model. 

We therefore repeated our \kappamu\ mapping, setting \epsilond\ to vary linearly as a function of $\Sigma_{\it ISM}$, with $\epsilon_{d} = 0.75$ at the point in each galaxy where $\Sigma_{\it ISM}$ is highest, and $\epsilon_{d} = 0.25$ at the point where where $\Sigma_{\it ISM}$ is lowest. This specific choice of relationship is effectively arbitrary, but approximates the trend reported by \citet{Chiang2018A} within M\,101, whilst also matching the range of \epsilond\ values reported by \citet{DeVis2019B} (although both of these sets of \epsilond\ values were calculated with FIR--submm data, using an assumed value of \kappad, limiting scope for direct comparison).

The \kappamu\ maps produced using the $\epsilon_{d} \propto \Sigma_{\it ISM}$ model are shown in the left panels of Figures~\ref{Fig:NGC0628_Kappa_Variations} and \ref{Fig:NGC5236_Kappa_Variations}, for M\,74 and M\,83 respectively. The trend of \kappamu\ being depressed in the denser environments of the spiral arms remains. In fact, the anticorrelation between \kappamu\ against $\Sigma_{\it ISM}$ is even more exaggerated than was the case for our fiducial model, as can be seen in the left panel of Figure~\ref{Fig:Kappa_vs_Sigma-ISM_Variations}. The Kendall rank correlation coefficients for the $\epsilon_{d} \propto \Sigma_{\it ISM}$ results are more strongly negative than those of the fiducual version, being $\tau = -0.66$ for both M\,74 and M\,83. The range of \kappamu\ values when using the  $\epsilon_{d} \propto \Sigma_{\it ISM}$ model increases to a factor 5 in M\,74, and to a factor of 20 in M\,83.

It appears that our choice of fixed $\epsilon_{d}$ in our fiducial model actually serves to {\it reduce} the variation in \kappamu, and that the (probably) more-physical $\epsilon_{d} \propto \Sigma_{\it ISM}$ model suggests a notably greater range of values. This increases our confidence that the variation in \kappamu\ we see is a real effect. Whilst we could, for instance, construct a model where \epsilond\ {\it decreases} with ISM density by a factor of \textgreater\,5.3, this would be completely unphysical, and would represent an entirely contrived attempt to minimise the \kappad\ variation we find. Similarly, we could construct a model where \epsilond\ {\it increases} with radius -- but whilst this would decrease the \kappad\ variation in M\,83, it would increase it for M\,74 (and would again be an unphysical contrivance).

\subsubsection{`Toy' Model} \label{Subsubsection:Toy_Model}

To establish the degree to which our results might simply be an artefact of our method, we again repeated our \kappamu\ mapping, using a `toy' model. For this repeat, metallicity was fixed at the Solar value of $12 + {\rm log}_{10} [\frac{\rm O}{\rm H}] = 8.69$, $\alpha_{\rm CO}$ was fixed at the standard Milky Way value of $3.2\,{\rm K^{-1}\,km^{-1}\,s\,pc^{-2}}$, $r_{2:1}$ was fixed at the local-Universe average of 0.7, $T_{d}$ was fixed at 20\,K, $\beta$ was fixed at 2, and \epsilond\ was fixed at 0.4. Although this toy model is unphysical, it strips out as many assumptions as possible -- allowing us to be confident that any trends that persist are not due to our GPR metallicity mapping, our SED fitting, our $r_{2:1}$ prescription, etc.

The \kappamu\ maps produced using the toy model are shown in the right panels of Figures~\ref{Fig:NGC0628_Kappa_Variations} and \ref{Fig:NGC5236_Kappa_Variations}, for M\,74 and M\,83 respectively. The corresponding plot of \kappamu\ against $\Sigma_{\it ISM}$ is shown in the right panel of Figure~\ref{Fig:Kappa_vs_Sigma-ISM_Variations}, where it can be seen that the scatter is markedly increased for both galaxies. For M\,83, the trend is nonetheless still present, with a Kendall rank correlation test giving $\mathcal{P}_{\it null} < 10^{-5}$; the lowest values of \kappamu\ are still visibly associated with the largest values of $\Sigma_{\it ISM}$, and vice-a-versa.  For M\,74, the correlation of \kappamu\ with $\Sigma_{\it ISM}$ is lost; however the far smaller dynamic range in ISM density for this galaxy made it more susceptible to the trend being removed by the toy model's increase in scatter. The fact the trend with ISM density persists for M\,83 despite the use of the toy model is extremely informative. It implies that the basic negative correlation is being driven by the interplay between the 21\,cm data, CO data, and 500\,\micron\ data -- not by the specifics of our method.

\begin{figure}
\centering
\includegraphics[width=0.23\textwidth]{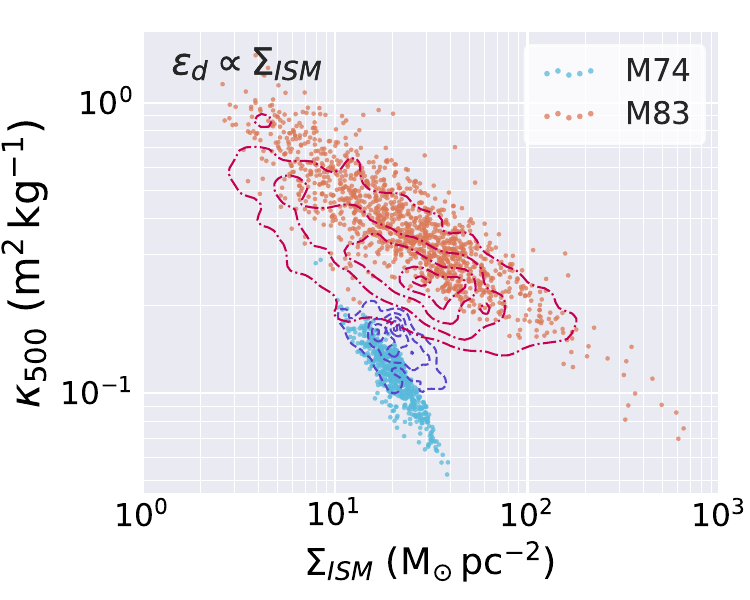}
\includegraphics[width=0.23\textwidth]{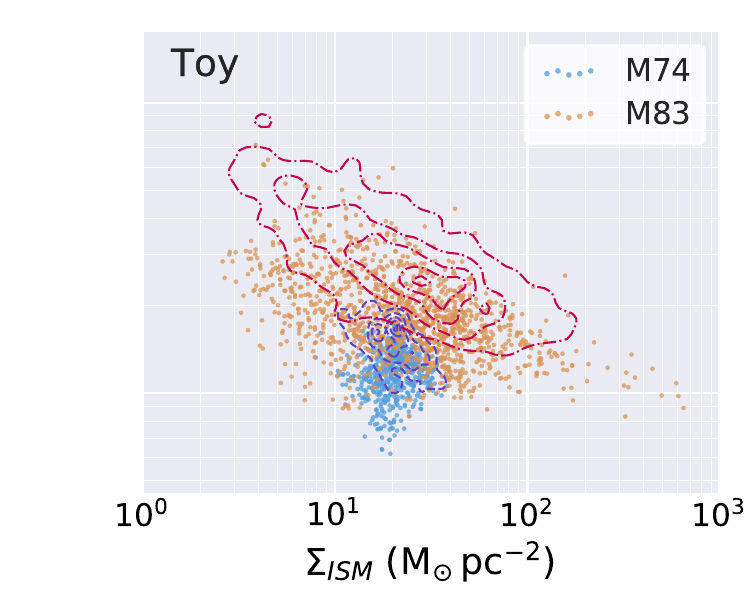}
\caption{Alternate versions of Figure~\ref{Fig:Kappa_vs_Sigma-ISM}, again plotting \kappamu\ against ISM surface density for M\,74 and M\,83, but for \kappamu\ calculated using different model assumptions than for our fiducial method. Model descriptions the same as for Figure~\ref{Fig:NGC0628_Kappa_Variations}. For comparison, the distributions for our fiducial maps, as plotted in Figure~\ref{Fig:Kappa_vs_Sigma-ISM}, are indicated with contours (showing the 5\textsuperscript{th}, 25\textsuperscript{th}, 50\textsuperscript{th}, 75\textsuperscript{th}, and 95\textsuperscript{th} percentiles); M\,74 as blue dashed, M\,83 as red dot-dashed.}
\label{Fig:Kappa_vs_Sigma-ISM_Variations}
\end{figure}

\subsubsection{Other Alternate Models} \label{Subsubsection:Other_Variations}

To provide further methodological checks, we produced additional alternate \kappamu\ maps. In Appendix~\ref{AppendixSection:Kappa_Z_Variations}, we present \kappamu\ maps generated using metallicities calculated via different strong-line prescriptions than the one employed for our fiducial \kappamu\ maps. in Appendix~\ref{AppendixSection:Kappa_2MBB}, we present \kappamu\ maps generated fitting a two-component MBB model to the FIR--submm fluxes, as opposed to the one-component MBB model used for our fiducial \kappamu\ maps. In all cases the resulting \kappamu\ maps display the same general morphology as our fiducial ones, with lower values of \kappamu\ associated with denser regions.



\subsection{Implications of Findings} \label{Subsection:Implications_Discussion}

Our finding that \kappamu\ shows a strong negative correlation with ISM density is in direct contradiction to standard models of dust emission, which predict that the densest regions of the ISM should exhibit the highest values of \kappad\ \citep{Ossenkopf1994B,Li2003C,Jones2018A}\comments{Krugel1994C,Pollack1994B,Kohler2015A,Ysard2018A}. This expectation arises from the fact that dust grains in the densest parts of the ISM are predicted to be larger, due to the coagulation of grains and the growth of (icy) mantles on their surfaces, and that larger grains should be more emissive per unit mass \citep{Kohler2012A,Jones2013C,Ysard2018A}. The apparent incompatibility of our results with these predictions presents one of two possibilities.

The first possibility is that our method has some fundamental flaw that has systematically affected the results. We have made an effort to construct our method so that it only relies upon standard, widely-used assumptions. If one (or more) of these assumptions breaks down systematically, in a manner such that the bias is a function of ISM density, and the bias is a factor of \textgreater\,5, then this could give rise to the results we see. We have tried to inoculate our findings against even this scenario (for instance, by trying the toy model where all possible variables were kept fixed). However if, for example, dark gas represents 75\% of the total gas mass in inter-arm space (artificially suppressing our assumed $M_{\rm H_{2}}$), this could negate our results for M\,74. If dark gas represents 75\% of the total gas mass in inter-arm space if  {\it and} if H{\sc ii} region oxygen depletion were a factor of 2 lower in inter-arm space versus other metals (compromising its use as a metallicity tracer), then our results for M\,83 could be negated -- however scenarios this extreme are unlikely, being unprecedented in the literature, and would have significant implications for extragalactic studies in general. 

The second possibility is that \kappamu\ truly does decrease in denser ISM. This would help explain some observational results. For instance, an excess in submm emission has been found in lower-density areas within galaxies \citep{Relano2018A}, and within galaxies dominated by diffuse regions (Lamperti  et al. {\it accepted}; De Looze et al. {\it in prep.}); if \kappamu\ is indeed elevated in low-density regions, it could give rise to this effect.

It is hard to explain decreasing \kappamu\ in denser ISM in the context of current dust physics. However it is possible to construct scenarios where it is not entirely unreasonable. \citet{Ysard2018A} present a detailed exploration of how changes in various physical parameters of dust should affect \kappad. For instance, spherical grains are predicted to have lower \kappamu\ than oblate or prolate grains, by up to a factor of $\sim$\,1.5 (see their Figure~5); and hydrogenated amorphous carbon grains are expected to have much lower \kappamu\ than amorphous silicate or unhydrogenated amorphous carbon grains (by up to an order magnitude). Whilst we do not suggest that this (or any other) specific physical scenario is the cause of our observed trend, it demonstrates that it is at least possible to envisage an evolution in dust properties that does not entail an uninterrupted monotonic increase in \kappad\ with ISM density.

On that theme, we also note that our data only provides physical resolution of 590\,pc\,pix$^{-1}$ in M\,74 and 330\,pc\,pix$^{-1}$ in M\,83. As such, we can do no better than distinguish between arm and inter-arm pixels. This will have `smeared out' the properties of the denser clouds within the spiral arms. When studies discuss grain growth in the dense ISM, and the associated increases in \kappad, the dense medium in question is typically described as having at least $1500$--$10000$\,$n_{\rm H}\,{\rm cm^{-3}}$, compared to 20--50\,$n_{\rm H}\,{\rm cm^{-3}}$ in the diffuse ISM \citep{Ferriere2001C,Kohler2015A,Jones2018A} -- a difference in density of at least a factor of 30. However, our data only traces a dynamic range in density of a factor of 5 in M\,74, and a factor of 50 in M\,83 (discounting pixels within 1 beam of the nuclear starburst, where our \kappamu\ values become unreliable, as per Section~\ref{Section:Results}). So whilst we are probing a wide range of ISM conditions, we are unable to perform a `clean' sampling of the densest grain-growth environments. Likewise, we only performed our analysis for pixels with sufficient SNR for all data -- thereby excluding regions of particularly low ISM density, especially at the outskirts of the target galaxies. As such, it seems likely that, in practice, we are effective probing intermediate density environments.

Some grain models do indeed predict that \kappamu\ should drop at intermediate densities, before increasing again at the highest densities. For example, the \citet{Kohler2015A} description of the THEMIS model finds that the grain-mixture-average \kappamu\ should {\it fall} by a factor of 2.3 (relative to the diffuse ISM) for grains undergoing accretion at intermediate densities ($1500\,n_{\rm H}\,{\rm cm^{-3}}$) -- with \kappamu\ falling by a factor of 26 for amorphous carbon grains in particular. Then at even higher densities, as grains start to aggregate, \kappamu\ will increase again, becoming even higher in the densest regions where icy mantles can form. Again, we do not argue that these specific effects are what are responsible for the relationship we find (as we lack the density resolution, and {\it volume} density information, necessary to test this). However, THEMIS does demonstrate that it is possible to construct a physical dust framework where \kappamu\ falls as $\Sigma_{\it ISM}$ increases, over some intermediate transition regime.

It is also worth considering why our distribution of \kappamu\ values for M\,74 is offset from that of M\,83, by about 0.3\,dex. The most obvious difference in the properties of the two galaxies is the greater ISM surface density of M\,83; but given the apparent anti-correlation of \kappamu\ with $\Sigma_{\it ISM}$, this seems unlikely to be the driver of the in \kappamu. M\,83 has almost 3 times the star formation rate of M\,73 \citep{Nersesian2019A}, despite being physically more compact (see Table~\ref{Table:Galaxy_Properties}), giving it an average star formation rate surface density that is \textgreater\,6 times greater. Despite this, M\,74 has bluer colours, and the relative scale-lengths of the dust and stars in M\,74 and M\,83, as reported in \citet{Casasola2017A}, differ considerably -- in M\,74, the dust and gas have very different scale lengths (2.35\arcmin\ vs 1.04\arcmin), whereas in M\,83, the dust and gas scale lengths are effectively identical (1.66\arcmin\ vs 1.68\arcmin). So there is clearly a difference in the relative geometries of the stars and ISM in these galaxies. When comparing resolved observations of spiral galaxies, it is well established that there can be appreciable differences in ISM properties, even at a given surface density \citep{Usero2015A,Gallagher2018A,Sun2018A}. Therefore, it is not necessarily surprising that \kappad\ may also have different values in different galaxies, at a given surface density.

X-ray observations of M\,74 and M\,83 indicate that their interstellar media contain diffuse hot gas components \citep{Owen2009A} that span much of their discs. Such gas could process the dust in a galaxy, sputtering the grains, and (in standard models) therefore decreasing the grains' \kappad\ \citep{Galliano2018C}. That said, we find decreased \kappad\ in the denser ISM, where grains should be more shielded from X-ray gas. Nonetheless, it is possible that the trends we find may not be applicable to galaxies with less prominent X-ray gas content.

\section{Conclusion} \label{Section:Conclusion}

Using a homogenous dataset assembled as part of the DustPedia project \citep{Davies2017A}, we have produced the first maps of the dust mass absorption coefficient, \kappad, within two nearby galaxies: M\,74 (NGC\,628) and M\,83 (NGC\,5236).

Our method for finding \kappad\ is empirical, and avoids making any assumptions about the composition or radiative properties of the dust. Instead, our approach exploits the fact that the ISM dust-to-metals ratio seems to exhibit minimal variation at high metallicity. With this one assumption, we can use gas and metallicity data to determine dust masses {\it a priori}; by comparing these masses to observed dust emission, we are able to calibrate values for \kappad. Given that the value of the dust-to-metals ratio is much less uncertain than the value of \kappad, we are able to leverage the one to explore the other.

As a proof-of-concept demonstration, we have applied this method on a resolved, pixel-by-pixel basis to M\,74 and M\,83, two nearby face-on spiral galaxies, that have well-suited atomic gas, molecular gas, dust emission, and ISM metallicity data available. We have produced gas-phase metallicity maps for these galaxies, using the many hundreds of available spectra measurements, via a novel application of Gaussian process regression, with which we infer the underlying metallicity distribution.

We find strong evidence for significant variation in \kappamu\ within both galaxies -- by a factor of 2.3 within M\,74 (0.11--0.25\,${\rm m^{2}\,kg^{-1}}$), and by a factor of 5.3 within M\,83 (0.15--0.80\,${\rm m^{2}\,kg^{-1}}$).

We examine whether \kappad\ shows variation with other measured and derived properties of the target galaxies. We find that \kappad\ exhibits a distinct negative correlation with the surface density of the ISM, following a power law slope of  index $-0.36^{+0.26}_{-0.21}$ (although the power-laws for the two galaxies are offset by 0.3 dex). This trend appears to be dictated by the total ISM surface density, as opposed to the surface density of either its atomic, molecular, or dust components. This trend is the opposite of what is predicted by most dust models. However, the relationship is robust against a wide range of changes to our method -- only the adoption of unphysical or highly unusual assumptions would be able to suppress it. We discuss possible ways of reconciling this finding with the current understanding of dust physics -- such as the possibility that our combination of resolution and sensitivity means that we biased towards probing regimes of intermediate density where the broader expected correlation between density and \kappad\ may not hold true.

We also find tentative indications of correlation of \kappad\ with other properties, such as metallicity, NIR radiation field intensity, and dust emissivity slope $\beta$. However, the evidence for these is less conclusive (and some of these parameters were inputs to our \kappad\ calculations), so we are more cautious about the significance of these relationships.

This study lays the groundwork for a wide range of future work. An expanded study of resolved \kappad\ is possible with the DustPedia dataset, but at present the availability of well-resolved metallicity data would limit it to a sample of only 10--20 galaxies. But in future, large IFU surveys of highly-extended nearby galaxies, especially the SDSS-V Local Volume Mapper \citep{Kollmeier2017A} will dramatically improve this situation. Simultaneously, data now exists to apply the dust-to-metals method to large, statistical samples of galaxies on a global basis; in particular, the {\sc Jcmt} dust and gas In Nearby Galaxies Legacy Exploration (JINGLE, \citealp{Saintonge2018A}), which is assembling consistent high-quality CO, \HI, dust, and IFU data for almost 200 galaxies, would be well-suited to this task.

Most importantly, many of the questions raised could be tackled by conducting a similar analysis at improved spatial resolution. For this reason, we have begun work on applying this method as part of an analysis of several Local Group galaxies -- including the Large and Small Magellanic Clouds, where we enjoy particularly exquisite resolution. Most significantly, better resolution will allow us to cleanly probe a larger range of density, from the densest grain-grown regions, down to the most diffuse ISM. We will thereby test if the surprising anticorrelation between \kappamu\ and $\Sigma_{\it ISM}$ holds true. Another benefit to expanding our analysis to the Magellanic Clouds is that they are the subjects of ongoing work to perform the first extragalactic depletion analyses \citep{Jenkins2017A,Roman-Duval2019A}. Exploiting that data will allow us to use in-situ measurements of the dust-to-metal ratio, removing the single largest source of uncertainty we presently face, and allowing us to produce the most reliable empirical \kappad\ determinations available with current data.

\section*{Acknowledgements} \label{Section:Acknowledgements}

{\small The DustPedia project\footnote{\url{https://dustpedia.com/}} \citep{Davies2017A}) has received funding from the European Union’s Seventh Framework Programme (FP7) for research, technological development, and demonstration, under grant agreement 606824 (PI Jon Davies).

The authors thank the anonymous referee whose comments have materially improved the quality of this work.

CJRC acknowledges financial support from the National Aeronautics and Space Administration (NASA) Astrophysics Data Analysis Program (ADAP) grant 80NSSC18K0944. CJRC thanks Andreas Lundgren and Tommy Wiklind for providing reduced SEST CO data for M\,83 \citep{Lundgren2004A}.
CJRC also thanks Philip Wiseman, Bruce Draine, Julia Roman-Duval, Karl Gordon, Rosie Beeston, and Phil Cigan for helpful discussions \& input.

This research made use of {\sc {\sc astropy}}\footnote{\url{https://www.astropy.org/}}, a community-developed core {\sc python} package for Astronomy \citep{astropy2013,astropy2019}. This research made use of {\sc astroquery}\footnote{\url{https://astroquery.readthedocs.io}}, an {\sc astropy}-affiliated {\sc python} package for accessing remotely hosted
astronomical data \citep{Ginsburg2019B}. This research made use of {\sc reproject}\footnote{\url{https://reproject.readthedocs.io}}, an {\sc astropy}-affiliated {\sc python} package for image reprojection. This research has made use of {\sc numpy}\footnote{\url{https://numpy.org/}} \citep{VanDerWalt2011B}), {\sc scipy}\footnote{\url{https://scipy.org/}} \citep{SciPy2001}, and {\sc matplotlib}\footnote{\url{https://matplotlib.org/}} \citep{Hunter2007A}. This research made use of {\sc aplpy}\footnote{\url{https://aplpy.github.io/}}, an open-source plotting package for {\sc python} \citep{Robitaille2012B}. This research made use of the {\sc pandas}\footnote{\url{https://pandas.pydata.org/}} data structures package for {\sc python} \citep{McKinney2010}. This research made use of the {\sc scikit-image}\footnote{\url{https://scikit-image.org/}} image processing package for {\sc python} and the {\sc scikit-learn}\footnote{\url{https://scikit-learn.org}} machine learning package for {\sc python} \citep{Scikit-Learn2011}. This research made use of {\sc emcee}\footnote{\url{https://dfm.io/emcee/current/}}, the MCMC hammer for {\sc python} \citep{ForemanMackey2013B}. This research made use of the {\sc pymc3}\footnote{\url{https://docs.pymc.io/}; \citealp{Salvatier2016A}} MCMC package for {\sc python}. This research made use of the {\sc corner}\footnote{\url{https://corner.readthedocs.io}}  scatterplot matrix plotting package for {\sc python} \citep{ForemanMackey2016D}. This research made use of {\sc ipython}, an enhanced interactive {\sc python} \citep{Perez2007A}. This research made use of {\sc python} code for working in the luminance-chroma-hue colour space, written by Endolith\footnote{\url{https://gist.github.com/endolith/5342521}}, kindly made available free and open-source under the BSD License\footnote{\url{https://opensource.org/licenses/BSD-3-Clause}}, and copyright 2014 Endlolith.

This research has made use of {\sc topcat}\footnote{\url{http://www.star.bris.ac.uk/~mbt/topcat/}} \citep{Taylor2005A}, an interactive graphical viewer and editor for tabular data, which was initially developed under the UK Starlink project, and has since been supported by the Particle Physics and Astronomy Research Council (PPARC), the VOTech project, the AstroGrid project, the Astronomical Infrastructure for Data Access (AIDA) project, the Science and Technology Facilities Council (STFC), the German Astrophysical Virtual Observatory (GAVO) project, the European Space Agency (ESA), and the Gaia European Network for Improved data User Services (GENIUS) project. This research made use of {\sc ds9}, a tool for data visualisation supported by the Chandra X-ray Science Center (CXC) and the High Energy Astrophysics Science Archive Center (HEASARC) with support from the James Webb Space Telescope (JWST) Mission office at the Space Telescope Science Institute for 3D visualisation.

This research made use of \montage\footnote{\url{https://montage.ipac.caltech.edu/}}, which is funded by the National Science Foundation under Grant Number ACI-1440620, and was previously funded by the NASA's Earth Science Technology Office, Computation Technologies Project, under Cooperative Agreement Number NCC5-626 between NASA and the California Institute of Technology.

This research made use of the VizieR catalogue access tool\footnote{\url{https://vizier.u-strasbg.fr/viz-bin/VizieR}} \citep{Ochsenbein2000B}, operated at CDS, Strasbourg, France. This research has made use of the {\sc Nasa/ipac} Extragalactic Database\footnote{\url{https://ned.ipac.caltech.edu/}} (NED), operated by the Jet Propulsion Laboratory, California Institute of Technology, under contract with NASA.

This research made use of data from the SEST, which was operated jointly by the European Southern Observatory (ESO) and the Swedish National Facility for Radio Astronomy, Chalmers University of Technology.

Much of the model fitting performed in this work benefitted from the invaluable guidance provided in \citet{Hogg2010B}.}

\bibliographystyle{mnras}
\bibliography{ChrisBib}

\begin{appendix}

\section{Radial Metallicity Profile Fitting} \label{AppendixSection:Metallicity_Profile_Fitting}

The model we employed to fit the radial metallicity profiles of our target galaxies in Section~\ref{Subsection:Metallicity_Data} is described by the likelihood function:
\begin{multline}
\mathcal{L}(Z^{[\frac{\rm O}{\rm H}]} | R, \sigma, m_{Z}, c_{Z}, \psi) = \\
\prod^{n}_{i} \left( \frac{1}{\sqrt{2 {\rm \pi} (\sigma^{2}_{i} + \psi^{2})}}  \times \exp{\left(\frac{- (Z^{[\frac{\rm O}{\rm H}]}_{i} - m_{Z} R_{i} - c_{Z})^{2}}{(2 \sqrt{\sigma^{2}_{i} + \psi^{2}})^{2}}\right)} \right)
\label{AppendixEquation:Gradient_Likelihood}
\end{multline}

\noindent where $Z^{[\frac{\rm O}{\rm H}]}_{i}$ is the \logOH\ metallicitiy of the $i$\textsuperscript{th} datapoint, $R_{i}$ is the deprojected galactocentric radius of the $i$\textsuperscript{th} datapoint (as a fraction of the $R_{25}$), $m_{Z}$ is the metallicity gradient (in ${\rm dex\,R^{-1}_{25}}$), $c_{Z}$ is the central metallicity (in \logOH), $\psi$ is the intrinsic scatter (in dex), and $n$ is the number of datapoints.

We determined the posterior probability of our variables of interest -- $m_{Z}$, $c_{Z}$, and $\psi$ -- in a Bayesian manner, sampling the posterior Probability Distribution Functions (PDF) using the {\sc pymc3} \citep{Salvatier2016A} MCMC package for {\sc python}.

To inform the priors, we first performed a simple, preliminary least-squares fit, with only the gradient and central metallicity as free parameters. The priors on all three parameters then took the form of normal distributions. For $c_{Z}$, the mean of the prior was set to the central metallicity found by the preliminary least-squares fit, and the standard deviation on the prior was set to the standard deviation of all the input metallicity values. For $m_{Z}$, the mean of the prior was set to the gradient found by the preliminary least-squares fit, and standard deviation of the prior was set to the absolute value of the gradient found by the preliminary least-squares fit. For $\psi$, both the mean and standard deviation of the prior were set to the root-mean-square of the residuals between the input metallicity values and the preliminary least-squares fit.

\section{Uncertainties on GPR Metallicity Mapping} \label{AppendixSection:GPR_Uncertainties}

To determine the uncertainty of the GPR metallicity maps, we repeated the regression procedure 1000 times. For each iteration, we draw a random sample from the posterior PDF of our radial metallicity profile model, and used that sample to calculate the residual on each datapoint; we then applied the GPR to these residuals in the same manner as described above. For each iteration, the GPR produced a full posterior PDF for the predicted metallicity in each pixel (and by definition, Gaussian process regression yields Gaussian posterior PDFs). 

Having repeated this process for the 1000 iterations, we had 1000 posterior PDFs for each pixel; these are then combined to give each pixel's final metallicity PDF. To quantify the uncertainty in each pixel, we take the 63.8\% quantile around the posterior median; these are the uncertainty values plotted in the lower-left panels of Figures~\ref{Fig:NGC0628_Metallicity_Grid} and \ref{Fig:NGC5236_Metallicity_Grid}. As can be seen, the uncertainty on the regression is low (\textless\,0.05\,dex) for pixels that have plenty of spectra metallicities; whilst for pixels more distant from any spectra, making the predicted values more dependent upon extrapolation, the uncertainty is much larger (\textgreater\,0.25\,dex). Indeed, for pixels with few or no spectra metallicities in the immediate vicinity, relying upon the metallicity predicted by a 1-dimensional globally-fitted gradient could provide a false sense of confidence -- especially in M\,83, where the metallicity data is concentrated in a central band.  We therefore argue that in these areas, the larger uncertainties predicted by our GPR approach are likely to be more realistic.

\section{Validation of GPR Metallicity Mapping} \label{AppendixSection:GPR_Validation}

\begin{figure}
\centering
\includegraphics[width=0.475\textwidth]{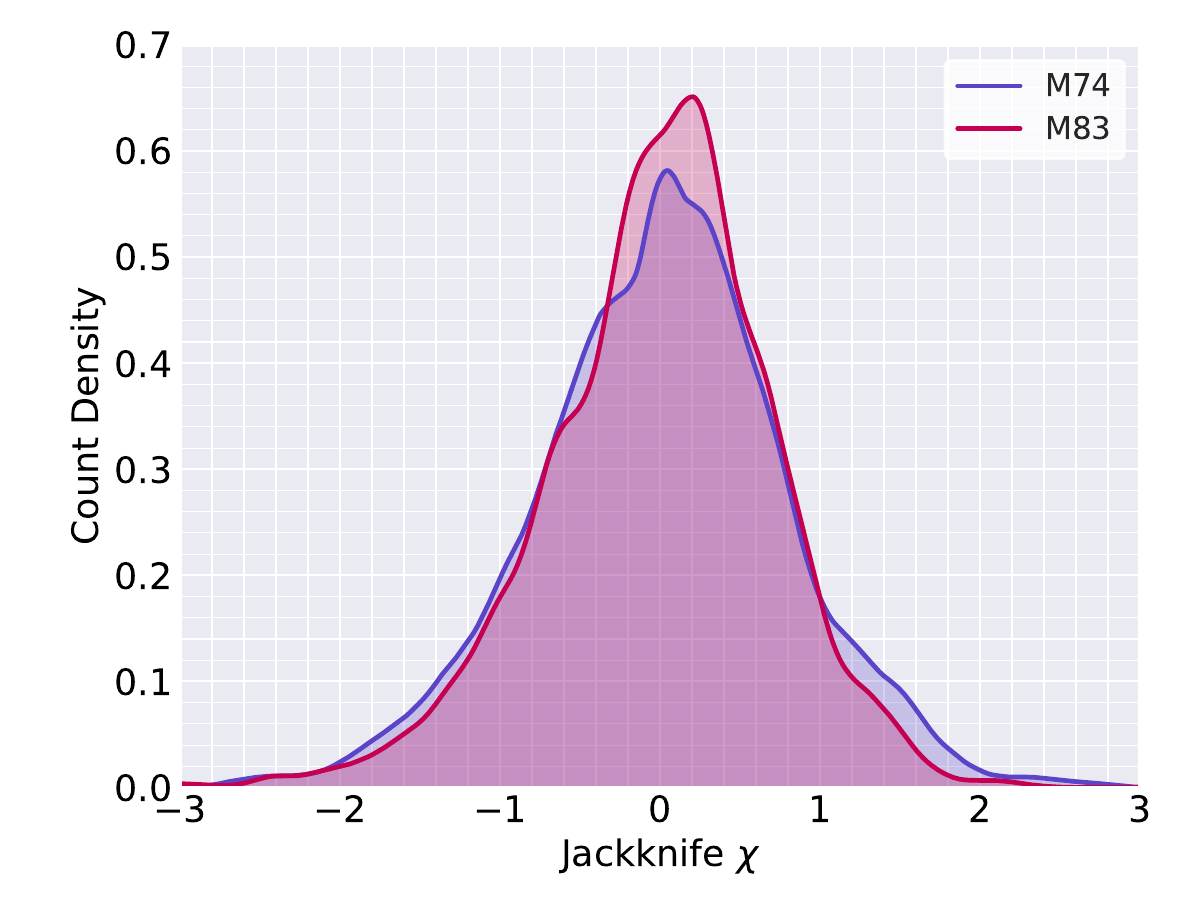}
\caption{Distribution of $\chi$ values found for our jackknife cross-validation of the GPR metallicity mapping. Distributions plotted as Kernel Density Estimates (KDEs), using an Epanechnikov kernel, with bandwidth calculated using the Sheather-Jones rule \citep{Sheather1991}.}
\label{AppendixFig:Jackknife_Chi}
\end{figure}

To verify that our GPR metallicity mapping technique is reliable, and not generating spurious features in the final metallicity maps, we used a Monte Carlo jackknife cross-validation analysis. For this, we performed 500 repeats of the GPR metallicity mapping; for each repeat, half of the spectra metallicity points were selected at random to be excluded from the fitting, to serve as a control sample for later reference. The GPR was then computed using the remaining half of the points (but otherwise following the modelling process as laid out above). By comparing the metallicity values of the masked spectra to the metallicities predicted by the GPR method at their positions, we can evaluate the accuracy of the generated metallicity maps.

For each of the 500 jackknife iterations, we found the deviations between the known metallicities of the control spectra, and the metallicity predicted by the GPR at those positions. We assessed the deviation at each position in terms of $\chi$, defined as:
\begin{equation}
\chi = \frac{Z^{[\frac{\rm O}{\rm H}]}_{\it GPR} - Z^{[\frac{\rm O}{\rm H}]}_{\it spec}}{\sqrt{ \left( {\sigma^{[\frac{\rm O}{\rm H}]}_{\it GPR}} \right)^{2} + \left( {\sigma^{[\frac{\rm O}{\rm H}]}_{\it spec}} \right)^{2} }}
\label{Equation:Chi}
\end{equation}

\noindent where $\smash{Z^{[O/H]}_{\it GPR}}$ is the metallicity predicted by the GPR at the position in question, $\smash{Z^{[O/H]}_{\it spec}}$ is the actual metallicity of the spectra, $\smash{\sigma^{[O/H]}_{\it GPR}}$ is the uncertainty on the GPR at the position in question, and $\smash{\sigma^{[O/H]}_{\it spec}}$ is the uncertainty on the spectra metallicity (all $Z$ and $\sigma$ terms expressed in ${12 + \log_{10} [\frac{\rm O}{\rm H}]}$ units). In short, $\chi$ expresses the deviation in terms of the mutual uncertainty on the spectra metallicity and the GPR.

If the metallicities predicted via GPR suffer from no systematic offset, then the mean $\chi$ should be $0 \pm n^{-\frac{1}{2}}$ (where n is the number of control spectra). Similarly, if the uncertainties on the GPR metallicities are Gaussian and accurate, then 68.3\% of the values of $\chi$ should lie in the range $-1 < \chi < 1$.

The distribution of jackknife $\chi$ values we find for both galaxies are shown in Figure~\ref{AppendixFig:Jackknife_Chi}. The distributions are symmetric, near-Gaussian, and centred close to zero. The mean jackknife $\chi$ values are $0.0079 \pm 0.0029$  and $-0.0057 \pm 0.0023$ for M\,74 and M\,83 respectively. These offsets are $> 2 \sigma$, suggesting that there tends to be a small systematic offset (positive for M\,74, and negative for M\,83) between the metallicity predicted by the GPR, and the actual metallicity of the spectra. But whilst technically significant, these systematic offsets are nonetheless vanishingly small in terms of actual metallicity -- the mean jackknife deviation in \logOH\ units is 0.00056 for M\,74, and -0.00037 for M\,83. We are satisfied that systematic effects at this scale are minute enough to have no appreciable impact on any of our results.

For M\,74, 80.5\% of the jackknife $\chi$ values lie in the $-1 < \chi < 1$ range; for M\,83 the fraction is 85.4\%. These are both somewhat larger than the expectation of 68.3\%, which suggests that our GPR maps are actually somewhat {\it more} precise than suggested by their uncertainties. In other words, it appears that the GPR uncertainties are overestimated by factors of approximately 1.18 and 1.25 (for M\,74 and M\,83 respectively) -- a small enough difference that we judge it unnecessary to attempt a {\it post-hoc} fine-tuning of the output uncertainties.

Additionally, see the discussion in Section~\ref{Subsection:Robustness_Discussion} of the effect of the metallicity maps upon our resulting maps of \kappad.

\section{Dust SED Priors} \label{AppendixSection:SED_Priors}

\begin{figure*}
\centering
\includegraphics[width=0.2425\textwidth]{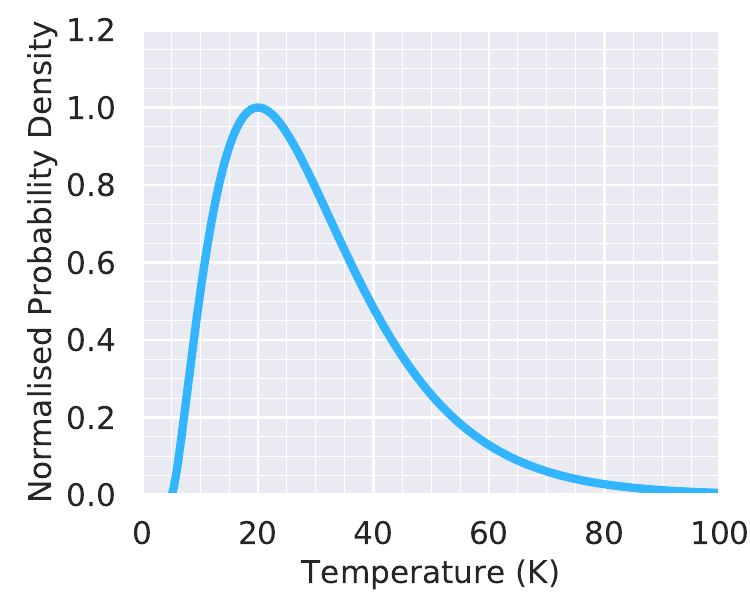}
\includegraphics[width=0.2425\textwidth]{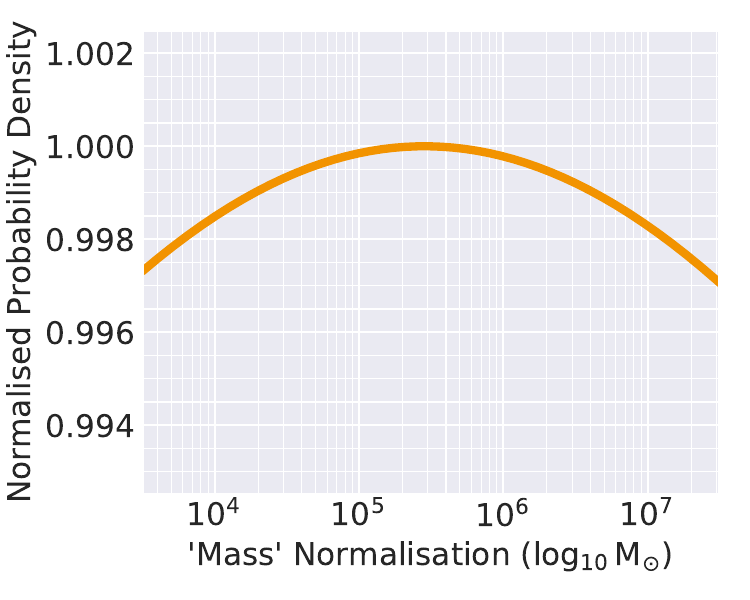}
\includegraphics[width=0.2425\textwidth]{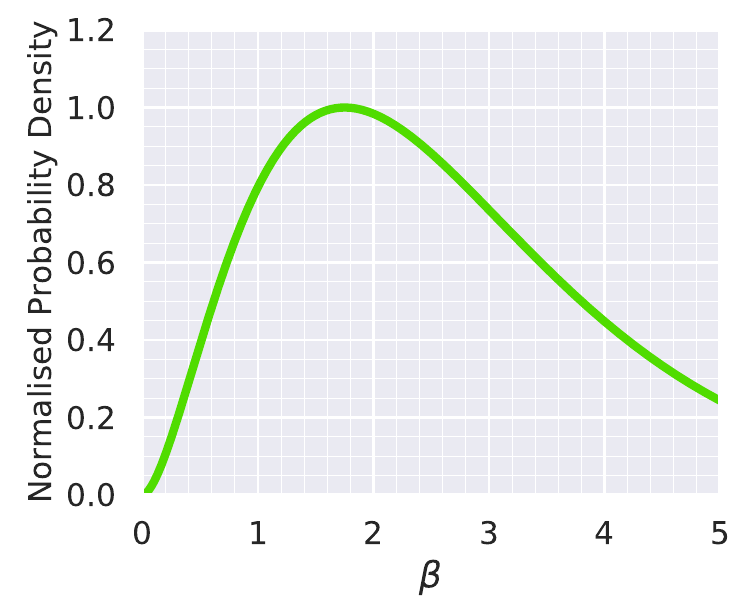}
\includegraphics[width=0.2425\textwidth]{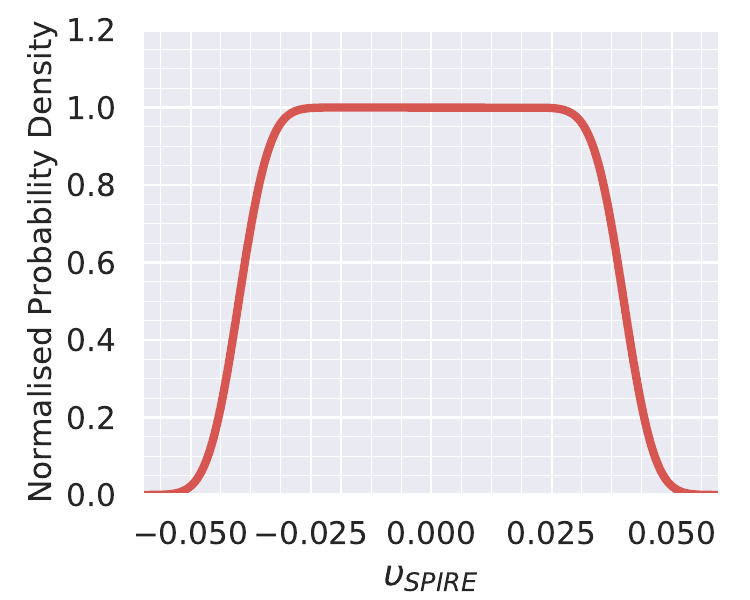}
\caption{Prior probability distributions for $T_{d}$, $M_{d}^{\rm (norm)}$, $\beta$, and $\upsilon_{\rm SPIRE}$. For ease of viewing and comparison, all distributions have been normalised so that they peak at a probability density of 1. The prior for $M_{d}^{\rm (norm)}$ is computed for each source (eg, pixel) based on its brightness and distance; the exemplar distribution displayed in the upper-right panel is for the pixel in M\,83 centred at $\alpha = 204.2841^{\circ}, \delta = -29.8559^{\circ}$.}
\label{AppendixFig:SED_Priors}
\end{figure*}

Our SED-fitting procedure, detailed in Section~\ref{Subsection:SED_Fitting}, has 6 free parameters: dust temperature, $T_{d}$; dust `mass' normalisation, $M_{d}^{\rm (norm)}$; emissivity slope, $\beta$; and correlated photometric error in the \hersc-SPIRE bands, $\upsilon_{\rm SPIRE}$. The prior probability distributions for all these free parameters are shown in Figure~\ref{AppendixFig:SED_Priors}.

\subsection{Temperature Prior} \label{AppendixSubsection:Mass_Prior}

The prior on temperature is given by a standardised\footnote{Standardised as per the SciPy \citep{SciPy2001} gamma distribution implementation: \url{https://docs.scipy.org/doc/scipy/reference/generated/scipy.stats.gamma.html}} gamma distribution of the form:
\begin{equation}
\mathcal{P}(T) = \frac{\left(\frac{T-l}{s}\right)^{\alpha-1} \exp{\left(\frac{T-l}{s}\right)}}{s\ \Gamma(\alpha)}
\label{AppendixEquation:Temp_Prior_1}
\end{equation}

\noindent where $T_{d}$ is the temperature, $l$ is the location parameter, $s$ is the scale parameter, and $\alpha$ is the shape parameter. The location parameter $l$ functions such that $\mathcal{P}(T<l) = 0$. We define the scale parameter in relation to the distribution mode $\check{T}$ (ie, the temperature with the peak prior probability), according to:
\begin{equation}
\begin{aligned}
s &= \frac{\check{T}-l}{\alpha-1}
\end{aligned}
\label{AppendixEquation:Temp_Prior_2}
\end{equation}

\noindent where, for our $T_{d}$ prior, these parameters take values of $\alpha = 2.5$, $l = 5$, and $\check{T} = 20$. 

For $T_{d}$, the modal value of 20\,K corresponds to the approximate average of the cold dust temperatures seen in nearby galaxies, in both global \citep{Galametz2012A,Clemens2013A,Ciesla2014A} and resolved \citep{MWLSmith2012B,Gordon2014B,Tabatabaei2014A} analyses. Across the 13--30\,K temperature range, $\mathcal{P}(T_{c}) > 0.8\mathcal{P}(\check{T}_{c})$; this corresponds to range spanned by the lower cold dust temperatures seen in blue dust- and gas-rich galaxies \citep{CJRClark2015A,Dunne2018A}, to the higher cold dust temperatures seen in dust-poor dwarf galaxies \citep{Remy-Ruyer2013,Izotov2014D}. In other words, temperatures across this `standard' temperature range are only slightly less favoured than $\check{T}_{c}$. Outside this range, there is an increasing penalty -- especially towards lower temperatures, where $\mathcal{P}(T_{c}<5) = 0$, to rule out unphysically cold dust.



\subsection{Mass Prior} \label{AppendixSubsection:Mass_Priors}

As described in Section~\ref{Subsection:SED_Fitting}, our SED fitting procedure uses an arbitrary placeholder value of \kappad\ (because the whole purpose of the SED fitting is to {\it find} values of \kappad); as a result, the `mass' variable being fitted simply serves as a normalisation parameter.

Our mass normalisation prior takes the form of a 1\textsuperscript{st}-order (ie, 1 degree of freedom) Student $t$ distribution, constructed in base-10 logarithmic space (see Figure~\ref{AppendixFig:SED_Priors}), with widths of $\sigma = 10\,{\rm dex}$. The peak of the mass normalisation prior is computed separately for each source (eg, pixel), based on its distance and brightness, according to the formula:
\begin{equation}
\log_{10}{(\check{M})} = \log_{10}{(S_{\it max} D^{2})} + \left(\frac{\check{T}-20}{-15}\right) + 4
\label{AppendixEquation:Mass_Prior_Mode}
\end{equation}
\noindent where $\check{M}^{\rm (norm)}$ is the modal mass of the prior probability distribution, $S_{\it max}$ is the brightest flux measured in the 150--1000\,\micron\ range (in Jy), and $D$ is the source distance (in Mpc). For MBB dust SEDs with temperatures in the 15--25\,K range, the brightest flux in the \spitz\ and \hersc\ bands will be the 160\,\micron\ measurement.

Equation~\ref{AppendixEquation:Mass_Prior_Mode} is a purely empirical relation, derived from the SED fitting of \hersc\ Reference Survey \citep{Boselli2010B} galaxies, as performed in \citet{CJRClark2015A} and \citet{CJRClark2016A}. 

The prior on $M^{\rm (norm)}$ shown in Figure~\ref{AppendixFig:SED_Priors} is for an example pixel from our M\,83 data (processed as per Section~\ref{Subsection:Data_Preparation}), centred at $\alpha = 204.2841^{\circ}, \delta = -29.8559^{\circ}$. The brightest band for this pixel is 160\,\micron, where the flux is 1.18\,Jy. Given a distance to M\,83 of 4.9\,Mpc, that corresponds to a priors centred at $\check{M}^{\rm (norm)} = 5.45\,{\rm log_{10}\,M_{\odot}}$, as per Equation~\ref{AppendixEquation:Mass_Prior_Mode}. 

The mass normalisation prior is designed to be very weak. This is because the strong $M \propto T^{4+\beta}$ dependence of mass on temperature (for a given luminosity) means that the fitted value of the mass normalisation term is often driven primarily by the fitted temperature. 

\subsection{$\beta$ Prior} \label{AppendixSubsection:Beta_Prior}

The prior on $\beta$ takes the form of a standardised gamma distribution, identical to Equations~\ref{AppendixEquation:Temp_Prior_1} and ~\ref{AppendixEquation:Temp_Prior_2}, except with $\beta$ replacing $T_{d}$, and $\check{\beta}$ replacing $\check{T}$. The parameters for our $\beta$ prior take values of $\alpha = 2.75$, $l = 0$, and $\check{\beta} = 1.75$.

For nearby galaxies and the Milky Way, $\beta$ is typically found to lie in the range 1.5--2.0, with resolved analyses finding values spanning 1.0--2.75 \citep{MWLSmith2012B,Kirkpatrick2013B,Planck2013XI}. We therefore construct our prior such that it peaks at $\check{\beta} = 1.75$, with $\mathcal{P}(\beta) > 0.8\mathcal{P}(\check{\beta})$ across the 1.0--2.75 range. To exclude dubiously-physical low $\beta$ values, $\mathcal{P}(\beta<0) = 0$.

\subsection{$\upsilon_{\rm SPIRE}$ Prior} \label{AppendixSubsection:Upsilon_Prior}

As discussed in Section~\ref{Subsection:SED_Fitting}, the calibration uncertainties on \hersc-SPIRE photometry has a correlated systematic error component, which we term $\upsilon_{\rm SPIRE}$, arising from uncertainty on the emission model of Neptune, the instrument's primary calibrator. $\upsilon_{\rm SPIRE}$ has a value of $\pm 4\%$; the true value of the systemic error is believed to be equally likely to lie anywhere in that range, with minimal likelihood ($\sim 5\%$) of the value lying outside it (\citealp{Bendo2013A}; A.\,Papageorgiou, {\it priv. comm.}; C.\,North, {\it priv. comm.}).

We therefore use a prior for $\upsilon_{\rm SPIRE}$ that takes the form of a boxcar function convolved with a Gaussian distribution. The boxcar function has a value of 1 over the range -0.04--0.04, with a value of 0 beyond this. The Gaussian with which it was smoothed has a standard deviation of 0.005. In the resulting prior, as shown in the lower-right panel of Figure~\ref{AppendixFig:SED_Priors}, 95\% of the probability density is contained within the -0.04--0.04 range.

\section{\kappad\ Maps at 160\,\micron} \label{AppendixSection:Kappa_160_Maps}

\begin{figure}
\centering
\includegraphics[width=0.23\textwidth]{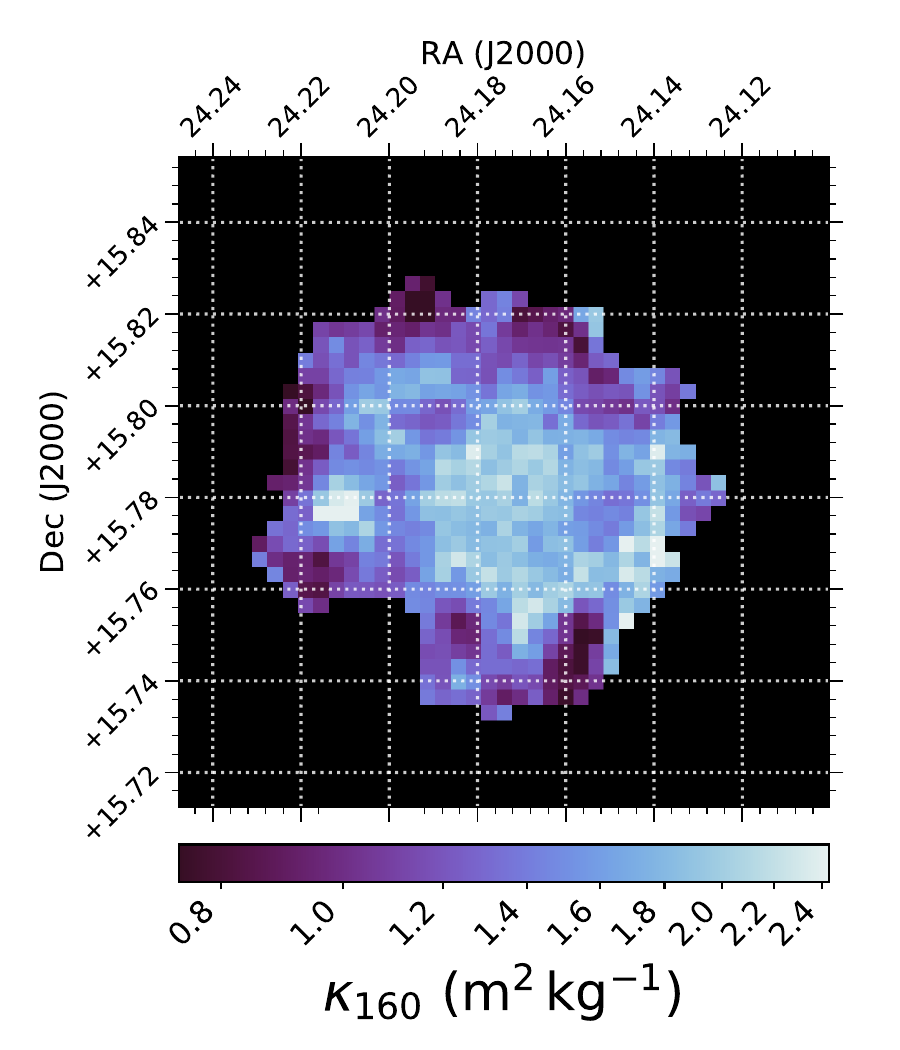}
\includegraphics[width=0.23\textwidth]{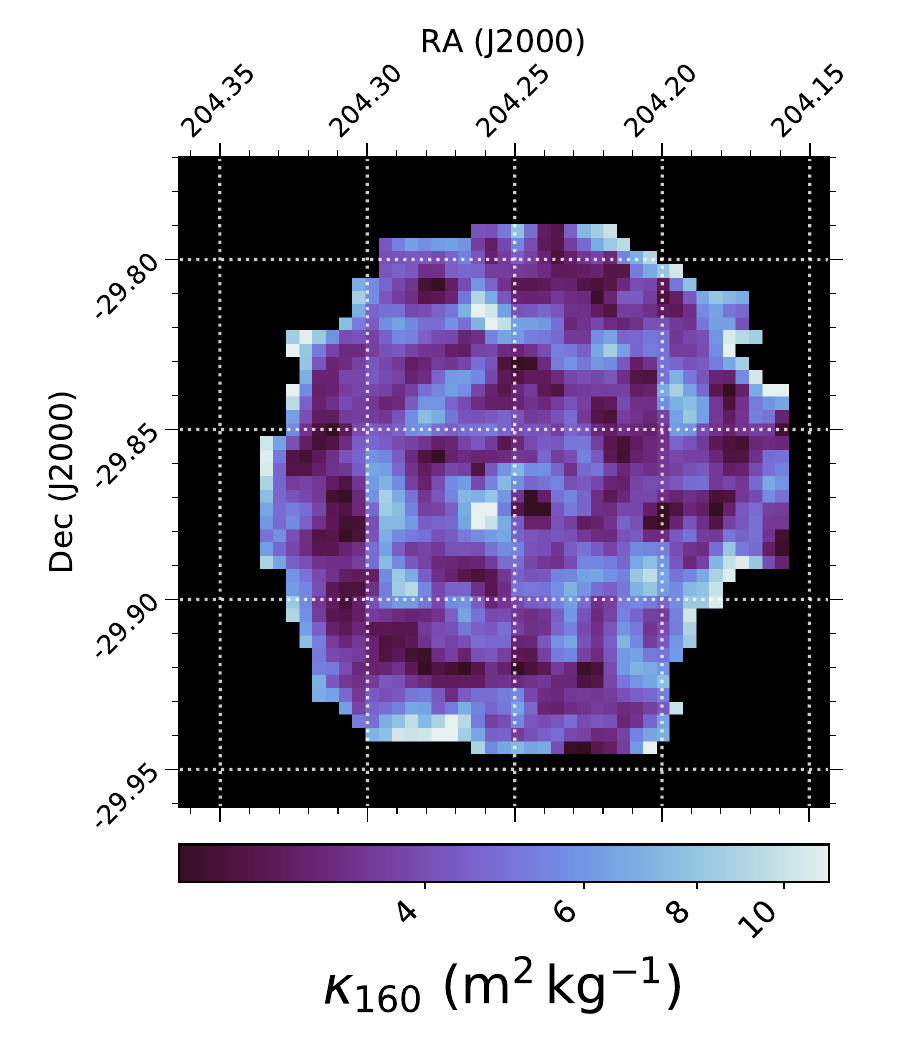}
\caption{Maps of $\kappa_{160}$ within M\,74 ({\it left}) and M\,83 ({\it right}).}
\label{AppendixFig:Kappa_160}
\end{figure}

As discussed in Section~\ref{Section:Results}, we calculated our \kappad\ maps at a reference wavelength of 500\,\micron. This is the longest wavelength at which we have data, making it less sensitive to uncertainties in temperature derived from the SED fitting. However, many authors opt to present \kappad\ at 160\,\micron, as this is the wavelength regime at which the \kappad\ of carbonaceous and silicate dust is most comparable.

For completeness, we therefore also produced $\kappa_{160}$ maps, which are shown in Figure~\ref{AppendixFig:Kappa_160}. The maps are noisier than those computed at 500\,\micron, but the overall morphology of $\kappa_{160}$ in both galaxies is nonetheless the same as that of \kappamu.

Using the same independent-pixel non-parametric bootstrap approach as in Section~\ref{Section:Results}, we find a median underlying range of $\kappa_{160}$ values of 0.74--2.4\,${\rm m^{2}\,kg^{-1}}$ for M\,74 (a factor of 3.2 variation), and 2.1--12\,${\rm m^{2}\,kg^{-1}}$ for M\,83 (a factor of 5.7 variation).

\section{\kappamu\ Maps Using Different Strong-Line Metallicity Prescriptions} \label{AppendixSection:Kappa_Z_Variations}

\begin{figure}

\centering
\includegraphics[width=0.23\textwidth]{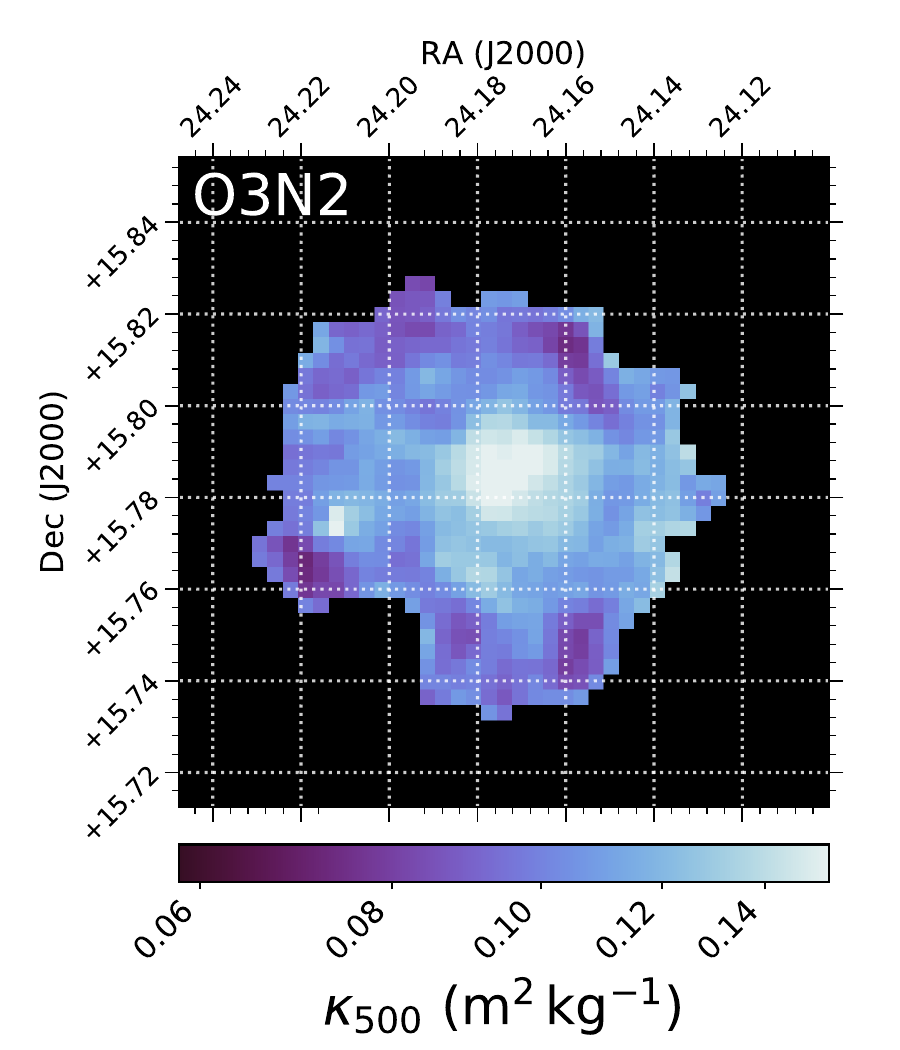}
\includegraphics[width=0.23\textwidth]{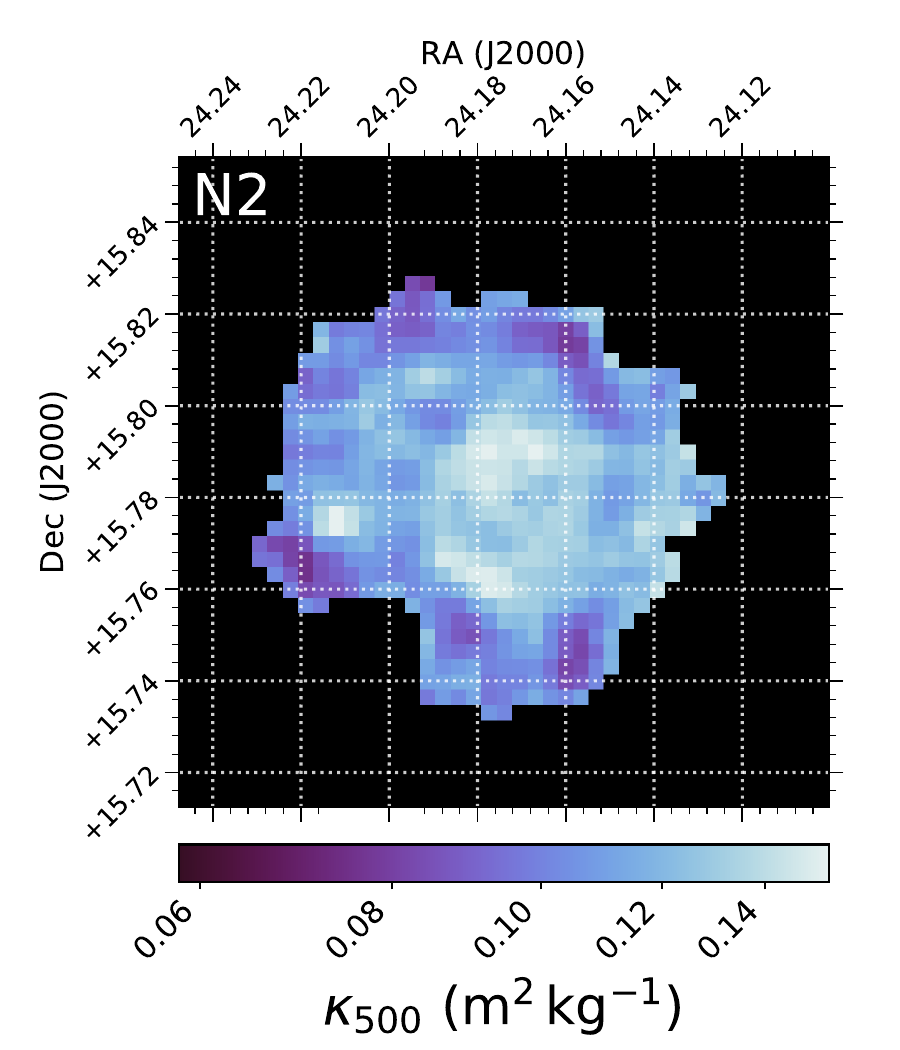}
\includegraphics[width=0.23\textwidth]{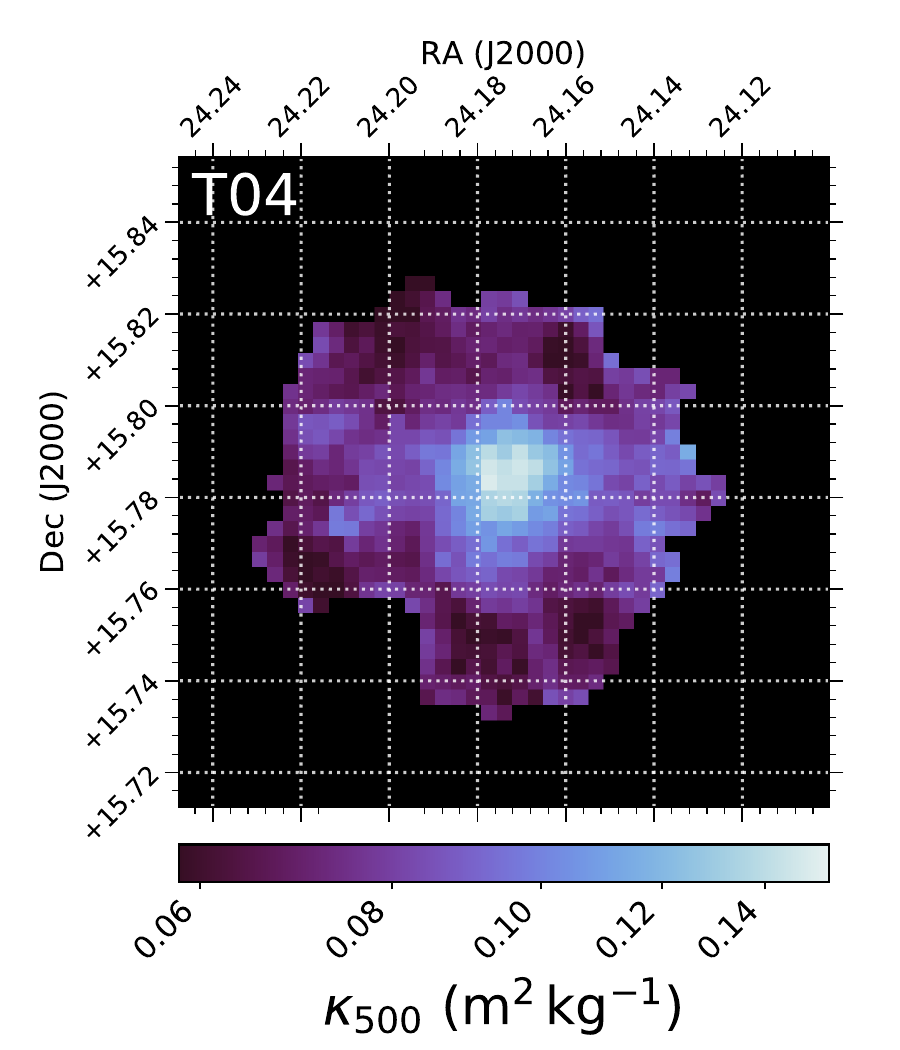}
\includegraphics[width=0.23\textwidth]{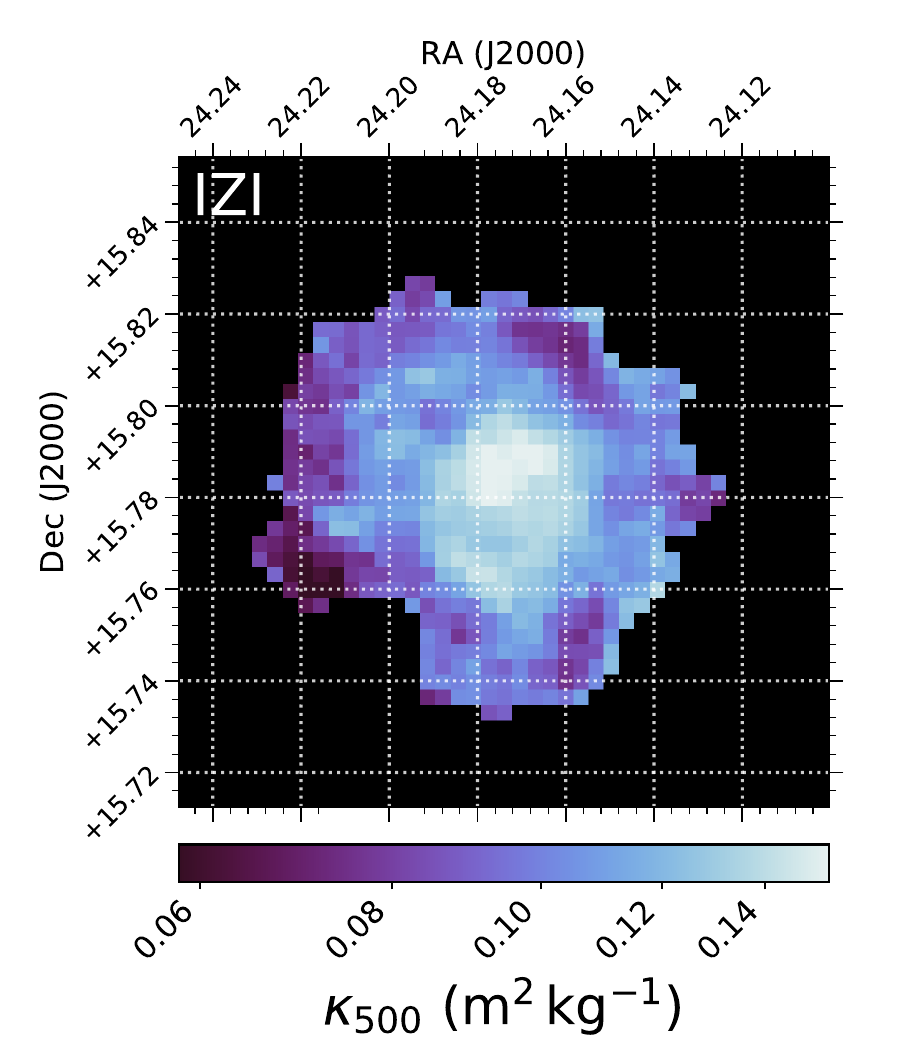}
\caption{Maps of \kappamu\ within M\,74, calculated using metallicities produced via different strong-line prescriptions. {\it Upper left:} the O3N2 prescription of \citet{Pettini2004B}. {\it Upper right:} The N2 prescription of \citet{Pettini2004B}. {\it Lower left:} The prescription of \citet{Tremonti2004A}. {\it Lower right:} the IZI prescription of \citet{Blanc2015A}.}
\label{AppendixFig:NGC0628_Kappa_Z_Variations}

\includegraphics[width=0.23\textwidth]{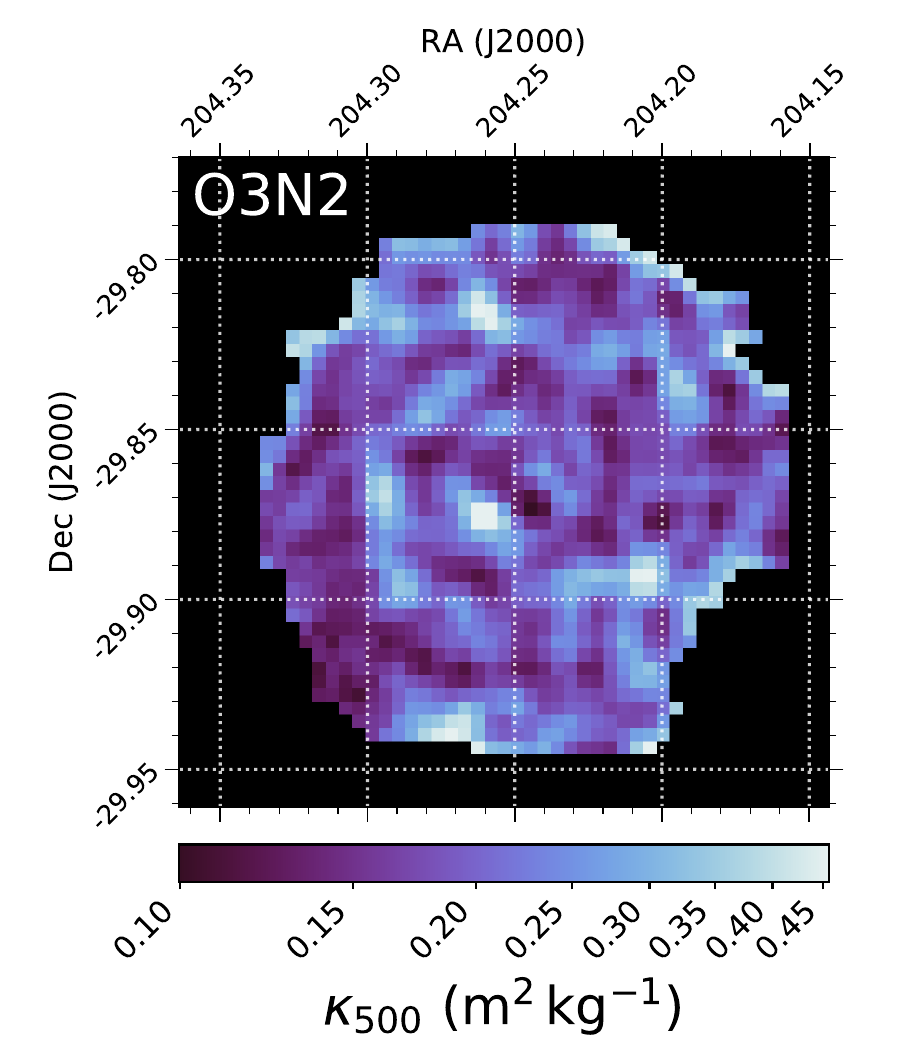}
\includegraphics[width=0.23\textwidth]{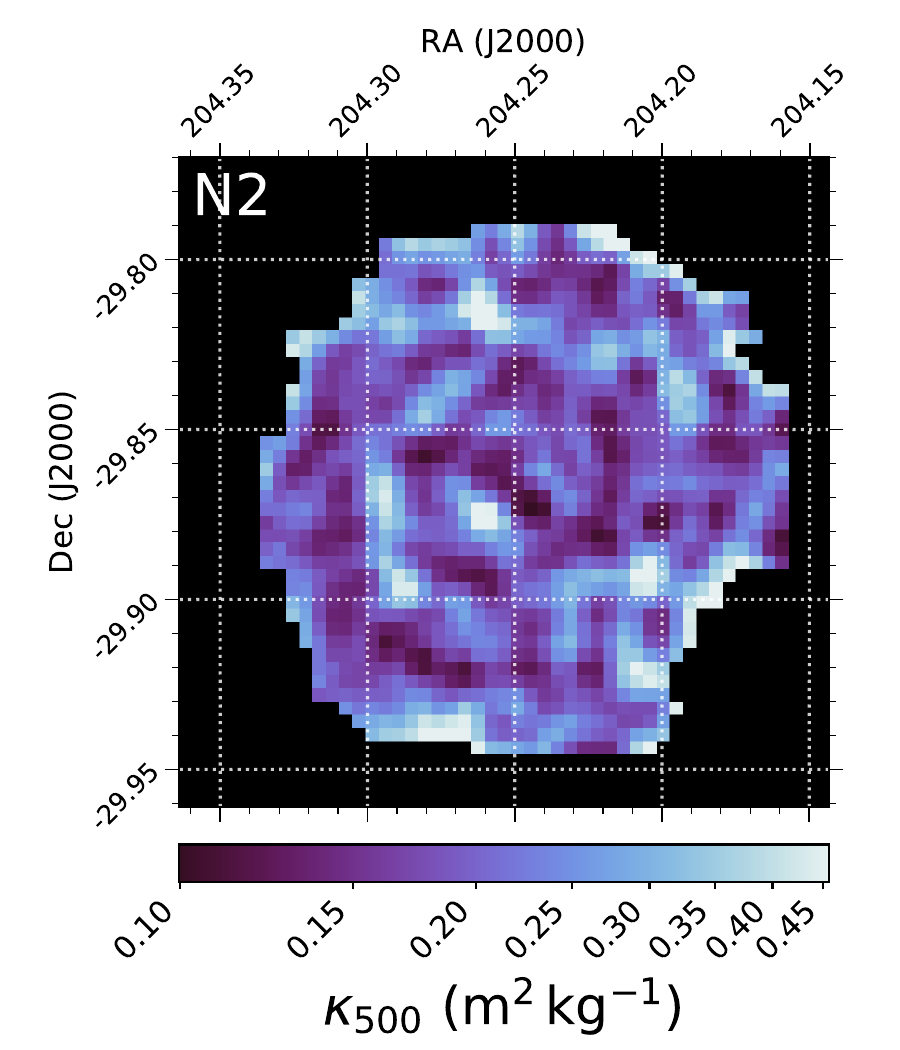}
\includegraphics[width=0.23\textwidth]{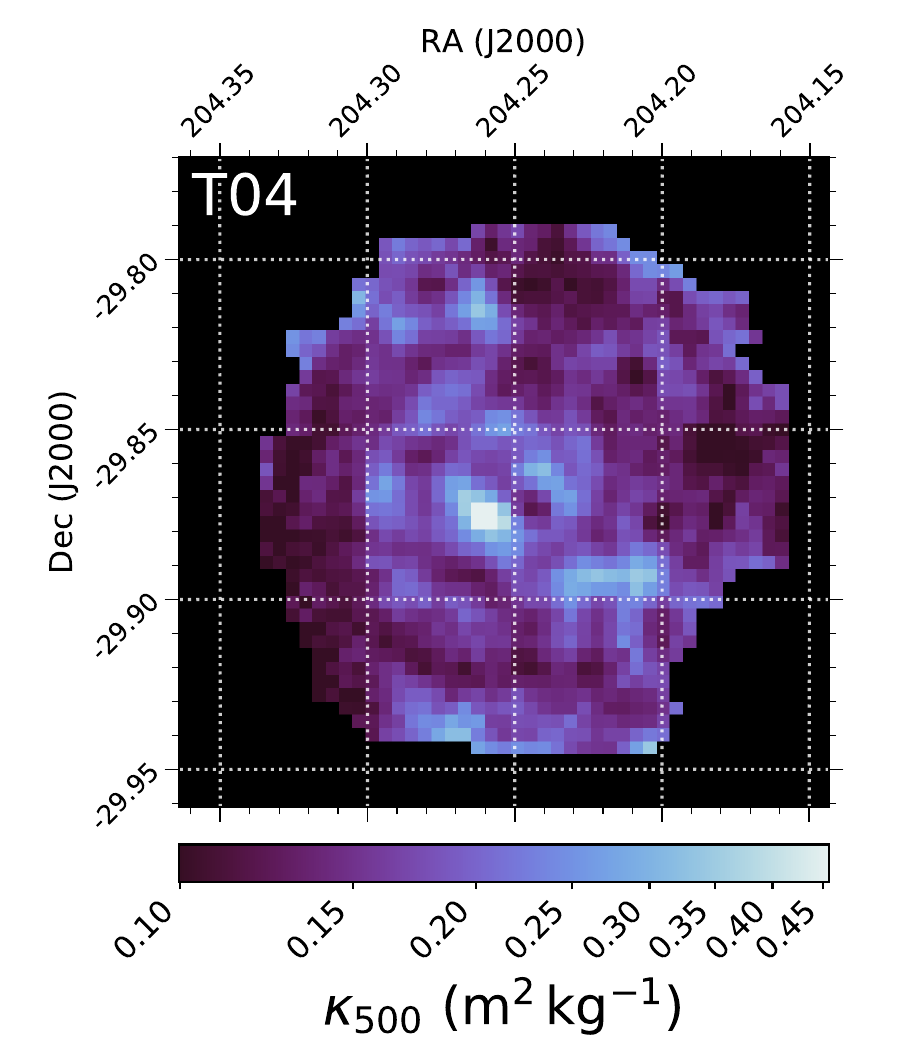}
\includegraphics[width=0.23\textwidth]{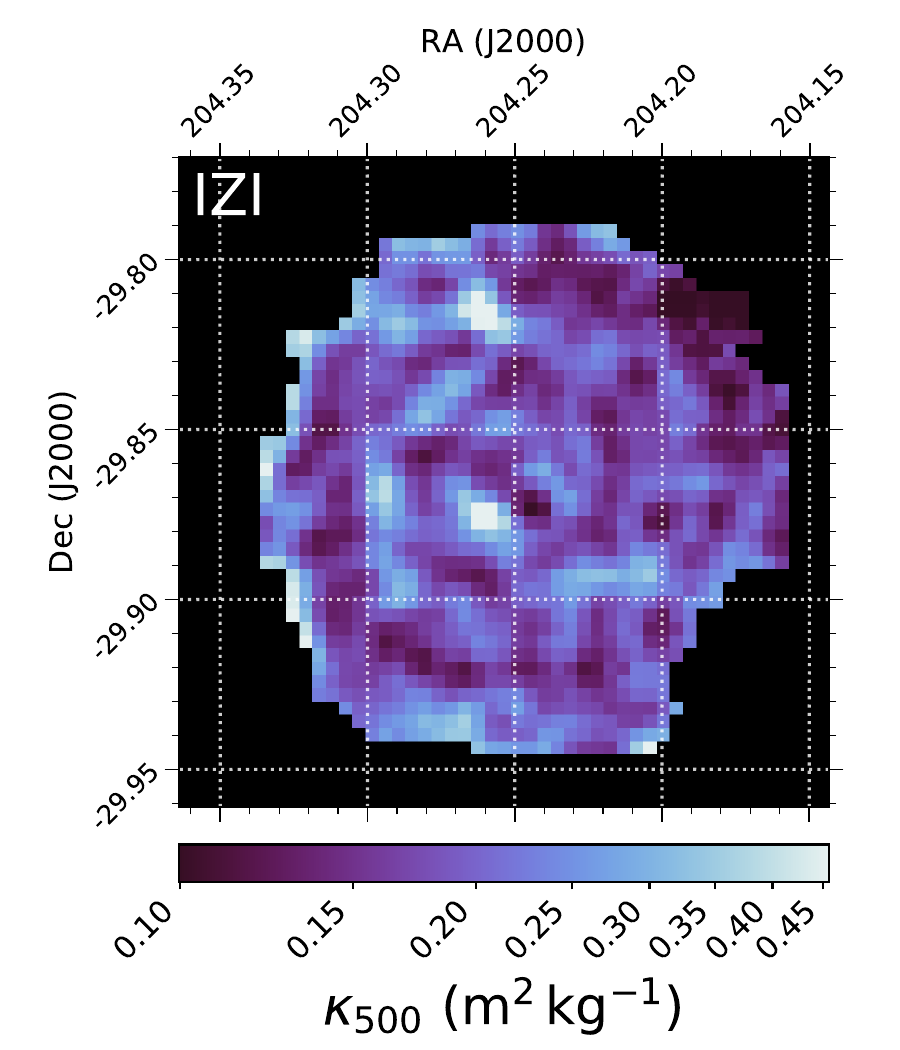}
\caption{Maps of \kappamu\ within M\,83, calculated using metallicities produced using different strong-line prescriptions, but otherwise following the same method as for our fiducial map in Figure~\ref{Fig:NGC5236_Kappa_Map}. Prescription descriptions the same as for Figure~\ref{AppendixFig:NGC0628_Kappa_Z_Variations}.}
\label{AppendixFig:NGC5236_Kappa_Z_Variations}

\end{figure}

As described in Section~\ref{Subsection:Metallicity_Data}, our metallicity maps were produced using metallicities calculated using the `S' strong-line prescription of \citet{Pilyugin2016A}. To ensure that our specific choice of metallicity prescription wasn't driving our results, we also repeated our \kappamu\ mapping using metallicity maps produced using 4 other strong-line prescriptions; the O3N2 prescription of \citet{Pettini2004B}, the N2 prescription of \citet{Pettini2004B}, the prescription of \citet{Tremonti2004A}, and the IZI prescription of \citet{Blanc2015A}. As with our fiducial \citet{Pilyugin2016A} `S' prescription values, these metallicities are all taken from the standardised database produced by \citet{DeVis2019B}. The resulting \kappamu\ maps for all 4 prescriptions for both galaxies are presented in Figures~\ref{AppendixFig:NGC0628_Kappa_Z_Variations} and \ref{AppendixFig:NGC5236_Kappa_Z_Variations}. These \kappamu\ maps all display the same general morphology as the fiducial maps in Figures~\ref{Fig:NGC0628_Kappa_Map} and \ref{Fig:NGC5236_Kappa_Map} -- with lower values of \kappamu\ associated with regions of denser ISM. The exceptions to this are the maps produced using the  \citet{Tremonti2004A} prescription, which causes a negative radial gradient in \kappamu\ to dominate over the density-anticorrelated variations; but nonetheless, at a given radius, areas of lowest \kappamu\ are associated with the same areas of denser ISM as seen in the other maps.

\section{\kappad\ Maps from Two-Component MBB SEDs} \label{AppendixSection:Kappa_2MBB}

\begin{figure}
\centering
\includegraphics[width=0.23\textwidth]{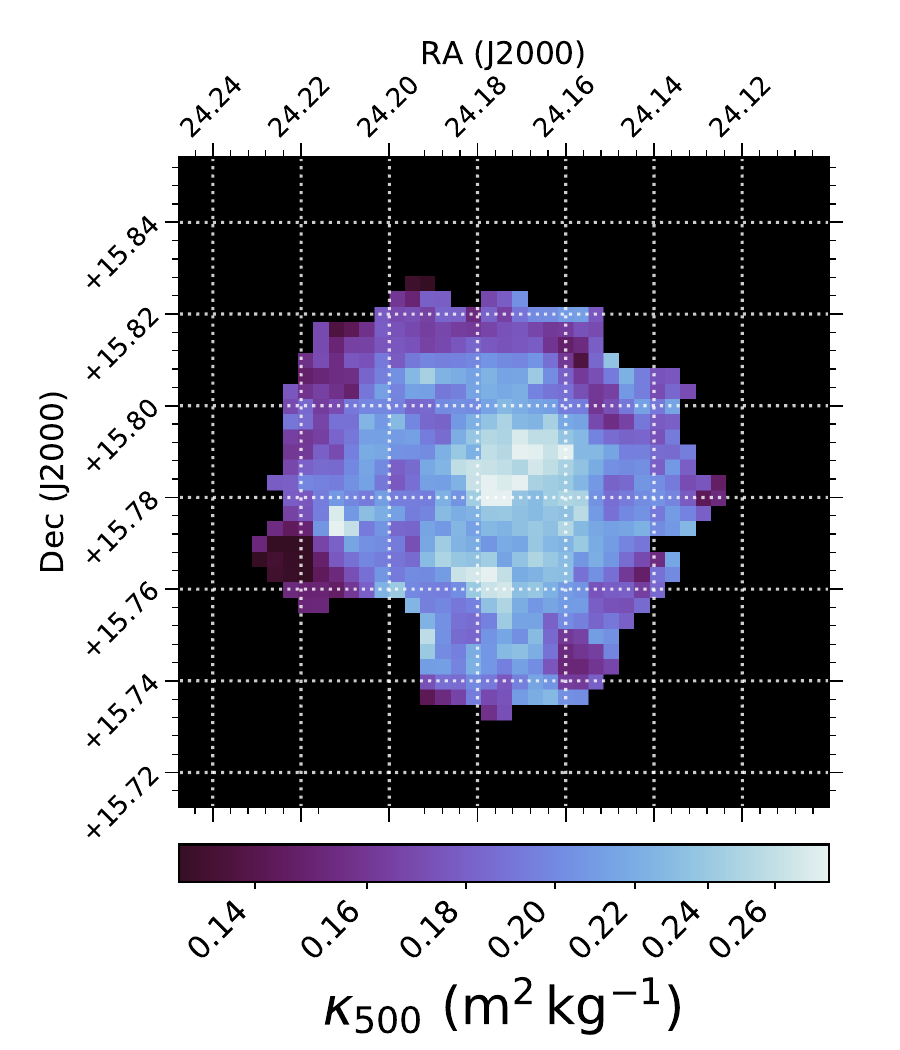}
\includegraphics[width=0.23\textwidth]{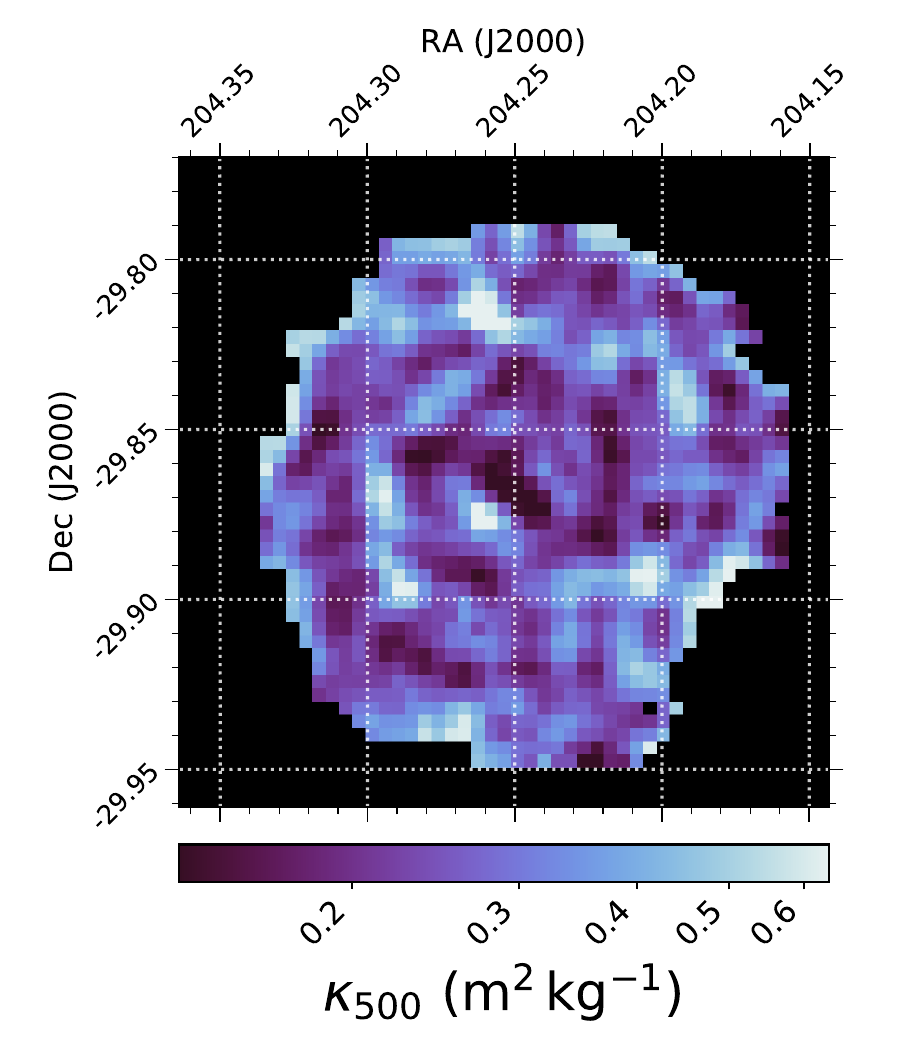}
\caption{Maps of \kappamu\ within M\,74 ({\it left}) and M\,83 ({\it right}), produced when the FIR--submm SED is modelled with a two-component MBB, as opposed to the one-component MBB used for our fiducial maps.}
\label{AppendixFig:Kappa_2MBB}
\end{figure}

As discussed in Section~\ref{Subsection:SED_Fitting}, we opt to use a one-component MBB model to fit the FIR--submm SEDs for our fiducial \kappamu\ maps. However, as a test, we also produced \kappamu\ maps using a two-component MBB model for the SED fitting. 
In practice, this entailed replacing Equation~\ref{Equation:SED_Dust_Flux} with:
\begin{equation}
S_{d_{i}} = \frac{\kappa_{0}}{D^{2}} \left( \frac{\lambda_{0}}{\lambda_{i}}\right)^{\beta} \left( M_{c}^{\rm (norm)} B(\lambda_{i},T_{c}) + M_{w}^{\rm (norm)} B(\lambda_{i},T_{w}) \right)
\label{AppendixEquation:2MBB_SED_Dust_Flux}
\end{equation}

where subscripts $c$ and $w$ denote the cold and warm dust components respectively. There are therefore 6 free parameters for the two-component MBB modelling: $T_{c}$, $M_{c}^{\rm (norm)}$,  $T_{w}$, $M_{w}^{\rm (norm)}$, $\beta$, and $\upsilon_{\rm SPIRE}$. Having performed this SED fitting,  computing the corresponding values of \kappamu\  simply requires setting $n = 2$ in Equation~\ref{Equation:Kappa} and providing $T$ and $S_{\lambda}$ for both MBB components.

Expanding the method to incorporate two dust components includes the tacit assumption that both dust components have the same dust-to-metals ratio. This is perhaps unlikely, as warmer dust will generally be associated with recent star formation and more intense ISRFs, where shocks and high-energy photons might destroy grains and return their metals to the gas phase. However, for dust SEDs with two distinct components at different temperatures, the total dust mass is invariably dominated by the colder component \citep{DaCunha2008A,Kirkpatrick2014B,CJRClark2015A}; therefore the resulting value of \kappad\ will primarily reflect the \kappad\ of the dominant component, insulating this approach against differences in \epsilond. 

The resulting maps are shown in Figure~\ref{AppendixFig:Kappa_2MBB}. The map for M\,74 shows some increase in \kappamu\ in the centre relative to the one-component MBB approach, whilst the map for M\,83 is practically identical.

The median for M\,74 is \kappamu\ = 0.20\,${\rm m^{2}\,kg^{-1}}$, and the median for M\,83 is \kappamu\ = 0.25\,${\rm m^{2}\,kg^{-1}}$. The ranges of values (estimated via same the non-parametric independent-pixel bootstrap method as used in Section~\ref{Section:Results}) are 0.13--0.28\,${\rm m^{2}\,kg^{-1}}$ for M\,74 (a factor of 2.2 variation), and  0.12--0.72\,${\rm m^{2}\,kg^{-1}}$ for M\,83 (a factor of 6.0 variation). The differences between these values and their counterparts for our fiducial maps are all much less than the average 0.15\,dex statistical uncertainty on each pixel's \kappamu\ value (and well within the 0.2\,dex systematic uncertainty).

\end{appendix}

\end{document}